\documentclass[11pt,a4paper]{article}
\pdfoutput=1
\usepackage{jheppub}
\usepackage{array}
\usepackage{amsmath}
\usepackage{epsfig}
\usepackage{amssymb}
\usepackage{graphics}
\usepackage[active]{srcltx}
\usepackage{epstopdf}
\usepackage{pdfsync}
\usepackage{shuffle}
\usepackage{color}
\usepackage[inline]{enumitem}
\usepackage{multirow}
\usepackage{subfig}
\usepackage{comment}

\usepackage[T1]{fontenc} 
\usepackage{tikz}
\usepackage{adjustbox}
\usetikzlibrary{calc,arrows.meta, decorations.markings,decorations.pathmorphing, backgrounds, positioning, fit, petri, automata,intersections}

\newcommand{\bea}{\begin{eqnarray}}
\newcommand{\eea}{\end{eqnarray}}
\newcommand{\bean}{\begin{eqnarray*}}
\newcommand{\eean}{\end{eqnarray*}}
\newcommand{\nn}{\nonumber \\}

\def\braket#1{\left\langle #1 \right\rangle}

\def\eref#1{(\ref{#1})}

\def\a{{\alpha}}

\def\b{{\beta}}

\def\vev{\braket}

\def\Spaa{\vev}

\def\Label#1{\label{#1}%
  \smash{\hbox to0pt{\raise1ex\hbox{\tiny[#1]}\hss}}}

\def\PT{{\mbox{PT}}}

\title{Permutation in the CHY Formulation}

\author[a]{Rijun Huang,}
\author[b]{Fei Teng\footnote{The corresponding author.}}
\author[c,d]{and Bo Feng\footnote{The unusual ordering of authors instead of the
standard alphabet ordering is for young researchers to get proper recognition of
contributions under the current out-dated practice in China. }}

\affiliation[a]{Institute of Theoretical Physics, School of Physics and Technology, Nanjing Normal University,\\
 No.1 Wenyuan Road, Nanjing 210023, P. R. China.}
\affiliation[b]{Department of Physics and Astronomy, Uppsala University,\\
 Box 516, SE-751 20 Uppsala, Sweden.}
\affiliation[c]{Zhejiang Institute of Modern Physics, Department of Physics,
 Zhejiang University,\\
 No.38 Zheda Road, Hangzhou 310027, P.R. China.}
\affiliation[d]{Center of Mathematical Science,
  Zhejiang University,\\
  No.38 Zheda Road, Hangzhou 310027, P.R. China.}

\emailAdd{huang@nbi.dk}
\emailAdd{fei.teng@physics.uu.se}
\emailAdd{fengbo@zju.edu.cn}

\abstract{The CHY-integrand of bi-adjoint cubic scalar theory is a product of two
PT-factors. This pair of PT-factors can be interpreted as defining a permutation.
We introduce the cycle representation of permutation in this paper for
the understanding of cubic scalar amplitude. We show that, given a permutation
related to the pair of PT-factors, the pole and vertex information of Feynman diagrams
of corresponding CHY-integrand is completely characterized by the cycle representation of permutation. Inversely, we also show that, given a set of Feynman diagrams, the cycle representation of corresponding PT-factor
can be recursively constructed. In this sense, there exists a deep connection between cycles of a permutation and amplitude. Based on these results, we have investigated the relations among different independent pairs of PT-factors in the context of cycle representation as well as the multiplication of cross-ratio factors.}


\begin{document}
\maketitle \flushbottom

\section{Introduction}
\label{secIntroduction}

It was first proposed by Cachazo, He and Yuan (CHY)~\cite{Cachazo:2013iaa,Cachazo:2013gna,Cachazo:2013hca,Cachazo:2013iea} that the tree-level scattering amplitudes of many massless theories are supported by the solutions to the scattering equations
\begin{equation}
\label{eq:SE}
f_i=\sum_{j=1, j\neq i}^{n}\frac{s_{ij}}{\sigma_{ij}}=0~,~~~~~~i=1,2\ldots n~,~~~
\end{equation}
where $s_{ij}=(p_i+p_j)^2$ are Mandelstam variables, and $\sigma_{ij}=\sigma_{i}-\sigma_{j}$ are moduli space variables. The CHY formulation consists of an integrand $\mathcal{I}_n$ that specifies the theory, and a measure on the moduli space that fully localizes the integration to the solutions of the scattering equations~\eqref{eq:SE}:
\begin{equation}
\int d\mu_{\text{CHY}}=\int (\sigma_{rs}\sigma_{st}\sigma_{tr})^2\prod_{i\neq r,s,t}d\sigma_i\delta(f_i)~.~~~
\end{equation}
The moduli space integration indicates that the CHY formulation should appear as a certain limit of the string amplitudes, which effectively reduces the path ordered string measure into the color ordered CHY measure. It has been shown that the CHY formulation naturally emerges as the infinite tension limit of ambitwistor strings~\cite{Mason:2013sva,Geyer:2014fka,Casali:2016atr}, chiral strings~\cite{Siegel:2015axg,Li:2017emw} and pure spinor formalism of superstrings~\cite{Berkovits:2013xba,Gomez:2013wza}. In the context of conventional string theory, the CHY-formulation can also appear as the zero tension limit of an alternative dual model~\cite{Bjerrum-Bohr:2014qwa}.

The scattering equations~\eqref{eq:SE} have $(n-3)!$ solutions, and to obtain the amplitudes, naively one needs to get all the solutions and sum up their contributions. However, solving the equations becomes computationally unavailable as the number of particles grows. This difficulty can be circumvented by using integration rules~\cite{Cachazo:2015nwa,Baadsgaard:2015voa,Baadsgaard:2015ifa,Baadsgaard:2015hia}. The idea behind this approach is that one can obtain the sum over algebraic combinations of the solutions in terms of the coefficients of the original polynomial equations without knowing each individual solution. Using the original integration rules, we can extract the correct amplitudes of those theories without the appearance of higher order poles, for example, the bi-adjoint scalar theory whose integrand consists of two Parke-Taylor (PT) factors,
\begin{align}
\label{eq:PTfactor}
\mathcal{I}_n=\PT(\pmb\alpha)\times\PT(\pmb\beta)&=\langle\alpha_1\alpha_2\cdots\alpha_n\rangle\times\langle\beta_1\beta_2\cdots\beta_n\rangle\nonumber\\
&=\frac{1}{\sigma_{\alpha_1\alpha_2}\sigma_{\alpha_2\alpha_3}\cdots\sigma_{\alpha_{n}\alpha_1}}\times\frac{1}{\sigma_{\beta_1\beta_2}\sigma_{\beta_2\beta_3}\cdots\sigma_{\beta_n\beta_1}}~,~~~
\end{align}
where $\pmb\alpha=\{\alpha_1,\ldots,\alpha_n\}$ and $\pmb\beta=\{\beta_1,\ldots,\beta_n\}$ are two permutations of external particles. On the other hand, theories with more complicated integrands usually involve (spurious) higher order poles. For example, by expanding the Yang-Mills integrand following the way of Lam and Yao~\cite{Lam:2016tlk}, one has to develop various techniques to evaluate the higher order poles and show that they indeed cancel towards the end~\cite{Huang:2016zzb,Cardona:2016gon,Bjerrum-Bohr:2016axv,Huang:2017ydz,Zhou:2017mfj}. Alternatively, one can expand the integrand in terms of linear combinations of bi-adjoint scalar ones with local coefficients, such that the calculation of higher order poles can be avoided. This approach has succeeded in Yang-Mills, Yang-Mills-scalar and nonlinear sigma model~\cite{Stieberger:2016lng,Nandan:2016pya,Fu:2017uzt,Teng:2017tbo,Du:2017kpo,Du:2017gnh}. The latter approach has an extra benefit that the expansion coefficients are automatically the Bern-Carrasco-Johansson numerators~\cite{Bern:2008qj,Bern:2010ue,Cachazo:2013iea}. It is not surprising that the amplitudes of various theories land on the bi-adjoint scalar ones after a recursive expansion. The reason is that these bi-adjoint scalar amplitudes capture exactly the physical poles associated with various diagrams that are planar under certain color ordering, while different theories just dress these diagrams with different kinematic numerators.

The above discussion shows the fundamental role of the bi-adjoint cubic scalar amplitudes in the understanding of CHY-formulation, so it is not surprising that many different approaches have been proposed towards the evaluation and better understanding of CHY-integrand with product of two PT-factors. While many of them focus on the rational function of complex variables $\sigma_i$'s, in paper~\cite{Arkani-Hamed:2017mur} the authors have related the PT-factors to the partial triangulations of a polygon with $n$ edges ($n$-gon), and the PT-factors are connected to the associahedron in a profound way. It also brings permutation group $S_n$ into the story, since each PT-factor is accompanied by a color trace with a definite ordering. We should find one-to-one correspondence between action of $S_n$ onto the PT-factors and partial triangulations of $n$-gon.
It would be a very natural idea to understand the CHY-integrand from the knowledge of permutations. Certain progress along this direction has been made in~\cite{Lam:2015sqb} by investigating pairing of external legs, whose results are presented in terms of illustrative objects like \emph{crystal} and \emph{defect}. In fact, for the two PT-factors in the CHY-integrand of bi-adjoint cubic scalar theory, if we set one PT-factor as the natural ordering $\Spaa{12\cdots n}$, corresponding to identity element in the permutation group, then the other PT-factor can be interpreted as a permutation acting on the identity. The physical information, i.e., the poles and vertices of the Feynman diagrams that this CHY-integrand evaluates to, should find its clue in the structure of permutations. 

In this note, we show that the aforementioned constructions of bi-adjoint scalar amplitudes have a unified description in terms of cycle structure of permutation group. We first demonstrate how the physical information is encoded in the so-called V-type and P-type cycle representations of a given permutation. We then explore how the relations between different PT-factors emerge from operations like merging and splitting of cycles. In terms of Feynman diagram, this corresponds to fixing/unfixing a certain pole. We also show how the same pattern arise from the use of cross-ratio factors. Thus not only does cycle representations provides a new efficient way to evaluate amplitudes, it also makes manifest the mathematical correspondence between amplitudes and cycle structures.



This paper is organized as follows. In~\S \ref{secSetup}, we set our convention and provide some necessary backgrounds. In \S \ref{secFeynman}, we show how the structure of those Feynman diagrams produced by a CHY-integrand can be extracted from the cycle representations of the corresponding PT-factor viewed as a permutation. In \S\ref{secPermutation}, we study the inverse problem on how to write out the PT-factor for an arbitrary given Feynman diagram. In the form of cycle representation, we propose a recursive method to construct an $n$-point PT-factor recursively from lower point PT-factors. In \S\ref{secRelation}, we investigate the relations among different PT-factors via the merging and splitting of cycle representations, as well as via multiplying cross-ratio factors. Conclusion is presented in \S\ref{secConclusion}, and in Appendix \ref{secAppendix}, we comment on an interesting interplay between the associahedron and cycle representations of permutation.

\section{The setup}
\label{secSetup}

In this section, we give the definitions of some important objects to be used in later.

\subsection{The canonical PT-factor}
\label{secSetup1}

Since the $2n$ PT-factors obtained by acting cyclic rotations and reflections evaluate to the same amplitude, despite an overall sign $(-1)^n$ for the latter case, all these $2n$ PT-factors form an \emph{equivalent class}.
Thus the number of
independent PT-factors is $n!/(2 n)$. We can represent each independent PT-factors by a \emph{canonical} ordering $\Spaa{\a_1 \a_2\cdots \a_n}$ that satisfy two conditions: \begin{enumerate*}[label=(\arabic*)]
	\item the first element $\a_1$ is fixed to be $1$ to eliminate the cyclic ambiguity,
	\item  the second element $\a_2$ should be smaller than the last element $\a_n$ to eliminate the reversing ambiguity.
\end{enumerate*}
The complete equivalent class can be generated from these independent PT-factors by acting the cyclic rotation and reversing.
For example, up to $n=5$, we can choose the independent PT-factors as follows,
\begin{align}
& n=3 & &\#=\frac{n!}{2n}=1: & &\Spaa{123}~,~~~\nn
& n=4 & &\#=\frac{n!}{2n}=3: & &\Spaa{1234}~~,~~\Spaa{1243}~~,~~\Spaa{1324}~,~~~\nn
& n=5 & &\#=\frac{n!}{2n}=12:& &\Spaa{12345}~~,~~\Spaa{12354}~~,~~\Spaa{12435}~~,~~\Spaa{12453}~~,~~\Spaa{12534}~~,~~\Spaa{12543}~,~~~\nn
& & & & &\Spaa{13245}~~,~~\Spaa{13254}~~,~~\Spaa{13425}~~,~~\Spaa{13524}~~,~~\Spaa{14235}~~,~~\Spaa{14325}~.~~~\nonumber
\end{align}
%
%
%
%
The CHY-integrand for bi-adjoint cubic scalar amplitudes is given by $\PT(\pmb{\a})\times \PT(\pmb{\b})$, as shown in Eq.~\eqref{eq:PTfactor}. A simultaneous permutation acting on $\pmb\a$ and $\pmb\b$ merely leads to the same result up to a relabeling of external legs. Hence we can fix one of the PT-factors to be the natural ordering $\PT(\pmb{\a})=\Spaa{12\cdots(n-1) n}$, and consider the other PT-factor $\PT(\pmb{\b})$ as a permutation acting on $\pmb{\a}$. Thus all the dynamical information is encoded in $\PT(\pmb{\b})$, from which one can read out the amplitude.

\subsection{Permutation, cycle representation and PT-factor}
\label{secSetup2}

Permutations, as group elements of the $n$-point symmetric group $S_n$, can be defined by their action onto the space spanned by the elements of $S_n$ themselves, for example,
\bea
\label{eq:groupaction}
\pmb\beta|\pmb{e}\rangle=|\pmb\beta\rangle ~~~,~~~ \pmb{\gamma\beta}|\pmb{e}\rangle=\pmb{\gamma}|\pmb\beta\rangle=|\pmb{\gamma\beta}\rangle~,~~~
\eea
where $\pmb\beta$ and $\pmb\gamma$ are two generic elements of $S_n$, and $\pmb{e}$ is the identity element defined as the natural ordering $\{1,2,\ldots, n\}$. Each permutation can be represented by a product of disjoint cycles $(i_1i_2\cdots i_s)$, which stands for the map $i_1\mapsto i_2$, $i_2\mapsto i_3$, $\ldots$, $i_s\mapsto i_1$. For example,
\begin{equation}
\label{eq:action123}
(123)(4)(5)\cdots(n)|1234\cdots n\rangle=|2314\cdots n\rangle
\end{equation}
stands for the permutation in Cauchy's two-line notation
\bea \left(\begin{array}{cccccc}
	1 & 2 & 3 & 4 & \cdots & n \\
	2 & 3 & 1 & 4 & \cdots  & n
\end{array}\right)~.
\eea
Cycles are defined up to a cyclic ordering, for example, $(123)=(231)=(312)$ gives the same permutation. It is also obvious that two disjoint cycles commute, i.e., $(123)(45)=(45)(123)$. Each permutation has a unique decomposition in terms of disjoint cycles, modulo the cyclicity of each cycle and the ordering of disjoint cycles.\footnote{In this statement, \emph{disjoint} is crucial to the uniqueness. Otherwise, we could have other decompositions like $(1)(2)=(12)(12)$, etc.} We call this unique decomposition the \emph{cycle representation} of a permutation in the rest of this paper. The number of disjoint cycles in a cycle representation is called the length of this cycle representation, and the number of elements in a cycle is called the length of cycle.


In our CHY-integrand~\eqref{eq:PTfactor}, we can treat $\PT(\pmb\b)=\langle \b_1\b_2\cdots \b_n\rangle$ also as a permutation. The equivalent class of $\PT(\pmb\beta)$ consists of all the $2n$ permutations obtained by cyclic rotations and reversing,
%
%
\bea \mathfrak{b}[\pmb\b]:=\left\{\begin{array}{l}
       \langle \beta_1\beta_2\cdots\beta_{n-1}\beta_n\rangle\,,\,\langle \beta_2\beta_3\cdots\beta_n\beta_1\rangle\,,\ldots\,,\,\langle \beta_n\beta_{1}\cdots\beta_{n-2}\beta_{n-1}\rangle \\
       \langle \beta_n \beta_{n-1}\cdots\beta_{2}\beta_1\rangle\,,\,\langle \beta_1\beta_n\cdots\beta_3\beta_2\rangle\,,\ldots\,,\,\langle \beta_{n-1}\beta_{n-2}\cdots\beta_{1}\beta_{n}\rangle
     \end{array}\right\}~.~~~\label{eq:PTequivclass}
\eea
We define the cyclic generator $\pmb{g}_c$ and reversing generator $\pmb{g}_r$ respectively as
\begin{align}
\label{eq:grgc}
\pmb{g}_c=(\beta_1\beta_2\cdots\beta_n)~~~,~~~ \pmb{g}_r=\left\{\begin{array}{lll}
(\beta_1\beta_n)(\beta_2\beta_{n-1})\cdots(\beta_{\frac{n}{2}}\beta_{\frac{n+2}{2}})&\quad &\text{for~even~}n \\
(\beta_1\beta_n)(\beta_2\beta_{n-1})\cdots (\beta_{\frac{n-1}{2}}\beta_{\frac{n+3}{2}})(\beta_{\frac{n+1}{2}})&\quad &\text{for~odd~}n
\end{array}\right.
~,~~~
\end{align}
which satisfy the relation $\pmb{g}_r\pmb{g}_c=\pmb{g}_c^{-1}\pmb{g}_r$ and $\pmb{g}_c^n=\pmb{g}_r^2=\pmb{e}$. Thus they generate the $n$-point dihedral group $D_n$.
For a permutation $\pmb\beta$, the equivalent class $\mathfrak{b}[\pmb\beta]$ is thus given by,
\bea
\label{eq:equivperm}
\mathfrak{b}[\pmb\beta]=\left\{~\pmb{\beta}~,~\pmb{\beta}\pmb{g}_c~,\ldots,~ \pmb{\beta}\pmb{g}_c^{n-1}~,~\pmb{\beta}\pmb{g}_r~,~\pmb{\beta}\pmb{g}_r\pmb{g}_c~,~ \ldots,~\pmb{\beta}\pmb{g}_r\pmb{g}_c^{n-1}~\right\}~.~~~
\eea
The elements in $\mathfrak{b}$ has one-to-one correspondence to the equivalent class of PT-factors (\ref{eq:PTequivclass}). There are in all $n!/(2n)$ non-equivalent permutations. However, they do not form a group in general, since $D_n$ is not a normal subgroup of $S_n$ for $n\geqslant 4$. For permutations in the same equivalent class, of course they have different cycle representations, since after all they are different group elements of $S_n$. In the next section, we are going to show that which set of Feynman diagrams $\PT(\pmb\beta)$ corresponds to is encoded collectively in the different cycle representations of the equivalent permutations in $\mathfrak{b}$.

\section{From permutations to Feynman diagrams}
\label{secFeynman}

One of our motivations is to explore the information encoded in the PT-factors, described in the form of permutations. As mentioned in the previous section, in our setup the amplitude result is determined by the second PT-factor $\PT(\pmb\beta)$, considered to be a permutation acting on the identity element. It determines an equivalent class containing $2n$ elements evaluating to the same amplitude. Thus we need to consider all the permutations in the equivalent class. For example, when working with the CHY-integrand $\PT(\pmb \a)\times \PT(\pmb \b)=\Spaa{1234}\times \Spaa{1243}$, we need the equivalent class of $\Spaa{1243}$, containing eight elements,
\bea\Big\{\Spaa{1243}~,~\Spaa{2431}~,~\Spaa{4312}~,~\Spaa{3124}~,~\Spaa{3421}~,~\Spaa{4213}~,~\Spaa{2134}~,~\Spaa{1342}\Big\}~,~~~\label{4p-F1-1-PermuOri}\eea
or in the form of cycle representations,
\bea \Big\{(1)(2)(34)~,~(124)(3)~,~(1423)~,~(132)(4)~,~(1324)~,~(143)(2)~,~(12)(3)(4)~,~(1)(234)\Big\}~.~~~\label{4p-F1-1-Permu} \eea
Our purpose is to relate these cycle representations to the Feynman diagrams contributing to the amplitude. All the above eight cycle representations in~\eqref{4p-F1-1-Permu}
can be used to  reconstruct the PT-factor $\PT(\pmb{\b})$ by acting
them on the natural ordering $\langle 1234\rangle$, so each one encodes
the complete information for evaluation. However, they have
different structures. In Eq.~\eqref{4p-F1-1-Permu}, two cycle representations are length-$1$,
four are length-$2$, while the remaining two are length-$3$. How can we read useful
information out of these different cycle structures?

To answer this question, let us recall the integral result of the
CHY-integrand  $\Spaa{1234}\times \Spaa{1243}$, which is $-\frac{1}{s_{12}}$. It corresponds to a Feynman diagram with two cubic vertices,
one connecting the legs $1$, $2$ and the internal propagator $P_{12}:= p_1+p_2$,
while the other connecting the legs $3$, $4$, and the internal propagator
$-P_{12}$. It is very plausible to conjecture that, the two
cycle representations $(1)(2)(34)$ and $(12)(3)(4)$ in fact describe
respectively the two cubic vertices. In $(1)(2)(34)$, the two cycles $(1)$ and $(2)$ describe respectively the external legs $1$ and $2$ attached to a
vertex, while the cycle $(34)$ describes the corresponding
internal propagator of that vertex. It also indicates that $\frac{1}{s_{34}}=\frac{1}{s_{12}}$ is an internal pole. Similar analysis can be carried out
for the cycle representation $(12)(3)(4)$.

Above discussion tells us that, although each cycle representation
contains the complete information of amplitude, the pole structure
is manifest in some of them but not all. The complete picture of
Feynman diagrams is determined collectively by all
cycle representations whose pole and/or vertex structures are
manifest. With this understanding, we only need to consider those
good cycle representations, i.e., the pole and/or vertex structure
are manifest. For bi-scalar theory, the physical pole would appear
only for the external legs with consecutive ordering. So let us
define the \emph{good cycle representations} as those satisfying the following criteria:
\begin{itemize}
	\item the cycles in the considered cycle representation can be separated into at least two parts, while the union of cycles in each part is consecutive (later called \emph{planar separation}).
	\item in case that the cycle representation \emph{can only be} separated into two parts, then each part should contain at least two elements.

\end{itemize}
Moreover, if the planar separation of a good cycle representation contains at least three parts, we call it a \emph{vertex type} (V-type) cycle representation. Otherwise, we call it a \emph{pole type }(P-type) cycle representation.
Let us give a few more examples. At six point, both $(12)(34)(56)$ and $(12)(35)(46)$ are good cycle representations. The former is a V-type one since it can be separated into three parts $(12)$, $(34)$ and $(56)$, while the latter is a P-type one since it can only be separated into two parts $(12)$ and $(35)(46)$.
On the other hand, both $(14)(25)(36)$ and $(1)(23456)$ are bad cycle representations according to the above criteria, since the former one has no planar separation at all while the separated part $(1)$ in the latter contains only one element. With the above definitions, we can give an answer to the question raised at the end of the first paragraph of this section: we can reconstruct the Feynman diagrams by considering all the good cycle representations in the equivalent class of a PT-factor.

Now we illustrate how a  good cycle representation reflects the vertex and pole structure of the corresponding Feynman diagram by an eight point example $\PT(\pmb\b)=\Spaa{12846573}$, which gives four trivalent Feynman diagrams after the CHY integration,
%
\bea \frac{1}{s_{12} s_{56}s_{8123}} \left(\frac{1}{s_{812}}+\frac{1}{s_{123}}\right)\left(\frac{1}{s_{456}}+\frac{1}{s_{567}}\right)~.~~~\label{8p-bad-1-1}\eea
The result can be summarized into a single effective Feynman diagram,
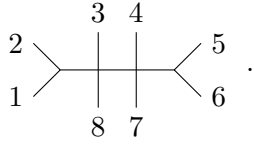
\begin{equation*}
\adjustbox{raise=0cm}{\begin{tikzpicture}
	\draw (135:0.5) node[left=0pt]{$2$} -- (0,0) -- (-135:0.5) node[left=0pt]{$1$};
	\draw (0,0) -- (1.5,0) -- ++(45:0.5) node[right=0pt]{$5$} (1.5,0) -- ++(-45:0.5) node[right=0pt]{$6$};
	\draw (0.5,-0.5) node[below=0pt]{$8$} -- (0.5,0.5) node[above=0pt]{$3$} (1,-0.5) node[below=0pt]{$7$} -- (1,0.5) node[above=0pt]{$4$};
	\node at (2.5,0) {$.$};
	\end{tikzpicture}}
\end{equation*}
There are $16$ equivalent cycle representations in $\mathfrak{b}[\pmb\b]$, collected as,
\bea \text{good}: & ~~~ & \text{V-type}:~~
(1)(2)(38)(4)(56)(7)~~,~~(12)(3)(47)(5)(6)(8)~,~~~\nn
& &
\text{P-type}:~~(132)(4875)(6)~~~,~~(128)(3467)(5)~~,~~(1)(2378)(456)~~,~~(1843)(2)(576)~,~~~\nn
\text{bad}: &~~~ &
(15274)(3)(68)~~,~~(14726)(35)(8)~~,~~(16385)(24)(7)~~,~~(17)(25836)(4)~,~~~\label{8p-bad-1-2}\\
& &
(176423)(58)~~,~~(182457)(36)~~,~~(135468)(27)~~,~~(14)(286753)~,~~~\nn
& & (1526)(3487)~~,~~(1625)(3784)~.~~~\nonumber\eea
In general, P-type cycle representations manifest certain poles contained in some of the Feynman diagrams. For example, $(132)\pmb{|}(4875)(6)$ is P-type since it can only be divided into two parts indicated by the vertical line. This separation indicates that the pole $s_{123}$ should appear in some Feynman diagrams.
Similarly, the other three P-type cycle representations correspond respectively to the pole $s_{812}$,
$s_{456}$ and $s_{567}$, as can be seen in Eq.~\eqref{8p-bad-1-1}.
In contrary, the V-type cycle representations contain both pole and vertex information. For example, the
cycle representation $(1)(2)(38)(4)(56)(7)$ allows two different planar separations,
%
\begin{subequations}
\bea \text{four parts}:&~~~ & (1)(2)(38)\pmb{|}(4)\pmb{|}(56)\pmb{|}(7)\,,\label{eq:4way}\\
\text{three parts}:&~~~ &(1)\pmb{|}(2)\pmb{|}(38)(4)(56)(7)\,.\label{eq:3way}\eea
\end{subequations}
These two separations indicate that the effective Feynman diagram
contains one quartic vertex with legs $\{P_{8123}, 4, P_{56}, 7\}$, and
one cubic vertex with legs $\{1, 2, P_{345678}\}$. Since the legs with more than one
elements also give pole information, we can read out the poles
$s_{8123}$, $s_{56}$, and $s_{12}$ from this V-type cycle representation. Similarly,
the cycle representation $(12)(3)(47)(5)(6)(8)$ gives one quartic
vertex with legs $\{P_{12}, 3, P_{4567}, 8\}$ and one cubic vertex
with legs $\{5, 6, P_{781234}\}$, and it gives the same pole structures as the previous one.
Combining these two, we do produce all the vertices in the effective Feynman diagram.

The above example shows that by collectively combining the information from all good cycle representations,
we can read out the complete Feynman diagram result. This provides one method of analysis. On the other hand, we can arrive at the complete final result by relying on only one good cycle representation, since each one should contain the complete information of PT-factor $\PT(\pmb{\b})$. Hence we should have another method of analysis.
Observations from practical computation show that,
\begin{enumerate}[label=(\Alph*)]
	
	\item All V-type cycle representations together manifest the vertex structure of the corresponding effective Feynman diagram.
	
	\item  It is also possible to reproduce the effective Feynman diagram from one V-type cycle representation if we recursively use the lower multiplicity results.
	
	\item One P-type cycle representation is not sufficient to reproduce the complete result, and in order to get the correct answer we should consider all P-type cycle representations.

\end{enumerate}
We use again the example~\eref{8p-bad-1-2} to demonstrate our observations. We start with the V-type cycle representation $(1)(2)(38)(4)(56)(7)$.
In the four-part separation~\eqref{eq:4way}, we first coarse grain the part $\{8123\}$ and $\{56\}$ by replacing them by a single propagator. This leads to an effective quartic vertex,
\begin{align}
&\PT(\pmb\b)=(1)(2)(38)\pmb{|}(4)\pmb{|}(56)\pmb{|}(7)~\rightarrow~(P_{8123})(4)(P_{56})(7)~,~~~\nonumber\\
&\PT(\pmb\a)=(8)(1)(2)(3)\pmb{|}(4)\pmb{|}(5)(6)\pmb{|}(7)~\rightarrow~(P_{8123})(4)(P_{56})(7)~,~~~\nonumber
\end{align}
which gives the contribution
\begin{equation}
	\adjustbox{raise=-0.65cm}{\begin{tikzpicture}
		\draw (-135:0.75) node[left=0pt]{$P_{8123}$} -- (45:0.75) node[right=0pt]{$P_{56}$};
		\draw (135:0.75) node[left=0pt]{$4$} -- (-45:0.75) node[right=0pt]{$7$};
		\end{tikzpicture}}=\frac{1}{s_{8123} s_{56}} \left( \frac{1}{s_{4P_{56}}}+\frac{1}{s_{P_{56} 7}}\right)=\frac{1}{s_{8123} s_{56}}\left(\frac{1}{s_{456}}+\frac{1}{s_{567}}\right)~.~~~\label{8p-bad-1-3}
\end{equation}
Next, we look into the substructures. In Eq.~\eqref{eq:4way}, the substructure $\{P_{56}, 5,6\}$ has the cycle representation $(P_{56})(56)$. In its equivalent class, we have only one good cycle representation
\begin{equation}
\label{eq:3pVertex}
	\text{V-type}:\qquad (P_{56})(5)(6)\;\Longrightarrow\;\adjustbox{raise=-0.65cm}{\begin{tikzpicture}
		\draw (-0.75,0) node[left=0pt]{$P_{56}$} -- (0,0) -- (45:0.75) node [right=0pt]{$5$} (-45:0.75) node[right=0pt]{$6$} -- (0,0);
		\end{tikzpicture}}~,~~~
\end{equation}
which gives the familiar cubic vertex. The substructure $\{P_{8123},8,1,2,3\}$ has the cycle representation $(1)(2)(38)(P_{8123})$ in Eq.~\eqref{eq:4way}. By acting on it with the cyclic generator $\pmb{g}_c=(P_{8123}8123)$ and the reversing generator $\pmb{g}_r=(P_{8123})(83)(12)$, we can reproduce all the ten permutations in the equivalent class,\footnote{We omit the subscript in the propagator $P$ when there is no possible confusion.}
\bea \text{V-type}: &~~& (P)(8)(12)(3)~,~~~  \nn
\text{P-type}: &~~ & (P8)(132)~~,~~(P3)(812)~,~~~\label{8p-bad-1-4}\\
\text{Bad}: &~~& (P)(38)(1)(2)~~,~~(P231)(8)~~,~~(P182)(3)~,~~~\nn
& & (P823)(1)~~,~~(P318)(2)~~,~~(P1)(832)~~,~~(P2)(813)~.~~~\nonumber\eea
From the V-type cycle representation $(P_{8123})(8)(12)(3)$, we see immediately
the quartic vertex structure
\begin{equation}
(P_{8123})(8)(12)(3)\;\Longrightarrow\;\adjustbox{raise=-0.65cm}{\begin{tikzpicture}
	\draw (-135:0.75) node[left=0pt]{$P_{8123}$} -- (45:0.75) node[right=0pt]{$P_{12}$};
	\draw (135:0.75) node[left=0pt]{$8$} -- (-45:0.75) node[right=0pt]{$3$};
	\end{tikzpicture}}~,~~~
\end{equation}
where the $\{P_{12},1,2\}$ part is another cubic vertex, following the analysis of~\eqref{eq:3pVertex}. Thus we get the contribution $\left(\frac{1}{s_{812}}+\frac{1}{s_{123}}\right)$. When combining with~\eref{8p-bad-1-3}, we do get the complete
result~\eref{8p-bad-1-1}. The above calculation shows that by recursively looking into the V-type cycle representations of each substructure, we can reproduce the full Feynman diagram result.

Now we move to the P-type cycle representation, for
example, the planar separation $(132)\pmb{|}(4875)(6)$. We need to analyze the two
substructures given by cycle representations $(P_{123})(132)$ and
$(P_{123})(4875)(6)$. Using the algorithm given in~\eqref{eq:equivperm}, the substructure $(P_{123})(132)$ gives the
following eight equivalent cycle representations
\bean \text{V-type}: &~~& (1)(2)(3P)~~,~~(12)(3)(P)~,~~~ \\
\text{Bad}: &~~&
(132)(P)~~,~~(1P23)~~,~~(124)(3)~~,~~(1P3)(2)~~,~~(132P)~~,~~(1)(23P)~,\eean
in which either $(1)(2)(3P_{123})$ or $(12)(3)(P_{123})$ manifests the pole structure $\frac{1}{s_{12} s_{123}}$. Similarly, the substructure
$(P_{123})(4875)(6)$ gives the following $12$ equivalent cycle representations
\bean \text{V-type}: &~~& (8P)(4)(56)(7)~~,~~(8)(P)(47)(5)(6)~,~~~ \\
\text{P-type}: &~~& (78P)(456)~~,~~(4P8)(765)~,~~~\\
\text{Bad}: &~~&
(P)(6)(4875)~~,~~(P764)(58)~~,~~(P5)(47)(68)~~,~~(P6)(4578)~~,~~(P467)(5)(8)~,~~~\\
& & (P48675)~~,~~(P6)(4)(58)(7)~~,~~(P54687)~.~~~\eean
From both the V-type cycle representation, we can read out the contribution
$\frac{1}{s_{56}s_{1238}}\left(\frac{1}{s_{456}}+\frac{1}{s_{567}}\right)$. Putting two substructures together, we get only two terms
\begin{equation*}
 \frac{1}{s_{12}s_{56}s_{1238}}\frac{1}{s_{123}}\left(\frac{1}{s_{456}}+\frac{1}{s_{567}}\right)~,
\end{equation*}
compared to the full result~\eqref{8p-bad-1-1}. One can check that only by combining with another P-type cycle representation $(128)(3476)(5)$ can we obtain the full result.

With the above example in mind, let us move on to the systematic investigation of four, five and six point PT-factors. For presentation purpose, we shall organize the independent PT-factors into categories according to the topology of corresponding Feynman diagrams. In the same category, different PT-factors are related by group actions, and can be analyzed in the same manner. Concretely, we can define the group action as follows. In the space of $\frac{n!}{2n}$ equivalent classes
$\mathfrak{b}[\pmb\b]$, we define the permutation action
\begin{equation}
\mathcal{C}(\mathfrak{b}[\pmb\b])=\mathfrak{b}\left[\left.\pmb\b\right|_{i\mapsto i+1}\right]~~~~,~~~~\mathcal{R}(\mathfrak{b}[\pmb\b])=\mathfrak{b}\left[\left.\pmb\b\right|_{i\mapsto n+1-i}\right]~,~~~
\end{equation}
where $\mathcal{C}$ and $\mathcal{R}$ also generate a dihedral group
$\mathcal{D}_n$.\footnote{We note that this $\mathcal{D}_n$ is
different from the $D_n$ defined in \S\ref{secSetup} that generates
the equivalent class $\mathfrak{b}[\pmb\b]$, since their actions on
the permutations are different.} The action of $\mathcal{D}_n$
further separates the space of $\mathfrak{b}[\pmb\b]$  into
different orbits. The number of elements inside each orbit depends
on the symmetric property of such orbit.   For example, the identity
permutation $\PT(\pmb\b)=\langle 12\cdots n\rangle$ is invariant
under $\mathcal{D}_n$ action such that it forms a one dimensional
orbit by itself. A nontrivial example is that by acting
$\mathcal{D}_n$ onto $\PT(\pmb\b)=\langle 12846573\rangle$, we get
an orbit with four elements:
\begin{equation}
\langle 12846573\rangle~~~,~~~\langle 13248675\rangle~~~,~~~\langle 15342687\rangle~~~,~~~\langle 17354628\rangle~.
\end{equation}
The above discussion is useful because, all PT-factors in the same
orbit of $\mathcal{D}_n$ share the same structure of the cycle
representations. Most importantly, starting from the V-type and
P-type cycle representations of one PT-factor in the orbit, we can get the V-type and P-type cycle representations of
all the other PT-factors in this orbit simply by the
mapping $i\mapsto i+1$ or $i\mapsto n+1-i$. Later on we will only study one PT-factor for each orbit.

After above general discussion, we present more example to further elaborate our algorithm. In appendix~\ref{secAppendix}, we will give some further discussions on the cycle structure of PT-factors and Feynman diagrams.

\subsection{The three and four point cases}
\label{secFeynmanSub1}

At three point, there is only one independent PT-factor $\PT(\pmb{\beta})=\Spaa{123}$. Among the six
cycle representations of the equivalent class, only $(1)(2)(3)$ is a
good one. The planar separation $(1)\pmb{|}(2)\pmb{|}(3)$ indicates
pictorially that the three external legs are attached to a single cubic
vertex. This Feynman diagram evaluates to $1$, which agrees with the CHY integration result.

At four point, there are three independent PT-factors $\PT(\pmb{\beta})$. Each of them corresponds to an equivalent class with $8$ permutations. Their good cycle representations are summarized in the following table as
\begin{center}
	\begin{tabular}{|c|c|c|}
		\hline
		$\PT(\pmb{\b})$ & V-type & P-type \\  \hline
		$\Spaa{1234}$  &  ~~$(1)(2)(3)(4)$~~& ~~$(41)(23)$~~,~~$(12)(34)$~~ \\ \hline
		$\Spaa{1243}$  &   ~~$(1)(2)(34)$~~,~~$(12)(3)(4)$~~ & \\ \hline
		$\Spaa{1324}$  &  ~~$(4)(1)(23)$~~,~~$(41)(2)(3)$~~ & \\
		\hline
	\end{tabular}~~~~\label{4point-Table}
\end{center}
For the PT-factor $\Spaa{1234}$,  we can read out the complete vertex information from the sole V-type cycle representation $(1)(2)(3)(4)$.
This PT-factor gives all trivalent four point Feynman diagrams whose four
external legs are connected at cubic vertices respecting the color ordering, and the result is simply $\frac{1}{s_{12}}+\frac{1}{s_{23}}$. In the language of planar separation, the V-type cycle representation can be separated into four
parts $(1)\pmb{|}(2)\pmb{|}(3)\pmb{|}(4)$, which can be explained as defining an effective quartic vertex with exactly the same meaning as mentioned above. This structure will be one of the building blocks for the analysis of higher point Feynman diagrams.

However,  the two P-type cycle representations alone only provide partial result. For example, the planar separation of
cycle representation $(41)\pmb{|}(23)$ indicates that, legs $2$ and $3$ are
connected to the same cubic vertex while legs $4$ and $1$ are connected to another, resulting in a contribution of $\frac{1}{s_{23}}$. This is half of the complete answer, and the remaining
part is given by the other P-type cycle representation $(12)\pmb{|}(34)$, leading to $\frac{1}{s_{12}}$.

The other two PT-factors $\Spaa{1243}$ and $\Spaa{1324}$ are related
through $i\mapsto i+1$, i.e.,  $\Spaa{1243}\mapsto
\Spaa{2314}=\Spaa{1324}$. In other words, they belong to the same
orbit under $\mathcal{D}_4$ action. Thus by knowing one, we can
obtain the other just by relabeling. For $\Spaa{1243}$, the V-type
cycle representation with planar separation
$(1)\pmb{|}(2)\pmb{|}(34)$ indicates a cubic vertex with three legs
$\{1, 2 ,P_{34}\}$, while $(12)\pmb{|}(3)\pmb{|}(4)$ indicates
the other cubic vertex with legs $\{P_{12}, 3 ,4\}$. Putting them
together, we get the $s$-channel Feynman diagram evaluated to
$\frac{1}{s_{12}}$. Alternatively, we can use just one V-type cycle
representation to reproduce the complete result. For example,
$(1)(2)(34)$ indicates a cubic vertex represented by
$(1)(2)(P_{34})$, and a three point substructure $(P_{34})(34)$.
Then by using the three point result, this substructure is nothing
but another cubic vertex $(P_{34})(3)(4)$. Thus $(1)(2)(34)$ indeed
gives the $s$-channel diagram.
For the V-type cycle representation $(12)(3)(4)$, the analysis is exactly the same, and we can show that both V-type cycle representations give the same answer.

\subsection{The five point case}
\label{secFeynmanSub2}

For five point case, there are $\frac{5!}{10}=12$ independent
PT-factors $\PT(\pmb{\b})$. They can be divided into following four
categories,
\begin{enumerate}[label=(\arabic*)]
	
	\item The PT-factor $\Spaa{12345}$ has only one  V-type cycle representation
	$(1)(2)(3)(4)(5)$ and five P-type cycle representations
	$(15)(24)(3)$, $(1)(25)(34)$, $(12)(35)(4)$, $(13)(2)(45)$ and
	$(14)(5)(23)$. The V-type one corresponds to the Feynman
	diagrams where the legs $\{1,2, 3, 4 ,5\}$ form all possible Feynman diagrams connected by cubic vertices, while respecting the color ordering. The result is simply
	\bea \adjustbox{raise=-0.9cm}{\begin{tikzpicture}
		\draw (0,0) -- (0,0.75) node[above=0pt]{$2$} (0,0) -- (18:0.75) node[right=0pt]{$3$} (0,0) -- (162:0.75) node[left=0pt]{$1$};
		\draw (0,0) -- (-54:0.75) node[right=0pt]{$4$} (0,0) -- (-126:0.75) node[left=0pt]{$5$};
		\end{tikzpicture}}\;\Longrightarrow\;\frac{1}{s_{12}s_{34}} +\frac{1}{s_{23}s_{45}}+\frac{1}{s_{34}s_{51}}+ \frac{1}{s_{45}s_{12}}+\frac{1}{s_{51}s_{23}}~.~~~\eea
    In the language of planar separations, $(1)\pmb{|}(2)\pmb{|}(3)\pmb{|}(4)\pmb{|}(5)$ indicates an
	effective five point vertex with the same meaning. As a comparison, each P-type cycle representation corresponds to only two trivalent Feynman diagrams. For example, the planar separation $(15)\pmb{|}(24)(3)$
	fixes the pole $s_{15}$. The substructure $(P_{15})(24)(3)$ has a V-type cycle representation $(P)(2)(3)(4)$ in its equivalent class, indicating an effective quartic vertex, which gives two trivalent Feynman diagrams.
	Only after combining all the five P-type cycle representations do we get the complete answer, where each Feynman diagram appears twice.
	
    \item The following five PT-factors $\Spaa{12354}$,
    $\Spaa{12435}$, $\Spaa{12543}$, $\Spaa{13245}$ and
    $\Spaa{14325}$ form an orbit under $\mathcal{D}_5$ action,
    and are related by the cyclic permutation $i\mapsto i+1$,
    for example, $\Spaa{12354}\mapsto
    \Spaa{23415}=-\Spaa{14325}$.\footnote{One can easily check
    that the result of $i\mapsto n+1-i$ is also in this orbit.}
    Thus we only need to analyze one of them, say,
    $\Spaa{12354}$. In its equivalent class, There are two
    V-type cycle representations $(1)(2)(3)(45)$ and
    $(13)(2)(4)(5)$, together with two P-type cycle
    representations $(12)(345)$ and $(154)(23)$. For the first
    V-type cycle representation, the planar separation
    $(1)\pmb{|}(2)\pmb{|}(3)\pmb{|}(45)$ indicates an effective
    quartic vertex, labeled as $V_1$ in Eq.~\eqref{eq:5pF2},
    while for the second V-type cycle representation, the planar
    separation $(13)(2)\pmb{|}(4)\pmb{|}(5)$ indicates a cubic
    vertex labeled as $V_2$ in Eq.~\eqref{eq:5pF2}. Putting them
    together, we do reproduce the unique effective Feynman
    diagram with five external legs,
	\begin{equation}
	\label{eq:5pF2}
	\adjustbox{raise=-1.3cm}{\begin{tikzpicture}
		\draw (0,0) node[left=0pt]{$2$} -- (2,0) -- ++(45:1) node[right=0pt]{$4$} (2,0) -- ++(-45:1) node[right=0pt]{$5$};
		\draw (1,-1) node[below=0pt]{$1$} -- (1,1) node[above=0pt]{$3$};
		\filldraw (1,0) circle (2pt) node[below left=0pt]{$V_1$};
		\filldraw (2,0) circle (2pt) node[below left=0pt]{$V_2$};
		\end{tikzpicture}}\;\Longrightarrow\;\frac{1}{s_{45}}\left(\frac{1}{s_{12}}+\frac{1}{s_{23}}\right)~.~~~
	\end{equation}	
Alternatively, we can obtain the same answer by using
	only one V-type cycle representation, for instance the planar separation $(1)\pmb{|}(2)\pmb{|}(3)\pmb{|}(45)$. We first replace the cycle $(45)$ by the propagator $(P_{45})$. This gives the effective quartic vertex $V_1$ marked in Eq.~\eqref{eq:5pF2} and a three point substructure $(P_{45})(45)$. Then following the three point analysis presented at the beginning of \S\ref{secFeynmanSub1}, we reproduce the vertex $V_2$ in Eq.~\eqref{eq:5pF2}. The complete result can be arrived by combining the two vertices along the propagator $P_{45}$, which is just Eq.~\eqref{eq:5pF2}.

    \item The following five PT-factors $\Spaa{12453}$,
    $\Spaa{12534}$, $\Spaa{13254}$, $\Spaa{13425}$ and
    $\Spaa{14235}$ form another orbit under $\mathcal{D}_5$
    action, and are related by the cyclic permutation $i\mapsto
    i+1$.
    As before, we only consider the PT-factor $\Spaa{12453}$ as example. It
    contains three V-type cycle representations, and the planar separations $(1)\pmb{|}(2)\pmb{|}(345)$, $(12)\pmb{|}(3)\pmb{|}(45)$ and
    $(321)\pmb{|}(4)\pmb{|}(5)$ manifest three cubic vertices.
    After combining them together, we get the effective Feynman diagram
	\begin{equation}
	\adjustbox{raise=-0.65cm}{\begin{tikzpicture}
		\draw (135:0.75) node[left=0pt]{$2$} -- (0,0) -- (-135:0.75) node[left=0pt]{$1$} (0,0) -- (1.5,0) -- ++(45:0.75) node[right=0pt]{$4$} (1.5,0) -- ++(-45:0.75) node[right=0pt]{$5$};
		\draw (0.75,0) -- (0.75,0.75) node[above=0pt]{$3$};
		\filldraw (0,0) circle (2pt) node[below=1pt]{$V_1$} (0.75,0) circle (2pt) node[below=1pt]{$V_2$} (1.5,0) circle (2pt) node[below=1pt]{$V_3$};
		\end{tikzpicture}}\;\Longrightarrow\;\frac{1}{s_{12}s_{45}}~.~~~
	\label{5point-re1}
	\end{equation}
	Next, we show how to reproduce above result by using only one V-type cycle representation, for example, $(12)(3)(45)$. The planar separation $(12)\pmb{|}(3)\pmb{|}(45)$ indicates a cubic vertex $(P_{12})(3)(P_{45})$, which is the vertex $V_2$ in~\eqref{5point-re1}. We also get two three point substructures $(P_{12})(12)$ and $(P_{45})(45)$, leading to the vertex $V_1$ and $V_3$ respectively in~\eqref{5point-re1}. The Feynman diagram thus contains only cubic vertices, and there are exactly two internal propagators $P_{12}$ and $P_{45}$, evaluating to $\frac{1}{s_{12}s_{45}}$. Also, the same result can be obtained by using the planar separation $(1)\pmb{|}(2)\pmb{|}(345)$, where the vertex $V_1$ in~\eqref{5point-re1} is manifest. For the substructure $(P_{345})(345)$, we need to look into its equivalent class. Now using~\eqref{eq:equivperm}, we find two V-type cycle representations $(P)(3)(45)$ and $(P3)(4)(5)$. According to the four point analysis in \S\ref{secFeynmanSub1}, they both lead to a pole $\frac{1}{s_{45}}$. Thus we get again the result $\frac{1}{s_{12}s_{45}}$.

	\item The cycle representations for the last PT-factor $\Spaa{13524}$ are
\bea \begin{array}{l}
       (1)(2354)~~,~~(2)(3415)~~,~~(3)(4521)~~,~~(4)(5132)~~,~~(5)(1243)~,~~~ \\
       (1)(2453)~~,~~(2)(3514)~~,~~(3)(4125)~~,~~(4)(5231)~~,~~(5)(1342)~.~~~
     \end{array}~~~\label{eq:5p0F}
\eea
There is no good cycle representation at all, so the contribution is zero, which
is indeed the case.

\end{enumerate}
%

\subsection{The six point case}
\label{secFeynmanSub3}

There are in total $\frac{6!}{12}=60$ independent PT-factors for the six point case. According to the number of trivalent Feynman diagrams they
evaluate to, we can distribute them into different groups, with the
number of PT-factors in each group as
\begin{equation}
\setlength{\arraycolsep}{10pt}
\begin{array}{|c|c|c|c|c|c|c|}\hline
\text{\# of trivalent Feyn. diagrams} & 14 & 5 & 4 & 2 & 1 & 0 \\ \hline
\text{\# of PT-factors} & 1 & 6 & 3 & 21 & 14 & 15 \\ \hline
\end{array}
\label{6pt-counting}
\end{equation}
We will study them group by group in the following paragraphs.

\paragraph{With $14$ Feynman diagrams:} There is only one
PT-factor $\PT(\pmb{\beta})=\Spaa{123456}$ that evaluates to 14
Feynman diagrams. In the equivalent class, the good cycle representations are
\begin{subequations}
\begin{align}
&\text{V-type}~:~~~(1)(2)(3)(4)(5)(6)~,~~~\\
\label{eq:6pPtype}
&\text{P-type}~:~~~ (61)(25)(34)~~,~~(12)(36)(45)~~,~~ (23)(41)(56)~,~~~\nonumber\\
&  ~~~~~~~~~~~~~~~(1)(26)(35)(4)~~,~~ (2)(31)(46)(5)~~,~~ (3)(42)(15)(6)~.~~~
\end{align}
\end{subequations}
The sole V-type one indicates that the six
external legs form all possible cubic diagrams respecting the color
ordering, contributing to $14$ terms,
\begin{align}
\adjustbox{raise=-0.7cm}{\begin{tikzpicture}
	\draw (-0.75,0) node[left=0pt]{$1$} -- (0.75,0) node[right=0pt]{$4$};
	\draw (120:0.75) node[left=0pt]{$2$} -- (-60:0.75) node[right=0pt]{$5$};
	\draw (60:0.75) node[right=0pt]{$3$} -- (-120:0.75) node[left=0pt]{$6$};
	\end{tikzpicture}}\;\Longrightarrow &\;\frac{1}{s_{16} s_{23} s_{45}}+\frac{1}{s_{12} s_{34} s_{56}}+\frac{1}{s_{12} s_{45} s_{123}}+\frac{1}{s_{23} s_{45} s_{123}}+\frac{1}{s_{12} s_{56} s_{123}}\nonumber\\
&+\frac{1}{s_{23} s_{56} s_{123}}+\frac{1}{s_{12} s_{34} s_{126}}+\frac{1}{s_{16} s_{34} s_{126}}+\frac{1}{s_{12} s_{45}
s_{126}}+\frac{1}{s_{16} s_{45} s_{126}}\label{6p-re5}\\
&+\frac{1}{s_{16} s_{23}
s_{156}}+\frac{1}{s_{16} s_{34} s_{156}}+\frac{1}{s_{23} s_{56}
s_{156}}+\frac{1}{s_{34} s_{56} s_{156}}~.~~~\nonumber
\end{align}
%
Again, the planar separation $(1)\pmb{|}(2)\pmb{|}(3)\pmb{|}(4)\pmb{|}(5)\pmb{|}(6)$ tells us that the above $14$ terms in~\eref{6p-re5} can be effectively represented by a six point vertex, which becomes a building block for higher point analysis.

The P-type cycle representations have two different structures, collected respectively in the first and second row of~\eqref{eq:6pPtype}.
Among the three cycle representations in the first row, we study $(61)(25)(34)$ as an example.
First, the planar separation $(61)\pmb{|}(25)(34)$ gives a cubic vertex with legs $\{6, 1, P_{61}\}$, and a five point substructure $(P_{61})(25)(34)$. In its equivalent class, we have a V-type cycle representation $(P_{61})(2)(3)(4)(5)$, indicating that the substructure is just an effective five point vertex. Thus from the planar separation $(61)\pmb{|}(25)(34)$, we reproduce five terms in Eq.~\eqref{6p-re5} that contain the pole $s_{61}$.
Similarly, the planar separation $(61)(25)\pmb{|}(34)$ gives a propagator $s_{34}$ and a substructure $(P_{34})(61)(25)$, which has a V-type cycle representation $(P_{34})(5)(6)(1)(2)$. Thus this substructure is another effective five point vertex, which means that the planar separation $(61)(25)\pmb{|}(34)$ gives another five terms in Eq.~\eqref{6p-re5} that contain the pole $s_{34}$.
They have two common terms to the result of the first planar separation, so the P-type cycle representation $(61)(25)(34)$ gives eight terms in Eq.~\eqref{6p-re5}. If we combine the contributions from all the three P-type ones in the first row of~\eqref{eq:6pPtype} and remove the overlaps, we get the complete answer~\eqref{6p-re5}.

Finally, each of the three cycle representations in the second row of~\eqref{eq:6pPtype} gives four terms in Eq.~\eqref{6p-re5}.
For example, the planar separation $(1)(26)\pmb{|}(35)(4)$ gives a propagator $s_{612}$, together with two four point substructures $(P)(1)(26)$ and $(P)(35)(4)$.
In their equivalent classes, the former has a V-type cycle representation $(P)(6)(1)(2)$ while the latter has $(P)(3)(4)(5)$, both of which are effective quartic vertices, so that their contribution is
\begin{equation}
(1)(26)\pmb{|}(35)(4)\;\Longrightarrow\;\frac{1}{s_{612}}\left(\frac{1}{s_{12}}+\frac{1}{s_{61}}\right)\left(\frac{1}{s_{34}}+\frac{1}{s_{45}}\right)~.~~~
\end{equation}
Summing over the contributions of these three P-type cycle representations, we reproduce the twelve terms in Eq.~\eqref{6p-re5} that contain a three-particle pole $s_{ijk}$.

\paragraph{With $5$ Feynman diagrams:} There are six PT-factors in this category,
\bea
\Spaa{123465}~~,~~\Spaa{123546}~~,~~\Spaa{124356}~~,~~\Spaa{126543}~~,~~\Spaa{132456}~~~,~~\Spaa{154326}~.~~~
\eea
The last five PT-factors can be generated from the first one by
cyclic  permutation $i\mapsto i+1$. They actually form an orbit
under $\mathcal{D}_6$ action. In the equivalent class of
$\Spaa{123465}$, the good cycle representations are
%
\bea \text{V-type}: &~~~& (1)(2)(3)(4)(56)~~,~~(14)(23)(5)(6)~,~~~\nn
\text{P-type}: &~~~ &
(1526)(34)~~,~~(3546)(12)~~,~~(13)(2)(456)~~,~~(24)(3)(165)~.~~~\eea
We first derive the result by combining the information from all the V-type cycle representations.
From previous examples, we can easily tell that the planar separation $(1)\pmb{|}(2)\pmb{|}(3)\pmb{|}(4)\pmb{|}(56)$ indicates an effective five point vertex while $(14)(23)\pmb{|}(5)\pmb{|}(6)$ indicates a cubic vertex. Thus we have fixed the effective Feynman diagram and
obtain following result,
\begin{equation} \adjustbox{raise=-1.1cm}{\begin{tikzpicture}
	\draw (150:0.75) node[left=0pt]{$3$} -- (0,0) -- (0.75,0) -- ++(45:0.75) node[right=0pt]{$5$} (0.75,0) -- ++(-45:0.75) node[right=0pt]{$6$};
	\draw (-150:0.75) node[left=0pt]{$2$} -- (0,0) -- (0,0.75) node[above=0pt]{$4$} (0,0) -- (0,-0.75) node[below=0pt]{$1$};
	\end{tikzpicture}}\;\Longrightarrow\;\frac{1}{s_{56}}\left(\frac{1}{s_{12}s_{34}}+\frac{1}{s_{12}s_{123}}+\frac{1}{s_{23}s_{123}}+\frac{1}{s_{23}s_{156}}+\frac{1}{s_{34}s_{156}}\right)~.~~~
\end{equation}
As an alternative approach, we repeat the result by using only one V-type cycle representation and recursively those of substructures.
For example, besides the cubic vertex, the separation $(14)(23)\pmb{|}(5)\pmb{|}(6)$ also indicates a substructure $(P)(14)(23)$, which has the V-type cycle representation $(P)(1)(2)(3)(4)$ in its equivalent class generated by~\eqref{eq:equivperm}. Thus we again recover the effective five point vertex.

Alternatively, let us use the P-type cycle representations to find  the result. There are two different structures $(1526)\pmb{|}(34)$ and $(3546)\pmb{|}(12)$ manifest a two-particle pole, while $(13)(2)\pmb{|}(456)$ and $(24)(3)\pmb{|}(165)$ manifest a three-particle pole. In the first class, the substructure $(P)(1526)$ gives the V-type cycle representation $(P)(56)(1)(2)$ after using~\eqref{eq:equivperm}, so  we have
\begin{equation}
(1526)\pmb{|}(34)\;\Longrightarrow\;\frac{1}{s_{34}s_{56}}\left(\frac{1}{s_{12}}+\frac{1}{s_{561}}\right)~.~~~
\end{equation}
Similarly, we can derive that
\begin{equation}
(3546)\pmb{|}(12)\;\Longrightarrow\;\frac{1}{s_{12}s_{56}}\left(\frac{1}{s_{34}}+\frac{1}{s_{123}}\right)~.~~~
\end{equation}
For $(13)(2)\pmb{|}(456)$ and $(24)(3)\pmb{|}(165)$, similar analysis applies for each substructure, we find that
\bea
(13)(2)\pmb{|}(456)\;\Longrightarrow\;\frac{1}{s_{456}s_{56}}\left(\frac{1}{s_{12}}+\frac{1}{s_{23}}\right)~~~,~~~(24)(3)\pmb{|}(165)\;\Longrightarrow\;\frac{1}{s_{561}s_{56}}\left(\frac{1}{s_{23}}+\frac{1}{s_{34}}\right)~.~~~
\eea
The complete result can only be recovered by combining all four P-type contributions.

\paragraph{With $4$ Feynman diagrams:} There are three PT-factors $\Spaa{123654}$, $\Spaa{125436}$ and
$\Spaa{143256}$ that evaluate to four Feynman diagrams. They are
related by the cyclic permutation $i\mapsto i+1$, and form an orbit
under $\mathcal{D}_6$ action. Let us take $\Spaa{123654}$ as an
example. The good cycle representations are
\bean \text{V-type}: &~~~ & (1)(2)(3)(46)(5)~~,~~(13)(2)(4)(5)(6)~,\\
\text{P-type}: &~~~& (1432)(56)~~,~~(1236)(45)~~,~~(12)(3456)~~,~~(23)(1654)~.\eean
In the V-type cycle representations, the planar separation $(1)\pmb{|}(2)\pmb{|}(3)\pmb{|}(46)(5)$
indicates an effective quartic vertex with legs $\{1, 2, 3 ,P_{456}\}$, while
the planar separation $(13)(2)\pmb{|}(4)\pmb{|}(5)\pmb{|}(6)$  indicates another effective quartic vertex with legs $\{4, 5, 6 ,P_{123}\}$. By combining them, we get an effective Feynman diagram with two quartic vertices, evaluated to
\begin{equation}
\adjustbox{raise=-1.1cm}{\begin{tikzpicture}
	\draw (-0.75,0) node[left=0pt]{$2$} -- (1.5,0) node[right=0pt]{$5$};
	\draw (0,-0.75) node[below=0pt]{$1$} -- (0,0.75) node[above=0pt]{$3$};
	\draw (0.75,-0.75) node[below=0pt]{$6$} -- (0.75,0.75) node[above=0pt]{$4$};
	\end{tikzpicture}}\;\Longrightarrow\;\frac{1}{s_{12}s_{123}s_{45}}+\frac{1}{s_{12}s_{123}s_{56}}+\frac{1}{s_{23}s_{123}s_{45}}+\frac{1}{s_{23}s_{123}s_{56}}~.~~~\label{6p-F4-1}
\end{equation}
We can also arrive at above result by analyzing just one V-type cycle representation with its substructures.
For $(1)\pmb{|}(2)\pmb{|}(3)\pmb{|}(46)(5)$, the substructure $(P)(46)(5)$ has an equivalent V-type cycle representation $(P)(4)(5)(6)$ in the equivalent class generated by~\eqref{eq:equivperm}, so we recover the result~\eqref{6p-F4-1}. If we start from $(13)(2)\pmb{|}(4)\pmb{|}(5)\pmb{|}(6)$, the analysis is similar.

For the P-type cycle representations, each one contributes two terms in~\eref{6p-F4-1}.
For example, the planar separation $(12)\pmb{|}(3456)$ gives a pole $s_{12}$ and a substructure $(P_{12})(3456)$, and the substructure has an equivalent V-type cycle representation $(P_{12})\pmb{|}(3)\pmb{|}(5)(46)$ in the equivalent class.
We can then read out the pole $s_{456}$ and another substructure $(P_{456})(5)(46)$. This substructure further gives a V-type cycle representation $(P_{456})(4)(5)(6)$, indicating a quartic effective vertex.
Putting them together, we get
\begin{equation}
(12)\pmb{|}(3456)\;\Longrightarrow\;\frac{1}{s_{12}s_{456}}\left(\frac{1}{s_{45}}+\frac{1}{s_{56}}\right)~.~~~
\end{equation}
Similar analysis can be done for other three P-type cycle representations, and the results are
\begin{align}
  (1432)\pmb{|}(56) & \Longrightarrow \frac{1}{s_{56}s_{123}}\left(\frac{1}{s_{12}}+\frac{1}{s_{23}}\right)~,\nn
(1236)\pmb{|}(45) & \Longrightarrow \frac{1}{s_{45}s_{123}}\left(\frac{1}{s_{12}}+\frac{1}{s_{23}}\right)~,\nn
(1654)\pmb{|}(23) & \Longrightarrow \frac{1}{s_{23}s_{456}}\left(\frac{1}{s_{45}}+\frac{1}{s_{56}}\right)~.
\end{align}
Again, we see that a single P-type cycle representation is not sufficient to provide the complete information, and we need to combine all of them.

\paragraph{With $2$ Feynman diagrams:} There are $21$ PT-factors that evaluate to two Feynman diagrams.
According to the action of $\mathcal{D}_6$, we can divide them into three orbits as
\begin{subequations}
\label{eq:6p2F}
\begin{align}
&\mathcal{S}_1:& &\Spaa{125643}~~,~~\Spaa{126534}~~,~~\Spaa{132546}~~,~~\Spaa{145326}~~,~~\Spaa{124365}~~,~~\Spaa{154236}~,~~~\\
&\mathcal{S}_2:& &\Spaa{126453}~~,~~\Spaa{132465}~~,~~\Spaa{153426}~,~~~\\
&\mathcal{S}_3:& &\Spaa{123645}~~,~~\Spaa{143265}~~,~~\Spaa{134526}~~,~~\Spaa{124563}~~,~~\Spaa{142356}~~,~~\Spaa{125346}~,~~~\nonumber\\
& & &\Spaa{123564}~~,~~\Spaa{152346}~~,~~\Spaa{126345}~~,~~\Spaa{132654}~~,~~\Spaa{134256}~~,~~\Spaa{124536}~.~~~
\end{align}
\end{subequations}
For the category $\mathcal{S}_1$ of~\eqref{eq:6p2F}, we take $\Spaa{125643}$ as an example. The good cycle representations are
\bean \text{V-type}: &~~~& (12)(3)(4)(56)~~,~~(1)(2)(3546)~~,~~(1423)(5)(6)~,\\
\text{P-type}: &~~~& (132)(456)~~,~~(15)(26)(34)~.\eean
The planar separations in the V-type cycle representations indicate the following vertex structures,
\begin{subequations}
\bea
(12)\pmb{|}(3)\pmb{|}(4)\pmb{|}(56)\; &\Longrightarrow&\; \text{quartic vertex with legs }P_{12}\,,3\,,4\,,\text{ and }P_{56}~,~~~\\
(1)\pmb{|}(2)\pmb{|}(3546)\;&\Longrightarrow&\; \text{cubic vertex with legs }1\,,2\,,\text{ and }P_{12}~,~~~\\
(1423)\pmb{|}(5)\pmb{|}(6)\;&\Longrightarrow&\; \text{cubic vertex with legs }5\,,6\,,\text{ and }P_{56}~.~~~
\eea
\end{subequations}
Combining them together, we get an effective Feynman diagram with two cubic vertices and one quartic vertex, and the result is
\bea \adjustbox{raise=-0.65cm}{\begin{tikzpicture}
	\draw (135:0.75) node[left=0pt]{$2$} -- (0,0) -- (-135:0.75) node[left=0pt]{$1$};
	\draw (0,0) -- (1.5,0) -- ++(45:0.75) node[right=0pt]{$5$} (1.5,0) -- ++(-45:0.75) node[right=0pt]{$6$};
	\draw (0.75,0) -- ++(135:0.75) node[above=0pt]{$3$} (0.75,0) -- ++(45:0.75) node[above=0pt]{$4$};
	\end{tikzpicture}}\;\Longrightarrow\;\frac{1}{s_{12}s_{56}}\left(\frac{1}{s_{123}}+\frac{1}{s_{34}}\right)~.~~~\label{6p-F2-A-1}\eea
We can also reproduce above result by considering only one V-type cycle representation and its substructures. For example, the planar separation $(1)\pmb{|}(2)\pmb{|}(3546)$ gives the substructure $(P)(3546)$, which has a V-type cycle representation $(P)(3)(4)(56)$ generated by~\eqref{eq:equivperm}. Thus the substructure contains a quartic vertex with legs $P_{12}$, $3$, $4$, $P_{56}$, and a cubic vertex with legs $5$, $6$, $P_{56}$.

Alternatively, let us discuss the contribution from P-type cycle representation. For the one $(132)\pmb{|}(456)$, we can derive two substructure of V-type cycle representations $(P)(12)(3)$ and $(P)(4)(56)$. Thus we get the contribution $\frac{1}{s_{12}s_{56}s_{123}}$. Similarly, the P-type cycle representation $(15)(26)\pmb{|}(34)$ has a substructure $(P)(15)(26)$ that has a V-type cycle representation $(P)(12)(56)$. Thus we get the contribution $\frac{1}{s_{12}s_{34}s_{56}}$. Again, we need to combine the information of all the P-type cycle representations to get the full result.

Next, we move to the category $\mathcal{S}_2$ of~\eqref{eq:6p2F}, and take $\Spaa{126453}$ as an example. The good cycle representations are
\bean \text{V-type}: &~~~& (1)(2)(36)(4)(5)~~,~~(12)(3)(45)(6)~,\\
\text{P-type}: &~~~& (132)(465)~~,~~(126)(345)~.\eean
The V-type cycle representation $(1)(2)(36)(4)(5)$ allows two different planar separations
$(1)\pmb{|}(2)\pmb{|}(36)(4)(5)$ and $(1)(2)(36)\pmb{|}(4)\pmb{|}(5)$, so it gives two cubic vertices. Meanwhile, $(12)(3)(45)(6)$
has only one planar separation $(12)\pmb{|}(3)\pmb{|}(45)\pmb{|}(6)$, so it gives an effective quartic vertex. Putting all these vertices
together, we get the effective Feynman diagram
%
\bea \adjustbox{raise=-1.1cm}{\begin{tikzpicture}
	\draw (135:0.75) node[left=0pt]{$2$} -- (0,0) -- (-135:0.75) node[left=0pt]{$1$};
	\draw (0,0) -- (1.5,0) -- ++(45:0.75) node[right=0pt]{$4$} (1.5,0) -- ++(-45:0.75) node[right=0pt]{$5$};
	\draw (0.75,-0.75) node[below=0pt]{$6$} -- (0.75,0.75) node[above=0pt]{$3$};
	\end{tikzpicture}}\;\Longrightarrow\;\frac{1}{s_{12} s_{45}} \left(\frac{1}{s_{123}}+\frac{1}{s_{612}}\right)~.~~~\label{6p-F2-B-1}\eea
Now we follow another approach by considering only one V-type cycle representation to reproduce the result~\eref{6p-F2-B-1}. If we focus on the cycle representation $(12)(3)(45)(6)$, the planar separation $(12)\pmb{|}(3)\pmb{|}(45)\pmb{|}(6)$ gives the final result directly. While if we focus on the V-type cycle representation $(1)(2)(36)(4)(5)$, the planar separation $(1)\pmb{|}(2)\pmb{|}(36)(4)(5)$ will manifest the cubic vertex with legs $1$, $2$, $P_{12}$, and the substructure $(P)(36)(4)(5)$ then generates a V-type cycle representation $(P)(3)(45)(6)$, which leads to a quartic and a cubic vertex. Hence we do reproduce the result ~\eqref{6p-F2-B-1}. Analysis of $(1)(2)(36)(4)(5)$ from the planar separation $(1)(2)(36)\pmb{|}(4)\pmb{|}(5)$ is exactly the same.
We note that the P-type cycle representation $(132)\pmb{|}(465)$ gives $\frac{1}{s_{12}s_{45}s_{123}}$ while $(126)\pmb{|}(345)$ gives $\frac{1}{s_{12}s_{45}s_{612}}$. Thus by combining them, we get the complete result.

Finally, we study the category $\mathcal{S}_3$ of~\eqref{eq:6p2F}, and take $\Spaa{123645}$ as example. The good cycle representations are
\bean \text{V-type}: &~~~& (1)(2)(3)(465)~~,~~(1236)(4)(5)~~,~~(13)(2)(45)(6)~,\\
\text{P-type}: &~~~& (12)(356)(4)~~,~~(23)(164)(5)~.\eean
Using the V-type cycle representations, we see that $(1)\pmb{|}(2)\pmb{|}(3)\pmb{|}(465)$ gives an effective quartic vertex, while $(1236)\pmb{|}(4)\pmb{|}(5)$ and $(13)(2)\pmb{|}(45)\pmb{|}(6)$ give two cubic vertices. Putting all of them
together, we get the effective Feynman diagram
\begin{equation}
  \adjustbox{raise=-1.1cm}{\begin{tikzpicture}
	\draw (-0.75,0) node[left=0pt]{$2$} -- (1.5,0) -- ++(45:0.75) node[right=0pt]{$4$} (1.5,0) -- ++(-45:0.75) node[right=0pt]{$5$};
	\draw (0,-0.75) node[below=0pt]{$1$} -- (0,0.75) node[above=0pt]{$3$} (0.75,0) -- (0.75,-0.75) node[below=0pt]{$6$};
  \end{tikzpicture}}\;\Longrightarrow\;\frac{1}{s_{123} s_{45}} \left( \frac{1}{s_{12}}+\frac{1}{s_{23}}\right)~.~~~\label{6p-F2-C-1}
\end{equation}
Alternatively, let us take just one single V-type cycle representation to reproduce above result. From the planar separation $(1)\pmb{|}(2)\pmb{|}(3)\pmb{|}(465)$, the substructure $(P)(465)$ generates a V-type cycle representation $(P)(45)(6)$ according to~\eqref{eq:equivperm}. Thus we get the same effective Feynman diagram as in~\eref{6p-F2-C-1}.
From the planar separation  $(1236)\pmb{|}(4)\pmb{|}(5)$, the substructure $(P)(1236)$ has a V-type cycle representation $(P6)(1)(2)(3)$, while from $(13)(2)\pmb{|}(45)\pmb{|}(6)$, the substructure $(P)(13)(2)$ gives a V-type cycle representation $(P)(1)(2)(3)$. Both of them recover the result~\eqref{6p-F2-C-1} respectively.

For the P-type cycle representations, we note that $(12)\pmb{|}(356)(4)$ gives the partial result $\frac{1}{s_{12}s_{123}s_{45}}$ following our algorithm, while $(23)\pmb{|}(164)(5)$ gives another piece $\frac{1}{s_{23}s_{123}s_{45}}$. They combine to give the full result~\eqref{6p-F2-C-1}.

\paragraph{With $1$ Feynman diagram:} There are 14 PT-factors that evaluate to one Feynman diagram. According to the action of $\mathcal{D}_6$, they can be distributed into three orbits,
\begin{subequations}
\label{eq:6p1F}
\begin{align}
&\mathcal{S}_1:& &\Spaa{132645}~~,~~\Spaa{134265}~~,~~\Spaa{135426}~~,~~\Spaa{124653}~~,~~\Spaa{153246}~~,~~\Spaa{126435}~,~~~\\
&\mathcal{S}_2:& &\Spaa{125463}~~,~~\Spaa{142365}~~,~~\Spaa{143526}~~,~~\Spaa{132564}~~,~~\Spaa{152436}~~,~~\Spaa{126354}~,~~~\\
&\mathcal{S}_3:& &\Spaa{125634}~~,~~\Spaa{145236}~.~~~
\end{align}
\end{subequations}
For the category $\mathcal{S}_1$ of~\eqref{eq:6p1F}, we analyze $\Spaa{132645}$ as example. Its good cycle representations are
\bea \text{V-type}: &~~~&
(1)(23)(465)~~,~~(123)(45)(6)~~,~~(136)(2)(4)(5)~~,~~(2)(3)(164)(5)~.~~~
\eea
There is no P-type cycle representation. This can be understood as follows. Since the final result only contains one term, so that we do not have partial result. From the V-type cycle representations, we see that each planar separation of $(1)\pmb{|}(23)\pmb{|}(465)$,
$(123)\pmb{|}(45)\pmb{|}(6)$, $(136)(2)\pmb{|}(4)\pmb{|}(5)$ and $(2)\pmb{|}(3)\pmb{|}(164)(5)$
indicates a cubic vertex. Thus we get a trivalent Feynman diagram evaluated to
\bea \adjustbox{raise=-0.8cm}{\begin{tikzpicture}
	\draw (-0.75,0) node[left=0pt]{$1$} -- (2.25,0) node[right=0pt]{$6$};
	\draw (0,0) -- (0,0.75) -- ++(45:0.75) node[above=0pt]{$3$} (0,0.75) -- ++(135:0.75) node[above=0pt]{$2$};
	\draw (1.5,0) -- (1.5,0.75) -- ++(45:0.75) node[above=0pt]{$5$} (1.5,0.75) -- ++(135:0.75) node[above=0pt]{$4$};
	\end{tikzpicture}}\;\Longrightarrow\;\frac{1}{s_{23} s_{45} s_{123}}~.~~~\label{6p-F1-A-1}\eea
Of course, we can get the same result by considering a single V-type cycle representation and its substructures. For example, besides the cubic vertex with legs $\{1, P_{23} ,P_{123}\}$, the planar separation $(1)\pmb{|}(23)\pmb{|}(465)$ also indicates a substructure $(P)(465)$, which contains a V-type cycle representation $(P)(45)(6)$, giving exactly the other two cubic vertices in Eq.~\eqref{6p-F1-A-1}.  Analysis for the other V-type cycle representations is the same.

For the category $\mathcal{S}_2$ of~\eqref{eq:6p1F}, we analyze $\Spaa{125463}$ as example. Its good cycle representations are
\bea \text{V-type}: &~~~&
(1)(2)(356)(4)~~,~~(132)(45)(6)~~,~~(1)(236)(4)(5)~~,~~(12)(3)(465)~.~~~\eea
Each planar separation of
$(1)\pmb{|}(2)\pmb{|}(356)(4)$, $(132)\pmb{|}(45)\pmb{|}(6)$, $(1)(236)\pmb{|}(4)\pmb{|}(5)$ and $(12)\pmb{|}(3)\pmb{|}(465)$ indicates a cubic vertex. After
combining them, we get the trivalent Feynman diagram evaluated to
\bea \adjustbox{raise=-0.8cm}{\begin{tikzpicture}
	\draw (-0.75,0) node[left=0pt]{$1$} -- (2.25,0) node[right=0pt]{$6$};
	\draw (0,0) -- (0,0.75) node[above=0pt]{$2$} (0.75,0) -- (0.75,0.75) node[above=0pt]{$3$};
	\draw (1.5,0) -- (1.5,0.75) -- ++(45:0.75) node[above=0pt]{$5$} (1.5,0.75) -- ++(135:0.75) node[above=0pt]{$4$};
	\end{tikzpicture}}\;\Longrightarrow\;\frac{1}{s_{12} s_{45} s_{123}}~.~~~\label{6p-F1-B-1}\eea
Finally, for the category $\mathcal{S}_3$ of~\eqref{eq:6p1F}, we analyze $\Spaa{125634}$ as example. Its good cycle representations are
\bea \text{V-type}: &~~~&
(1)(2)(35)(64)~~,~~(13)(24)(5)(6)~~,~~(15)(26)(3)(4)~~,~~(12)(34)(56)~.~~~\eea
Again, all the planar separations
$(1)\pmb{|}(2)\pmb{|}(35)(64)$, $(13)(24)\pmb{|}(5)\pmb{|}(6)$, $(15)(26)\pmb{|}(3)\pmb{|}(4)$ and
$(12)\pmb{|}(34)\pmb{|}(56)$ indicate cubic vertices, so that the final result is
\bea \adjustbox{raise=-0.8cm}{\begin{tikzpicture}
	\draw (-0.75,0) node[left=0pt]{$1$} -- (2.25,0) node[right=0pt]{$6$};
	\draw (0,0) -- (0,0.75) node[above=0pt]{$2$} (1.5,0) -- (1.5,0.75) node[above=0pt]{$5$};
	\draw (0.75,0) -- (0.75,0.75) -- ++(45:0.75) node[above=0pt]{$4$} (0.75,0.75) -- ++(135:0.75) node[above=0pt]{$3$};
	\end{tikzpicture}}\;\Longrightarrow\;\frac{1}{s_{12} s_{34} s_{56}}~.~~~\label{6p-F1-C-1}\eea
Both Eq.~\eqref{6p-F1-B-1} and~\eqref{6p-F1-C-1} can be reproduced by a single V-type cycle representation, and the analysis is almost the same to that of the category $\mathcal{S}_1$ shown above.

\paragraph{With no Feynman diagrams:} Finally, there are $15$ PT-factors that evaluate to zero. According
to the action of $\mathcal{D}_6$, they can be distributed into three
orbits,
\begin{subequations}
\label{eq:6p0F}
\begin{align}
&\mathcal{S}_1:& & \Spaa{124635}~~,~~\Spaa{146235}~~,~~\Spaa{134625}~~,~~\Spaa{136245}~~,~~\Spaa{135624}~~,~~\Spaa{135246}~,\\
&\mathcal{S}_2:& & \Spaa{125364}~~,~~\Spaa{146325}~~,~~\Spaa{143625}~~,~~\Spaa{136254}~~,~~\Spaa{136524}~~,~~\Spaa{142536}~,\\
&\mathcal{S}_3:& & \Spaa{135264}~~,~~\Spaa{136425}~~,~~\Spaa{142635}~.~~~
\end{align}
\end{subequations}
The category $\mathcal{S}_3$ of~\eqref{eq:6p0F} does not have any good cycle representations, and indeed it evaluates to zero.

For the category $\mathcal{S}_1$ of~\eqref{eq:6p0F}, we take $\Spaa{124635}$ as example. The good cycle representations are
\begin{align}
\label{eq:catA6p}
\text{V-type}:\quad(1)(2)(3465)~~~~~~,~~~~~~\text{P-type}:\quad(12)(3564)~.~~~
\end{align}
Both of them manifest a pole $P_{12}$, together with a substructure $(P)(3465)$ and $(P)(3564)$ respectively. However, both substructures are members of~\eqref{eq:5p0F}, which give zero contribution. This can be seen clearly if we replace $P$ by $2$, and then replace $i\rightarrow i-1$ for the rest.
Thus we see that the category $\mathcal{S}_1$ evaluates to zero because
it contains a substructure with zero contribution. Actually, the category $\mathcal{S}_2$ of~\eqref{eq:6p0F} evaluates to zero for the same reason. For example, $\Spaa{125364}$ has good cycle representations
\begin{align}
\label{eq:catB6p}
\text{V-type}:\quad(1)(2)(3564)~~~~~~,~~~~~~\text{P-type}:\quad(12)(3465)~,~~~
\end{align}
whose substructures are identical to the case of category $\mathcal{S}_1$.

This result can also be understood from another point of view. As we have shown with many examples, the
V-type cycle representations encode the vertex structure of corresponding effective Feynman diagram. In the  current case, there
is only one V-type cycle representation and it has only one planar separation of cubic vertex. If we use $v_m$ to denote the number of $m$-point vertices, we should have the following constraint for all the valid effective Feynman diagrams as
\bea \sum_{m= 3}^{n} (m-2) v_m= n-2~.~~~\label{V-cond}\eea
For the cases ~\eqref{eq:catA6p} and~\eqref{eq:catB6p}, we only have $v_3=1$ and all the other $v_i=0$, while for category $\mathcal{S}_3$, all $v_i$'s are zero. Thus the identity~\eref{V-cond} is violated, indicating that no such effective Feynman diagram exits. One can check that for all the cases with nonzero results, the identity~\eref{V-cond} is satisfied.

\subsection{Higher point examples}

Let us now consider an eight point example with PT-factor $\Spaa{12347856}$. It has 16 cycle representations, and we can classify them as
%
\begin{center}
\begin{tabular}{|c|c|}
  \hline
  V-type & $(1)(2)(3)(4)(57)(68)~~,~~(153)(264)(7)(8)~~,~~(137)(248)(5)(6)~~,~~(14)(23)(56)(78)$\\ \hline
  P-type & $(12)(367458)~~,~~(13)(2)(468)(5)(7)~~,~~(175)(24)(3)(6)(8)~~,~~(34)(185276)$\\ \hline
  Bad & $(16785432)~~,~~(1874)(25)(36)~~,~~(1735)(2846)~~,~~(1456)(27)(38)$ \\
   & $(12347658)~~,~~(1638)(2547)~~,~~(1)(2648)(357)~~,~~(1537)(286)(4)$\\
  \hline\end{tabular}
\end{center}
Among them, four are V-type cycle representations, and each one encodes the vertex information. For each cycle representation, there is only one planar separation, and from which we can directly work out
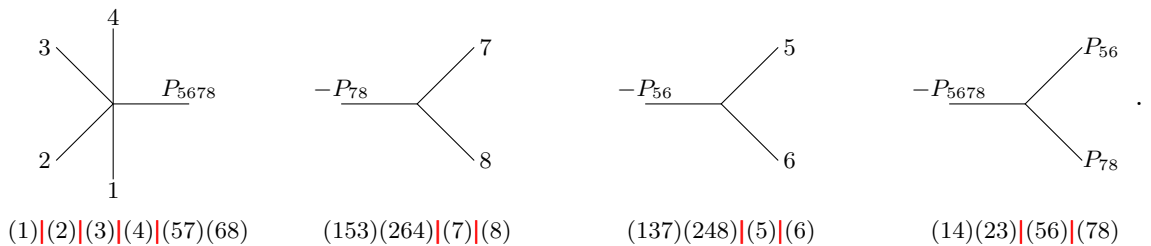
\begin{center}
\begin{tikzpicture}
  \draw[](1.25,0.75)--(2,0)--(1.25,-0.75);
  \draw[](2,1)--(2,-1);
  \draw[](2,0)--(3,0);
  \node at (2,-1.15) []{{\footnotesize $1$}};
  \node at (1.1,-0.75) []{{\footnotesize $2$}};
  \node at (1.1,0.75) []{{\footnotesize $3$}};
  \node at (2,1.15) []{{\footnotesize $4$}};
  \node at (3,0.2) []{{\footnotesize $P_{5678}$}};
  \node at (2.2,-1.7) []{{\footnotesize $(1){\color{red}\pmb{|}}(2){\color{red}\pmb{|}}(3){\color{red}\pmb{|}}(4){\color{red}\pmb{|}}(57)(68)$}};

  \draw[](5,0)--(6,0);
  \draw[](6.75,0.75)--(6,0)--(6.75,-0.75);
  \node at (6.9,-0.75) []{{\footnotesize $8$}};
  \node at (6.9,0.75) []{{\footnotesize $7$}};
  \node at (5,0.2) []{{\footnotesize $-P_{78}$}};
  \node at (6,-1.7) []{{\footnotesize $(153)(264){\color{red}\pmb{|}}(7){\color{red}\pmb{|}}(8)$}};

  \draw[](9,0)--(10,0);
  \draw[](10.75,0.75)--(10,0)--(10.75,-0.75);
  \node at (10.9,-0.75) []{{\footnotesize $6$}};
  \node at (10.9,0.75) []{{\footnotesize $5$}};
  \node at (9,0.2) []{{\footnotesize $-P_{56}$}};
  \node at (10,-1.7) []{{\footnotesize $(137)(248){\color{red}\pmb{|}}(5){\color{red}\pmb{|}}(6)$}};

  \draw[](13,0)--(14,0);
  \draw[](14.75,0.75)--(14,0)--(14.75,-0.75);
  \node at (15,-0.75) []{{\footnotesize $P_{78}$}};
  \node at (15,0.75) []{{\footnotesize $P_{56}$}};
  \node at (13,0.2) []{{\footnotesize $-P_{5678}$}};
  \node at (14,-1.7) []{{\footnotesize $(14)(23){\color{red}\pmb{|}}(56){\color{red}\pmb{|}}(78)$}};

  \node at (15.5,0) {$.$};
\end{tikzpicture}
\end{center}
Collectively considering all these four V-type cycle representations, and gluing them via propagators, we get the Feynman diagram as
\begin{equation}
\label{eq:8p5F}
\adjustbox{raise=-1.9cm}{\begin{tikzpicture}
	\draw (72:1) node[above=0pt]{$4$} -- (0,0) -- (144:1) node[above=0pt]{$3$};
	\draw (-72:1) node[below=0pt]{$1$} -- (0,0) -- (-144:1) node[below=0pt]{$2$};
	\draw (0,0) -- (0.75,0) -- ++(45:0.75) -- ++(0,1) node[above=0pt]{$5$};
	\draw (0.75,0) ++(45:0.75) -- ++(1,0) node[above=0pt]{$6$};
	\draw (0.75,0) -- ++(-45:0.75) -- ++(1,0) node[below=0pt]{$7$};
	\draw (0.75,0) ++(-45:0.75) -- ++(0,-1) node[below=0pt]{$8$};
\end{tikzpicture}}\;\Longrightarrow\;\frac{1}{s_{56}s_{78}s_{1234}}\left(\frac{1}{s_{12}s_{123}}+\frac{1}{s_{23}s_{123}}+\frac{1}{s_{23}s_{234}}+\frac{1}{s_{34}s_{234}}+\frac{1}{s_{12}s_{34}}\right)\,,
\end{equation}
which exactly produces the result of CHY-integrand $\Spaa{12345678}\times\Spaa{12347856}$.
Alternatively, we can reproduce the same result from only one V-type cycle representation by going into its substructures. We use the V-type cycle representation $(14)(23)(56)(78)$ as our example.
The planar separation $(14)(23)\pmb{|}(56)\pmb{|}(78)$ gives three substructures, namely $(14)(23)(P_{5678})$, $(56)(P_{56})$ and $(78)(P_{78})$. For each one, we can find a V-type cycle representation that manifests the vertex structure in the equivalent class,
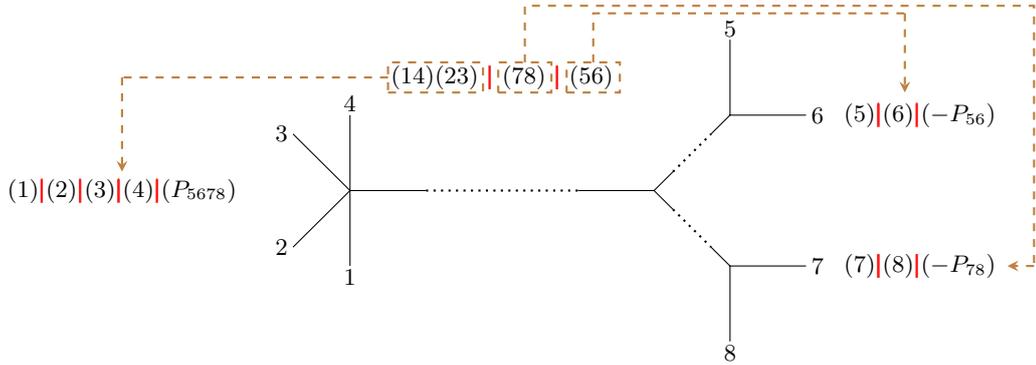
\begin{center}
\begin{tikzpicture}[scale=1]
  \draw[](-2.75,0.75)--(-2,0)--(-2.75,-0.75) (-2,1)--(-2,-1) (-2,0)--(-1,0) (1,0)--(2,0) (2.25,0.25)--(2,0)--(2.25,-0.25) (3,2)--(3,1)--(4,1) (3,1)--(2.75,0.75) (3,-2)--(3,-1)--(4,-1) (3,-1)--(2.75,-0.75);
  \draw[dotted,thick](-1,0)--(1,0) (2.25,0.25)--(2.75,0.75) (2.25,-0.25)--(2.75,-0.75);
  \node at (-2,-1.15) []{{\footnotesize $1$}};
  \node at (-2.9,-0.75) []{{\footnotesize $2$}};
  \node at (-2.9,0.75) []{{\footnotesize $3$}};
  \node at (-2,1.15) []{{\footnotesize $4$}};
  \node at (3,2.15) []{{\footnotesize $5$}};
  \node at (4.15,1) []{{\footnotesize $6$}};
  \node at (4.15,-1) []{{\footnotesize $7$}};
  \node at (3,-2.15) []{{\footnotesize $8$}};

  \node at (0,1.5) []{{\footnotesize $(14)(23)~{\color{red}\pmb{|}}~(78)~{\color{red}\pmb{|}}~(56)$}};
  \node at (-5,0) []{{\footnotesize $(1){\color{red}\pmb{|}}(2){\color{red}\pmb{|}}(3){\color{red}\pmb{|}}(4){\color{red}\pmb{|}}(P_{5678})$}};
  \node at (5.5,-1) []{{\footnotesize $(7){\color{red}\pmb{|}}(8){\color{red}\pmb{|}}(-P_{78})$}};
  \node at (5.5,1) []{{\footnotesize $(5){\color{red}\pmb{|}}(6){\color{red}\pmb{|}}(-P_{56})$}};

  \draw[dashed, brown, thick](-1.5,1.3)--(-0.25,1.3)--(-0.25,1.7)--(-1.5,1.7)--(-1.5,1.3);
  \draw[dashed, brown, thick](-1.5,1.5)--(-5,1.5);
  \draw[dashed,brown, thick, ->,>=stealth](-5,1.5)--(-5,0.25);
  \draw[dashed, brown, thick](0.85,1.3)--(1.55,1.3)--(1.55,1.7)--(0.85,1.7)--(0.85,1.3);
  \draw[dashed,brown, thick](1.2,1.7)--(1.2,2.35)--(5.3,2.35);
  \draw[dashed, brown, thick, ->,>=stealth](5.3,2.35)--(5.3,1.3);
  \draw[dashed, brown, thick](-0.05,1.3)--(0.65,1.3)--(0.65,1.7)--(-0.05,1.7)--(-0.05,1.3);
  \draw[dashed, brown, thick](0.3,1.7)--(0.3,2.45)--(7,2.45)--(7,-1);
  \draw[dashed, brown, thick, ->,>=stealth](7,-1)--(6.65,-1);

\end{tikzpicture}
\end{center}
By connecting the substructures together, we obtain the effective Feynman diagram as in~\eqref{eq:8p5F}.

There are also four P-type cycle representations. As mentioned previously, they should be considered collectively in order to produce the complete result, while each one only contributes a partial result. From the planar separations of these cycle representations and their substructures, we can work out the contribution of each P-type cycle representation. We will not repeat the detailed analysis here, but only give the result as,
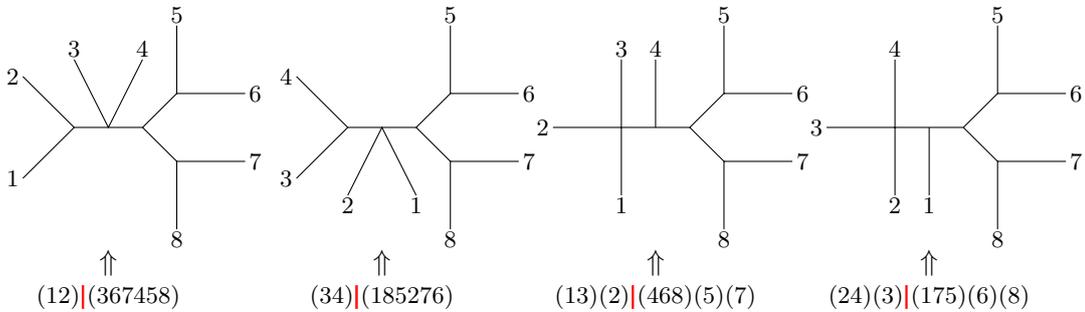
\begin{center}
\begin{tikzpicture}[scale=0.9]
  \draw[](1,0)--(2,0)--(2.5,0.5)--(2.5,1.5) (2.5,0.5)--(3.5,0.5) (2,0)--(2.5,-0.5)--(2.5,-1.5) (2.5,-0.5)--(3.5,-0.5);
  \draw[](0.25,0.75)--(1,0)--(0.25,-0.75) (1,1)--(1.5,0)--(2,1);
  \node at (0.1,-0.75) []{{\footnotesize $1$}};
  \node at (0.1,0.75) []{{\footnotesize $2$}};
  \node at (1,1.15) []{{\footnotesize $3$}};
  \node at (2,1.15) []{{\footnotesize $4$}};
  \node at (2.5,1.65) []{{\footnotesize $5$}};
  \node at (3.65,0.5) []{{\footnotesize $6$}};
  \node at (3.65,-0.5) []{{\footnotesize $7$}};
  \node at (2.5,-1.65) []{{\footnotesize $8$}};

  \draw[](5,0)--(6,0)--(6.5,0.5)--(6.5,1.5) (6.5,0.5)--(7.5,0.5) (6,0)--(6.5,-0.5)--(6.5,-1.5) (6.5,-0.5)--(7.5,-0.5);
  \draw[](4.25,0.75)--(5,0)--(4.25,-0.75) (5,-1)--(5.5,0)--(6,-1);
  \node at (4.1,-0.75) []{{\footnotesize $3$}};
  \node at (4.1,0.75) []{{\footnotesize $4$}};
  \node at (5,-1.15) []{{\footnotesize $2$}};
  \node at (6,-1.15) []{{\footnotesize $1$}};
  \node at (6.5,1.65) []{{\footnotesize $5$}};
  \node at (7.65,0.5) []{{\footnotesize $6$}};
  \node at (7.65,-0.5) []{{\footnotesize $7$}};
  \node at (6.5,-1.65) []{{\footnotesize $8$}};

  \draw[](9,0)--(10,0)--(10.5,0.5)--(10.5,1.5) (10.5,0.5)--(11.5,0.5) (10,0)--(10.5,-0.5)--(10.5,-1.5) (10.5,-0.5)--(11.5,-0.5);
  \draw[](8,0)--(9,0)--(9,1) (9,0)--(9,-1) (9.5,0)--(9.5,1);
  \node at (9,-1.15) []{{\footnotesize $1$}};
  \node at (7.85,0) []{{\footnotesize $2$}};
  \node at (9,1.15) []{{\footnotesize $3$}};
  \node at (9.5,1.15) []{{\footnotesize $4$}};
  \node at (10.5,1.65) []{{\footnotesize $5$}};
  \node at (11.65,0.5) []{{\footnotesize $6$}};
  \node at (11.65,-0.5) []{{\footnotesize $7$}};
  \node at (10.5,-1.65) []{{\footnotesize $8$}};

  \draw[](13,0)--(14,0)--(14.5,0.5)--(14.5,1.5) (14.5,0.5)--(15.5,0.5) (14,0)--(14.5,-0.5)--(14.5,-1.5) (14.5,-0.5)--(15.5,-0.5);
  \draw[](12,0)--(13,0)--(13,1) (13,0)--(13,-1) (13.5,0)--(13.5,-1);
  \node at (13,-1.15) []{{\footnotesize $2$}};
  \node at (11.85,0) []{{\footnotesize $3$}};
  \node at (13,1.15) []{{\footnotesize $4$}};
  \node at (13.5,-1.15) []{{\footnotesize $1$}};
  \node at (14.5,1.65) []{{\footnotesize $5$}};
  \node at (15.65,0.5) []{{\footnotesize $6$}};
  \node at (15.65,-0.5) []{{\footnotesize $7$}};
  \node at (14.5,-1.65) []{{\footnotesize $8$}};

  \node at (1.5,-2)[]{$\Uparrow$};
  \node at (1.5,-2.5)[]{{\footnotesize $(12){\color{red}\pmb{|}}(367458)$}};

  \node at (5.5,-2)[]{$\Uparrow$};
  \node at (5.5,-2.5)[]{{\footnotesize $(34){\color{red}\pmb{|}}(185276)$}};

  \node at (9.5,-2)[]{$\Uparrow$};
  \node at (9.5,-2.5)[]{{\footnotesize $(13)(2){\color{red}\pmb{|}}(468)(5)(7)$}};

  \node at (13.5,-2)[]{$\Uparrow$};
  \node at (13.5,-2.5)[]{{\footnotesize $(24)(3){\color{red}\pmb{|}}(175)(6)(8)$}};

\end{tikzpicture}
\end{center}
We see that, each P-type cycle representation gives a quartic vertex contained in the original five point vertex. Thus each P-type cycle representation produces two terms.
Only after summing up the above four contributions and removing the duplicates can we recover the complete result.

Our last example involves a nine point PT-factor $\PT(\pmb{\beta})=\Spaa{123857649}$. There are in all $18$ cycle representations and we can classify them as
\begin{center}
\begin{tabular}{|c|c|}
  \hline
  V-type & $(1)(2)(3)(48)(5)(67)(9)~,~(12)(39)(4)(567)(8)~,~(13)(2)(4985)(6)(7)$ \\ \hline
  P-type & $(19432)(586)(7)~,~(12389)(457)(6)~,~(19)(247368)(5)~,~(1)(29)(346578)~,~(18764)(23)(59)$ \\ \hline
  Bad & $(142968753)~,~(163978524)~,~(173495)(26)(8)~,~(159836)(27)(4)~,~(182546937)$\\
  & $(135647928)~,~(15)(286974)(3)~,~(1796)(25)(384)~,~(1627)(35)(489)~,~(1458)(2637)(9)$ \\
  \hline
\end{tabular}
\end{center}
There are three V-type cycle representations. Each planar separation of V-type cycle representation tells us about the vertex information. The V-type cycle representation $(1)(2)(3)(48)(5)(67)(9)$ allows two different planar separations while  each of the other two allow one planar separation. The vertex information of them are presented as follows,
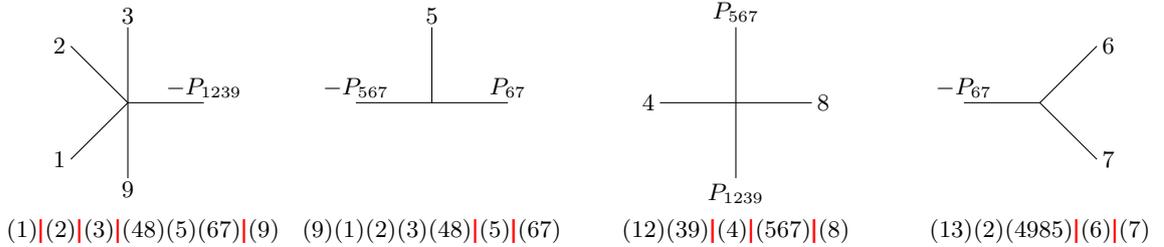
\begin{center}
\begin{tikzpicture}
  \draw[](1.25,0.75)--(2,0)--(1.25,-0.75);
  \draw[](2,1)--(2,-1);
  \draw[](2,0)--(3,0);
  \node at (2,-1.15) []{{\footnotesize $9$}};
  \node at (1.1,-0.75) []{{\footnotesize $1$}};
  \node at (1.1,0.75) []{{\footnotesize $2$}};
  \node at (2,1.15) []{{\footnotesize $3$}};
  \node at (3,0.2) []{{\footnotesize $-P_{1239}$}};
  \node at (2.2,-1.7) []{{\footnotesize $(1){\color{red}\pmb{|}}(2){\color{red}\pmb{|}}(3){\color{red}\pmb{|}}(48)(5)(67){\color{red}\pmb{|}}(9)$}};

  \draw[](5,0)--(7,0);
  \draw[](6,0)--(6,1);
  \node at (5,0.2) []{{\footnotesize $-P_{567}$}};
  \node at (7,0.2) []{{\footnotesize $P_{67}$}};
  \node at (6,1.15) []{{\footnotesize $5$}};
  \node at (6,-1.7) []{{\footnotesize $(9)(1)(2)(3)(48){\color{red}\pmb{|}}(5){\color{red}\pmb{|}}(67)$}};

  \draw[](9,0)--(11,0);
  \draw[](10,-1)--(10,1);
  \node at (8.85,0) []{{\footnotesize $4$}};
  \node at (11.15,0) []{{\footnotesize $8$}};
  \node at (10,1.2) []{{\footnotesize $P_{567}$}};
  \node at (10,-1.2) []{{\footnotesize $P_{1239}$}};
  \node at (10,-1.7) []{{\footnotesize $(12)(39){\color{red}\pmb{|}}(4){\color{red}\pmb{|}}(567){\color{red}\pmb{|}}(8)$}};

  \draw[](13,0)--(14,0);
  \draw[](14.75,0.75)--(14,0)--(14.75,-0.75);
  \node at (14.9,-0.75) []{{\footnotesize $7$}};
  \node at (14.9,0.75) []{{\footnotesize $6$}};
  \node at (13,0.2) []{{\footnotesize $-P_{67}$}};
  \node at (14,-1.7) []{{\footnotesize $(13)(2)(4985){\color{red}\pmb{|}}(6){\color{red}\pmb{|}}(7)$}};

\end{tikzpicture}
\end{center}
The first planar separation indicates a five point vertex, which corresponds to five possible terms, while the second planar separation indicates a quartic vertex which corresponds to two possible terms. So gluing them together we get the nine point effective Feynman diagram as
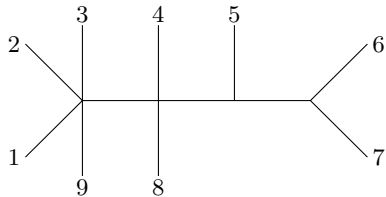
\begin{center}
\begin{tikzpicture}
  \draw[](1.25,0.75)--(2,0)--(1.25,-0.75);
  \draw[](2,1)--(2,-1);
  \draw[](2,0)--(5,0);
  \draw[](3,1)--(3,-1);
  \draw[](4,0)--(4,1);
  \draw[](5.75,0.75)--(5,0)--(5.75,-0.75);
  \node at (2,-1.15) []{{\footnotesize $9$}};
  \node at (1.1,-0.75) []{{\footnotesize $1$}};
  \node at (1.1,0.75) []{{\footnotesize $2$}};
  \node at (2,1.15) []{{\footnotesize $3$}};
  \node at (3,1.15) []{{\footnotesize $4$}};
  \node at (4,1.15) []{{\footnotesize $5$}};
  \node at (5.9,0.75) []{{\footnotesize $6$}};
  \node at (5.9,-0.75) []{{\footnotesize $7$}};
  \node at (3,-1.15) []{{\footnotesize $8$}};

\end{tikzpicture}
\end{center}
which is a collection of 10 trivalent Feynman diagrams, agreeing with the result of the CHY-integrand $\Spaa{123456789}\times\Spaa{123857649}$.

Alternatively, we derive the above result from one V-type cycle representation and its substructures. For instance, the planar separation $(1){\pmb{|}}(2){\pmb{|}}(3){\pmb{|}}(48)(5)(67){\pmb{|}}(9)$ indicates a five point vertex, while the substructure $(P_{1239})(48)(5)(67)$ has further structure. By working out the equivalent class of $(P_{1239})(48)(5)(67)$, we find a V-type cycle representation of this substructure that allows the planar separation $(4)\pmb{|}(567)\pmb{|}(8)\pmb{|}(P_{1239})$, indicating a quartic vertex and a four point substructure $(567)(P_{567})$. Then the equivalent V-type cycle representation $(5)\pmb{|}(67)\pmb{|}(P_{567})$ of this four point substructure indicates two cubic vertices with legs $5$, $P_{67}$, $P_{567}$ and $6$, $7$, $P_{67}$ respectively, according to our four point discussion in \S\ref{secFeynmanSub1}. The above recursive process can be graphically represented by
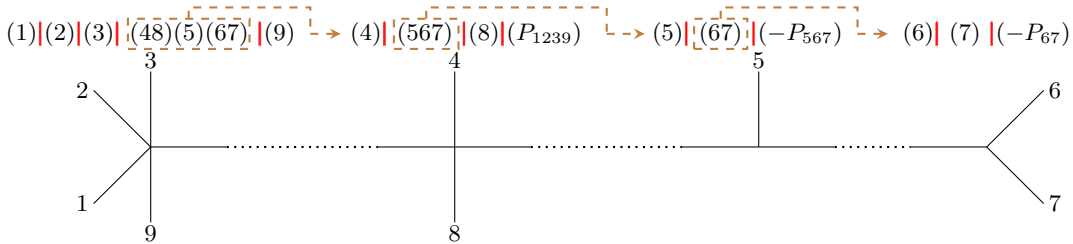
\begin{center}
\begin{tikzpicture}

  \node at (1,1.5) []{{\footnotesize $(1){\color{red}\pmb{|}}(2){\color{red}\pmb{|}}(3){\color{red}\pmb{|}}~(48)(5)(67)~{\color{red}\pmb{|}}(9)$}};
  \draw[](0.25,0.75)--(1,0)--(0.25,-0.75);
  \draw[](1,1)--(1,-1);
  \draw[](1,0)--(2,0)(4,0)--(5,0);
  \draw[dotted,thick](2,0)--(4,0);
  \draw[brown,dashed,thick](0.7,1.3)--(2.3,1.3)--(2.3,1.7)--(0.7,1.7)--(0.7,1.3);
  \draw[brown,dashed,thick](1.5,1.7)--(1.5,1.85)--(3.1,1.85)--(3.1,1.5);
  \draw[brown,dashed,thick,->,>=stealth](3.1,1.5)--(3.5,1.5);
  \node at (1,-1.15)[]{{\footnotesize $9$}};
  \node at (0.1,-0.75)[]{{\footnotesize $1$}};
  \node at (0.1,0.75)[]{{\footnotesize $2$}};
  \node at (1,1.15)[]{{\footnotesize $3$}};
  \node at (5,1.15)[]{{\footnotesize $4$}};
  \node at (9,1.15)[]{{\footnotesize $5$}};
  \node at (12.9,0.75)[]{{\footnotesize $6$}};
  \node at (12.9,-0.75)[]{{\footnotesize $7$}};
  \node at (5,-1.15)[]{{\footnotesize $8$}};

  \node at (5.15,1.5) []{{\footnotesize $(4){\color{red}\pmb{|}}~(567)~{\color{red}\pmb{|}}(8){\color{red}\pmb{|}}(P_{1239})$}};
  \draw[brown,dashed,thick](4.2,1.3)--(5.05,1.3)--(5.05,1.7)--(4.2,1.7)--(4.2,1.3);
  \draw[brown,dashed,thick](4.625,1.7)--(4.625,1.85)--(7,1.85)--(7,1.5);
  \draw[brown,dashed,thick,->,>=stealth](7,1.5)--(7.5,1.5);
  \draw[](5,1)--(5,-1);
  \draw[](5,0)--(6,0) (8,0)--(9,0);
  \draw[dotted,thick](6,0)--(8,0);

  \node at (8.85,1.5)[]{{\footnotesize $(5){\color{red}\pmb{|}}~(67)~{\color{red}\pmb{|}}(-P_{567})$}};
  \draw[brown, dashed, thick](8.15,1.3)--(8.85,1.3)--(8.85,1.7)--(8.15,1.7)--(8.15,1.3);
  \draw[brown, dashed,thick](8.5,1.7)--(8.5,1.85)--(10.3,1.85)--(10.3,1.5);
  \draw[brown,dashed,thick,->,>=stealth](10.3,1.5)--(10.7,1.5);
  \draw[](9,0)--(9,1);
  \draw[](9,0)--(10,0) (11,0)--(12,0);
  \draw[dotted,thick](10,0)--(11,0);

  \node at (12,1.5)[]{{\footnotesize $(6){\color{red}\pmb{|}}~(7)~{\color{red}\pmb{|}}(-P_{67})$}};
  \draw[](12.75,0.75)--(12,0)--(12.75,-0.75);

\end{tikzpicture}
\end{center}
Hence from one V-type cycle representation it is sufficient to obtain the complete result.

On the contrary, if taking only one P-type cycle representation, we will end up with a partial result. For instance, if we take the P-type cycle representation $(19432)(586)(7)$, it has only one planar separation $(19432)\pmb{|}(586)(7)$ that splits the external legs into two parts. For both the substructures $(19432)(P_{5678})$ and $(P_{5678})(586)(7)$, we can work out their contributions by recursively going into their substructures. We will not repeat the details but show the result as follows,
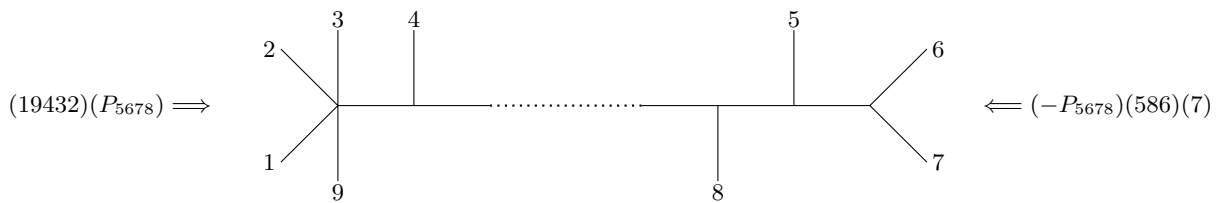
\begin{center}
\begin{tikzpicture}
  \draw[](4.25,0.75)--(5,0)--(4.25,-0.75) (5,1)--(5,-1) (5,0)--(6,0)--(6,1) (6,0)--(7,0);
  \draw[dotted,thick](7,0)--(9,0);
  \draw[](9,0)--(10,0)--(10,-1) (10,0)--(11,0)--(11,1) (11,0)--(12,0) (12.75,0.75)--(12,0)--(12.75,-0.75);

  \node at (5,-1.15) []{{\footnotesize $9$}};
  \node at (4.1,-0.75) []{{\footnotesize $1$}};
  \node at (4.1,0.75) []{{\footnotesize $2$}};
  \node at (5,1.15) []{{\footnotesize $3$}};
  \node at (6,1.15) []{{\footnotesize $4$}};
  \node at (11,1.15) []{{\footnotesize $5$}};
  \node at (12.9,0.75) []{{\footnotesize $6$}};
  \node at (12.9,-0.75) []{{\footnotesize $7$}};
  \node at (10,-1.15) []{{\footnotesize $8$}};

  \node at (2,0) []{{\footnotesize $(19432)(P_{5678})\Longrightarrow$}};
  \node at (15,0) []{{\footnotesize $\Longleftarrow (-P_{5678})(586)(7)$}};

\end{tikzpicture}
\end{center}
The corresponding effective Feynman diagram obtained by gluing these two subdiagrams contributes only half of the full result, since the leg $4$ and $8$ is connected in just one way of the two allowed by the quartic vertex $\{P_{9123},4,P_{567},8\}$ in the original Feynman diagram.
We need to include all the P-type cycle representations to reproduce the complete result.

\section{From Feynman diagrams to permutations}
\label{secPermutation}

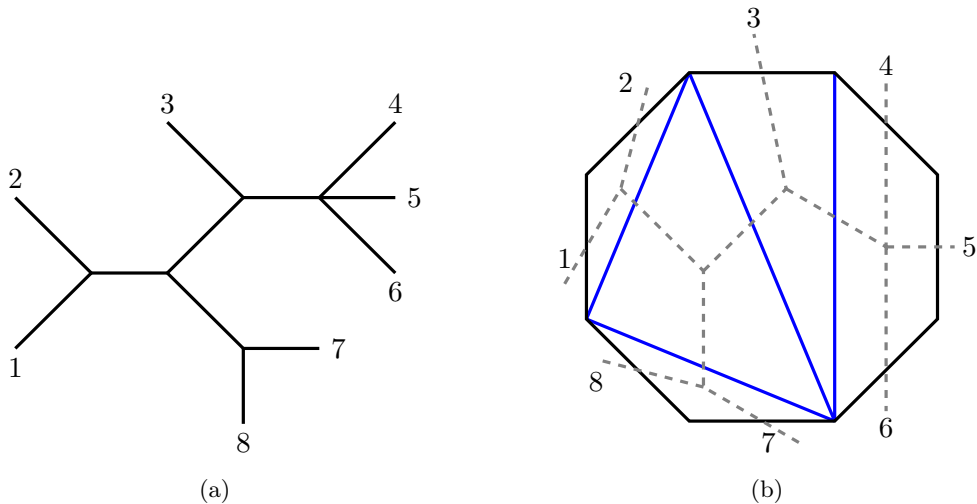
\begin{figure}
\centering
\subfloat[][]{\begin{tikzpicture}

    \draw[very thick](0,0)--(1,1)--(2,1)--(3,2)--(4,2)--(5,3);
    \draw[very thick](1,1)--(0,2);
    \draw[very thick](3,2)--(2,3);
    \draw[very thick](4,2)--(5,1);
    \draw[very thick](2,1)--(3,0)--(4,0);
    \draw[very thick](3,0)--(3,-1);
    \draw[very thick](4,2)--(5,2);

    \node (A1) at (0,-0.25)[]{$1$};
    \node (A2) at (0,2.25)[]{$2$};
    \node (A3) at (2,3.25)[]{$3$};
    \node (A4) at (5,3.25)[]{$4$};
    \node (A5) at (5.25,2)[]{$5$};
    \node (A6) at (5,0.75)[]{$6$};
    \node (A7) at (4.25,0)[]{$7$};
    \node (A8) at (3,-1.25)[]{$8$};
  \end{tikzpicture}\label{8pFeyn}}\qquad\qquad
\subfloat[][]{\begin{tikzpicture}[vertex/.style={inner sep=0pt}]
\coordinate (g1) at (22.5:2.5);
\coordinate (g2) at (67.5:2.5);
\coordinate (g3) at (112.5:2.5);
\coordinate (g4) at (157.5:2.5);
\coordinate (g5) at (-157.5:2.5);
\coordinate (g6) at (-112.5:2.5);
\coordinate (g7) at (-67.5:2.5);
\coordinate (g8) at (-22.5:2.5);
\draw [very thick,blue] (g2) -- (g7) (g7) -- (g3)  (g3) -- (g5) (g5) -- (g7);
\draw [very thick] (g1) -- (g2) -- (g3) -- (g4) -- (g5) -- (g6) -- (g7) -- (g8) -- cycle;
\coordinate (V12) at (barycentric cs:g3=1,g4=1,g5=1);
\coordinate (V1) at (barycentric cs:g3=-0.35,g4=1,g5=1);
\coordinate (V2) at (barycentric cs:g3=1,g4=1,g5=-0.35);
\coordinate (V456) at (barycentric cs:g7=1,g8=1,g1=1,g2=1);
\coordinate (V4) at (barycentric cs:V456=-0.5,g1=1,g2=1);
\coordinate (V5) at (barycentric cs:V456=-0.5,g1=1,g8=1);
\coordinate (V6) at (barycentric cs:V456=-0.5,g7=1,g8=1);
\coordinate (V78) at (barycentric cs:g5=1,g6=1,g7=1);
\coordinate (V7) at (barycentric cs:g5=-0.35,g6=1,g7=1);
\coordinate (V8) at (barycentric cs:g5=1,g6=1,g7=-0.35);
\coordinate (V7812) at (barycentric cs:g3=1,g5=1,g7=1);
\coordinate (V3456) at (barycentric cs:g2=1,g3=1,g7=1);
\coordinate (V3) at (barycentric cs:g2=1,g3=1,g7=-0.2);
\draw [very thick,gray,dashed] (V1) -- (V12) -- (V2) (V12) -- (V7812) -- (V78) -- (V7) (V78) -- (V8) (V7812) -- (V3456) -- (V3) (V3456) -- (V456) -- (V4) (V456) -- (V5) (V456) -- (V6);
\node at (V1) [above=2pt] {$1$};
\node at (V2) [left=2pt]{$2$};
\node at (V3) [above=-1pt] {$3$};
\node at (V4) [above=-1pt]{$4$};
\node at (V5) [right=-1pt]{$5$};
\node at (V6) [below=-1pt]{$6$};
\node at (V7) [left=5pt]{$7$};
\node at (V8) [below=1pt]{$8$};
\end{tikzpicture}\label{8pngon}}
\caption{An eight point Feynman diagram and its dual $n$-gon diagram.}\label{Feyndiagram}
\end{figure}

In \S \ref{secFeynman}, we have addressed the problem that given a PT-factor as a permutation acting on the identity element, how we can determine the Feynman diagrams the CHY-integrand evaluated. In this section, we will consider the inverse problem, namely, given an effective Feynman diagram, how to obtain directly the corresponding good cycle representations.
We will show that there is a recursive construction to produce the good cycle representations of a given Feynman diagram from the relation between subdiagrams and planar separations. Later, we will use the eight point example given in Fig.~\ref{Feyndiagram} to illustrate general discussions.
%
We remind the readers that an $m$-point vertex in the Feynman diagram stands for the sum of all the $m$-point trivalent diagrams. For example, a quartic vertex gives the sum of $s$ and $t$ channel trivalent diagrams.
The Feynman diagram in Fig.~\ref{8pFeyn} corresponds to the CHY-integrand $\Spaa{12345678}\times\Spaa{12783654}$, which evaluates to
\bea \frac{1}{s_{12}s_{78}s_{1278}s_{456}}\left(\frac{1}{s_{45}}+\frac{1}{s_{56}}\right)~.~~~\eea
As we have mentioned in previous section, the PT-factor $\PT(\pmb{\b})$ determines a permutation acting on identity element and it encodes the pole structure of the Feynman diagram. Conversely, the pole structure in the Feynman diagram also encodes the information of permutation.
To describe the way of reading out the PT-factor and cycle representations, we find it is more convenient to introduce the polygon with $n$ edges ($n$-gon) that is dual to the $n$-point Feynman diagram under inspection.

An $n$-point effective Feynman diagram can be described as \emph{partial triangulation} of $n$-gon diagram~\cite{Gao:2017dek,delaCruz:2017zqr,Arkani-Hamed:2017mur}, and an example of our considered eight point Feynman diagram is presented in Fig.~\ref{Feyndiagram}. Each external leg is dual to an edge of the $n$-gon, and each vertex is dual to a subpolygon in the interior. A triangulation line inside the $n$-gon, which cut it into two subpolygons, is dual to a propagator. If the Feynman diagram considered is trivalent diagram with only cubic vertices, the corresponding $n$-gon is completely triangulated, while if it is an effective Feynman diagram with also higher point vertices, the $n$-gon diagram is partially triangulated. If we use $E_i$ to denote the number of edges of a subpolygon inside the original $n$-gon, then the number of terms in the final result is given by
\bea \prod_{i\in {\text{all~polygons}}}C(E_i)=\prod_{i\in {\text{all~polygons}}}\frac{2^{E_i-2}(2E_i-5)!!}{(E_i-1)!}~,~~~\eea
where $C(n)$ is also the number of all possible $n$-point color ordered trivalent Feynman diagrams. The blue line in Fig.~\ref{8pngon} represents the triangulation of our considered example, and the dashed gray line gives the Feynman diagram dual to the partial triangulation of $n$-gon. Discussion on the Feynman diagram can as well be applied to the $n$-gon diagram, and the latter is naturally enrolled in the associahedron~\cite{delaCruz:2017zqr,Arkani-Hamed:2017mur}.

\subsection{The zig-zag path and cycle-representation of permutation}
\label{secPermutation1}

Now let us return to the problem of reading out the PT-factor from Feynman diagram. A solution for this problem has been provided in paper \cite{Baadsgaard:2015ifa}, where a pictorial method has been proposed to write the PT-factor for a given Feynman diagram. Based on their discussion, we will rephrase it by the language of \emph{zig-zag path}.\footnote{More discussions on the zig-zag path can be found in~\cite{Kenyon,Hanany:2005ss,Feng:2005gw}.} The basic idea comes as follows,
\begin{itemize}
\item Any tree-level Feynman diagram can be placed as a planar diagram, while the external legs lying in the plane apparently define an ordering, identified as PT-factor $\PT(\pmb{\alpha})$.
\item Starting from any external leg, we can draw a zig-zag path along the boundary of diagram, which crosses each internal line it meets and closes at the starting point.
The ordering of legs along the direction of zig-zag path is identified as PT-factor $\PT(\pmb{\beta})$.
\end{itemize}
The corresponding CHY-integrand $\PT(\pmb{\alpha})\times\PT(\pmb{\beta})$ then evaluates to the given Feynman diagram.

\begin{figure}
  \centering
\subfloat[][]{\begin{tikzpicture}

    \draw[very thick](0,0)--(1,1)--(2,1)--(3,2)--(4,2)--(5,3);
    \draw[very thick](1,1)--(0,2);
    \draw[very thick](3,2)--(2,3);
    \draw[very thick](4,2)--(5,1);
    \draw[very thick](2,1)--(3,0)--(4,0);
    \draw[very thick](3,0)--(3,-1);
    \draw[very thick](4,2)--(5,2);

    \node (A1) at (0,-0.25)[]{$1$};
    \node (A2) at (0,2.25)[]{$2$};
    \node (A3) at (2,3.25)[]{$3$};
    \node (A4) at (5,3.25)[]{$4$};
    \node (A5) at (5.25,2)[]{$5$};
    \node (A6) at (5,0.75)[]{$6$};
    \node (A7) at (4.25,0)[]{$7$};
    \node (A8) at (3,-1.25)[]{$8$};

    \tikzstyle{line}=[red,very thick, dashed]
    \tikzstyle{arrow}=[red,->,>=latex,shorten >=-4pt, very thick,dashed]

    \draw[arrow](0,-0.5)--(-0.25,-0.25);
    \draw[line](-0.25,-0.25)--(-0.5,0);
    \draw[arrow](-0.5,0)--(0,0.5);
    \draw[line](0,0.5)--(0.5,1);
    \draw[arrow](0.5,1)--(0,1.5);
    \draw[line](0,1.5)--(-0.5,2);
    \draw[arrow] (-0.5,2)--(-0.25,2.25);
    \draw[line](-0.25,2.25)--(0,2.5);
    \draw[arrow](0,2.5)--(1,1.5);
    \draw[line](1,1.5)--(2,0.5);
    \draw[arrow](2,0.5)--(3.5,0.5);
    \draw[line](3.5,0.5)--(4.5,0.5);
    \draw[arrow](4.5,0.5)--(4.5,0);
    \draw[line](4.5,0)--(4.5,-0.5);
    \draw[arrow](4.5,-0.5)--(4,-0.5);
    \draw[line](4,-0.5)--(3.5,-0.5);
    \draw[arrow](3.5,-0.5)--(3.5,-1);
    \draw[line](3.5,-1)--(3.5,-1.5);
    \draw[arrow](3.5,-1.5)--(3,-1.5);
    \draw[line](3,-1.5)--(2.5,-1.5);
    \draw[arrow](2.5,-1.5)--(2.5,-0.5);
    \draw[line](2.5,-0.5)--(2.5,0.5);
    \draw[arrow](2.5,0.5)--(2.5,1);
    \draw[line](2.5,1)--(2.5,2);
    \draw[arrow](2.5,2)--(2,2.5);
    \draw[line](2,2.5)--(1.5,3);
    \draw[arrow](1.5,3)--(1.75,3.25);
    \draw[line](1.75,3.25)--(2,3.5);
    \draw[arrow](2,3.5)--(2.75,2.75);
    \draw[arrow](2.75,2.75)--(4.25,1.25);
    \draw[line](4.25,1.25)--(5,0.5);
    \draw[arrow](5,0.5)--(5.25,0.75);
    \draw[line](5.25,0.75)--(5.5,1);

    \draw[arrow](5.5,1)--(5.125,1.375);
    \draw[line](5.125,1.375)--(4.75,1.75);
    \draw[arrow](4.75,1.75)--(5.25,1.75);
    \draw[line](5.25,1.75)--(5.5,1.75);
    \draw[arrow](5.5,1.75)--(5.5,2);
    \draw[line](5.5,2)--(5.5,2.25);
    \draw[arrow](5.5,2.25)--(5.25,2.25);
    \draw[line](5.25,2.25)--(4.75,2.25);
    \draw[arrow](4.75,2.25)--(5.125,2.625);
    \draw[line](5.125,2.625)--(5.5,3);

    \draw[arrow](5.5,3)--(5.25,3.25);
    \draw[line](5.25,3.25)--(5,3.5);
    \draw[arrow](5,3.5)--(4,2.5);
    \draw[line](4,2.5)--(3,1.5);
    \draw[arrow](3,1.5)--(2.25,1.5);
    \draw[line](2.25,1.5)--(2,1.5);
    \draw[arrow](2,1.5)--(0.75,0.25);
    \draw[line](0.75,0.25)--(0,-0.5);




  \end{tikzpicture}\label{pathFeyn}}\qquad\qquad
\subfloat[][]{\begin{tikzpicture}
	\coordinate (g1) at (22.5:2.5);
	\coordinate (g2) at (67.5:2.5);
	\coordinate (g3) at (112.5:2.5);
	\coordinate (g4) at (157.5:2.5);
	\coordinate (g5) at (-157.5:2.5);
	\coordinate (g6) at (-112.5:2.5);
	\coordinate (g7) at (-67.5:2.5);
	\coordinate (g8) at (-22.5:2.5);
	\draw [very thick,blue] (g2) -- (g7) (g7) -- (g3)  (g3) -- (g5) (g5) -- (g7);
	\draw [very thick] (g1) -- (g2) -- (g3) -- (g4) -- (g5) -- (g6) -- (g7) -- (g8) -- cycle;
	\path [name path=p12] ($(g1)!0.15cm!90:(g2)$) -- ++(135:2);
	\path [name path=p23] ($(g2)!0.15cm!90:(g3)$) -- ++(-2,0);
	\path [name path=p81] ($(g8)!0.15cm!90:(g1)$) -- ++(0,2);
	\path [name path=p78] ($(g7)!0.15cm!90:(g8)$) -- ++(45:2);
	\path [name path=p72a] ($(g2)!0.15cm!90:(g7)$) -- ++(0,-5);
	\path [name path=p72b] ($(g2)!-0.15cm!90:(g7)$) -- ++(0,-5);
	\path [name path=p73a] ($(g3)!0.15cm!90:(g7)$) -- ++(-67.5:5);
	\path [name path=p73b] ($(g3)!-0.15cm!90:(g7)$) -- ++(-67.5:5);
	\path [name path=p34] ($(g3)!0.15cm!90:(g4)$) -- ++(-135:2);
	\path [name path=p45] ($(g4)!0.15cm!90:(g5)$) -- ++(0,-2);
	\path [name path=p35a] ($(g3)!0.15cm!90:(g5)$) -- ++(-112.5:5);
	\path [name path=p35b] ($(g3)!-0.15cm!90:(g5)$) -- ++(-112.5:5);
	\path [name path=p57a] ($(g5)!0.15cm!90:(g7)$) -- ++(-22.5:5);
	\path [name path=p57b] ($(g5)!-0.15cm!90:(g7)$) -- ++(-22.5:5);
	\path [name path=p56] ($(g5)!0.15cm!90:(g6)$) -- ++(-45:2);
	\path [name path=p67] ($(g6)!0.15cm!90:(g7)$) -- ++(2,0);
	\path [name path=c72] ($(g2)!0.5!(g7)$) circle (0.3cm);
	\path [name path=c73] ($(g3)!0.5!(g7)$) circle (0.3cm);
	\path [name path=c35] ($(g3)!0.5!(g5)$) circle (0.3cm);
	\path [name path=c57] ($(g7)!0.5!(g5)$) circle (0.3cm);
	
	\path [name intersections={of=p12 and p81,name=a}] (a-1);
	\path [name intersections={of=p78 and p81,name=b}] (b-1);
	\path [name intersections={of=p78 and p72a,name=c}] (c-1);
	\path [name intersections={of=p12 and p72a,name=d}] (d-1);
	\path [name intersections={of=p23 and p72b,name=e}] (e-1);
	\path [name intersections={of=p23 and p73a,name=f}] (f-1);
	\path [name intersections={of=p72b and p73a,name=g}] (g-1);
	\path [name intersections={of=p73b and p35a,name=h}] (h-1);
	\path [name intersections={of=p34 and p35b,name=i}] (i-1);
	\path [name intersections={of=p34 and p45,name=j}] (j-1);
	\path [name intersections={of=p35b and p45,name=k}] (k-1);
	\path [name intersections={of=p57a and p35a,name=l}] (l-1);
	\path [name intersections={of=p57a and p73b,name=m}] (m-1);
	\path [name intersections={of=p57b and p56,name=n}] (n-1);
	\path [name intersections={of=p67 and p56,name=o}] (o-1);
	\path [name intersections={of=p57b and p67,name=p}] (p-1);
	
	\path [name intersections={of=c72 and p72a,name=q}] (q-1);
	\path [name intersections={of=c72 and p72a,name=q}] (q-2);
	
	\path [name intersections={of=c72 and p72b,name=r}] (r-1);
	\path [name intersections={of=c72 and p72b,name=r}] (r-2);
	
	\path [name intersections={of=c73 and p73a,name=s}] (s-1);
	\path [name intersections={of=c73 and p73a,name=s}] (s-2);
	
	\path [name intersections={of=c73 and p73b,name=t}] (t-1);
	\path [name intersections={of=c73 and p73b,name=t}] (t-2);
	
	\path [name intersections={of=c35 and p35a,name=u}] (u-1);
	\path [name intersections={of=c35 and p35a,name=u}] (u-2);
	
	\path [name intersections={of=c35 and p35b,name=v}] (v-1);
	\path [name intersections={of=c35 and p35b,name=v}] (v-2);
	
	\path [name intersections={of=c57 and p57a,name=w}] (w-1);
	\path [name intersections={of=c57 and p57a,name=w}] (w-2);
	
	\path [name intersections={of=c57 and p57b,name=x}] (x-1);
	\path [name intersections={of=c57 and p57b,name=x}] (x-2);
	
	\draw [red,dashed,thick,rounded corners=3pt] (a-1) -- (b-1) -- (c-1) -- (q-2) -- (r-1) -- (e-1) -- (f-1) -- (s-1) -- (t-2) -- (m-1) -- (w-1) -- (x-1) -- (n-1) -- (o-1) -- (p-1) -- (x-2) -- (w-2) -- (l-1) -- (u-2) -- (v-1) -- (i-1) -- (j-1) -- (k-1) -- (v-2) -- (u-1) -- (h-1) -- (t-1) -- (s-2) -- (g-1) -- (r-2) -- (q-1) -- (d-1) -- cycle;
	
	\node at ($(g1)!0.5!(g2)$) [above right=-1.5pt]{$4$};
	\node at ($(g2)!0.5!(g3)$) [above=0pt]{$3$};
	\node at ($(g3)!0.5!(g4)$) [above left=-1.5pt]{$2$};
	\node at ($(g4)!0.5!(g5)$) [left=0pt]{$1$};
	\node at ($(g5)!0.5!(g6)$) [below left=-1.5pt]{$8$};
	\node at ($(g6)!0.5!(g7)$) [below=0pt]{$7$};
	\node at ($(g7)!0.5!(g8)$) [below right=-1.5pt]{$6$};
	\node at ($(g8)!0.5!(g1)$) [right=0pt]{$5$};
	
	\node at ($(a-1)!0.5!(b-1)$) [rotate=0,red] {\tikz\draw[-Latex](0,0);};
	\node at ($(b-1)!0.5!(c-1)$) [rotate=-45,red] {\tikz\draw[-Latex](0,0);};
	\node at ($(c-1)!0.5!(q-2)$) [rotate=180,red] {\tikz\draw[-Latex](0,0);};
	\node at ($(q-1)!0.5!(d-1)$) [rotate=180,red] {\tikz\draw[-Latex](0,0);};
	\node at ($(d-1)!0.5!(a-1)$) [rotate=45,red] {\tikz\draw[-Latex](0,0);};
	\node at ($(f-1)!0.5!(e-1)$) [rotate=-90,red] {\tikz\draw[-Latex](0,0);};
	\node at ($(e-1)!0.5!(r-2)$) [rotate=180,red] {\tikz\draw[-Latex](0,0);};
	\node at ($(r-2)!0.5!(g-1)$) [rotate=180,red] {\tikz\draw[-Latex](0,0);};
	\node at ($(g-1)!0.5!(s-2)$) [rotate=22.5,red] {\tikz\draw[-Latex](0,0);};
	\node at ($(s-1)!0.5!(f-1)$) [rotate=22.5,red] {\tikz\draw[-Latex](0,0);};
	
	\node at ($(t-1)!0.5!(h-1)$) [rotate=22.5,red] {\tikz\draw[-Latex](0,0);};
	\node at ($(h-1)!0.5!(u-1)$) [rotate=157.5,red] {\tikz\draw[-Latex](0,0);};
	\node at ($(u-2)!0.5!(l-1)$) [rotate=157.5,red] {\tikz\draw[-Latex](0,0);};
	\node at ($(l-1)!0.5!(w-2)$) [rotate=-112.5,red] {\tikz\draw[-Latex](0,0);};
	\node at ($(w-1)!0.5!(m-1)$) [rotate=-112.5,red] {\tikz\draw[-Latex](0,0);};
	\node at ($(m-1)!0.5!(t-2)$) [rotate=22.5,red] {\tikz\draw[-Latex](0,0);};
	\node at ($(k-1)!0.5!(j-1)$) [rotate=0,red] {\tikz\draw[-Latex](0,0);};
	\node at ($(j-1)!0.5!(i-1)$) [rotate=-45,red] {\tikz\draw[-Latex](0,0);};
	\node at ($(p-1)!0.5!(o-1)$) [rotate=90,red] {\tikz\draw[-Latex](0,0);};
	\node at ($(o-1)!0.5!(n-1)$) [rotate=45,red] {\tikz\draw[-Latex](0,0);};
	\end{tikzpicture}\label{pathngon}}
  \caption{The zig-zag path in an eight point Feynman diagram and $n$-gon diagram.}\label{FigZigzag}
\end{figure}
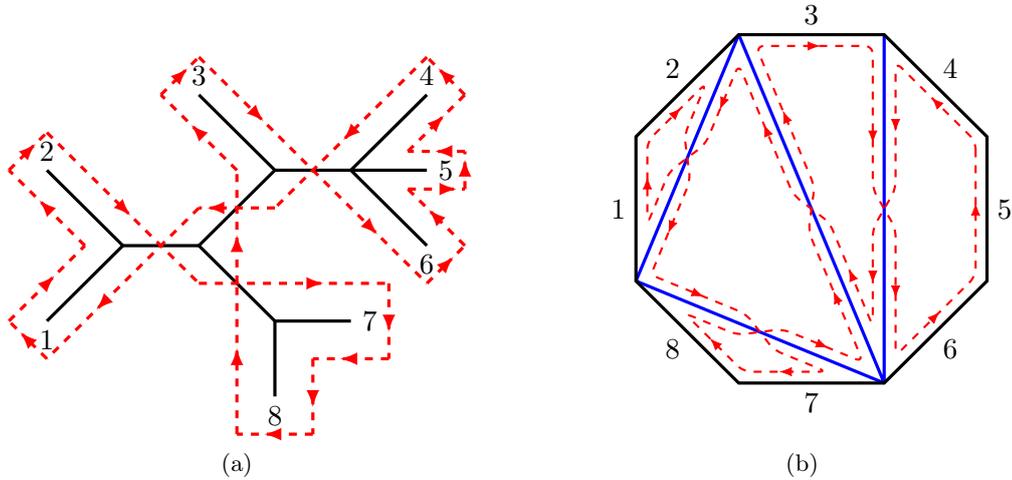
The zig-zag path for our considered example is shown in Fig.~\ref{FigZigzag}, both in the Feynman diagram and $n$-gon diagram. In
the Feynman diagram, the zig-zag path is along the external legs,
while in the $n$-gon diagram, it is along the interior of edges. In
both diagrams, the path crosses the lines whenever they are
propagators. It is easy to tell that, for the Feynman diagram shown
in Fig.~\ref{FigZigzag}, we have
$\PT(\pmb{\alpha})=\Spaa{12345678}$, while along the arrows of
zig-zag path, we can read out $\PT(\pmb{\beta})=\Spaa{12783654}$, as
it should be. However, there are two subtleties we should pay
attention to. The PT-factor $\PT(\pmb{\beta})=\Spaa{12783654}$ is
obtained under the condition that we read out the zig-zag path from
leg 1 towards a specific direction. If starting from the same leg
$1$ but with opposite direction, we will get
$\PT(\pmb{\beta})=\Spaa{14563872}$.  If we start from another leg
and a chosen direction, we will get another PT-factor, though they
are in the same equivalent class. Special attention should be paid
to the orientation of zig-zag path. It is a local but not global
property, and it only make sense with respect to a vertex. For
example, for the cubic vertex where legs $\{1, 2\}$ attached to, the
zig-zag path around it is clockwise, while for the quartic vertex
where legs $\{4, 5, 6\}$ attached to, the zig-zag path around it is
anti-clockwise. The orientation of zig-zag path is more obvious in
the $n$-gon diagram, as shown in Fig.~\ref{pathngon}. For each
polygon inside the $n$-gon, the zig-zag path can be considered as a
closed loop with definite orientation. If the zig-zag path in the
triangle with edge 1, 2 is clockwise, then the zig-zag paths in the
triangles with legs 7, 8 and with legs 3 are also clockwise, while
zig-zag paths in the triangle in middle and the quadrangle are anti-clockwise. To the whole Feynman diagram or $n$-gon diagram, we do not need to worry about the ambiguity of the zig-zag path orientation, since our canonical definition of PT-factors in~\S\ref{secSetup} fixes which leg to start and which direction it should ahead.
However, as we would discuss soon, the orientation of zig-zag path
is important in the recursive construction of PT-factors.

\begin{figure}
\centering
\subfloat[][]{\begin{tikzpicture}

    \draw[very thick](0,0)--(1,1)--(2,1)--(2.5,1.5);
    \draw[very thick](1,1)--(0,2);
    \draw[very thick](2,1)--(3,0)--(4,0);
    \draw[very thick](3,0)--(3,-1);

    \draw[very thick](3.5,1.5)--(4,2)--(5,2)--(6,3);
    \draw[very thick](4,2)--(3,3);
    \draw[very thick](5,2)--(6,1);
    \draw[very thick](5,2)--(6,2);

    \node (A1) at (0,-0.25)[]{$1$};
    \node (A2) at (0,2.25)[]{$2$};

    \node (A3) at (3,3.25)[]{$3$};
    \node (A4) at (6,3.25)[]{$4$};
    \node (A5) at (6.25,2)[]{$5$};
    \node (A6) at (6,0.75)[]{$6$};

    \node (A7) at (4.25,0)[]{$7$};
    \node (A8) at (3,-1.25)[]{$8$};

    \tikzstyle{line}=[red,very thick, dashed]
    \tikzstyle{arrow}=[red,->,>=latex,shorten >=-4pt, very thick,dashed]

    \draw[arrow](0,-0.5)--(-0.25,-0.25);
    \draw[line](-0.25,-0.25)--(-0.5,0);
    \draw[arrow](-0.5,0)--(0,0.5);
    \draw[line](0,0.5)--(0.5,1);
    \draw[arrow](0.5,1)--(0,1.5);
    \draw[line](0,1.5)--(-0.5,2);
    \draw[arrow] (-0.5,2)--(-0.25,2.25);
    \draw[line](-0.25,2.25)--(0,2.5);
    \draw[arrow](0,2.5)--(1,1.5);
    \draw[line](1,1.5)--(2,0.5);
    \draw[arrow](2,0.5)--(3.5,0.5);
    \draw[line](3.5,0.5)--(4.5,0.5);
    \draw[arrow](4.5,0.5)--(4.5,0);
    \draw[line](4.5,0)--(4.5,-0.5);
    \draw[arrow](4.5,-0.5)--(4,-0.5);
    \draw[line](4,-0.5)--(3.5,-0.5);
    \draw[arrow](3.5,-0.5)--(3.5,-1);
    \draw[line](3.5,-1)--(3.5,-1.5);
    \draw[arrow](3.5,-1.5)--(3,-1.5);
    \draw[line](3,-1.5)--(2.5,-1.5);
    \draw[arrow](2.5,-1.5)--(2.5,-0.5);
    \draw[line](2.5,-0.5)--(2.5,1);
   \draw[arrow](2.5,1)--(2.6875,1.1875);
    \draw[line](2.6875,1.1875)--(2.875,1.375);
    \draw[arrow](2.875,1.375)--(2.625,1.625);
   \draw[line](2.625,1.625)--(2.375,1.875);
   \draw[arrow](2.375,1.875)--(0.75,0.25);
    \draw[line](0.75,0.25)--(0,-0.5);

    \node (P1) at (2.5,1.25)[]{{\footnotesize $P$}};
     \node (P2) at (3.4,1.75)[]{{\footnotesize $-P$}};
     \node (V1) at (2,0.75)[]{{\footnotesize $V_1$}};
     \node (V2) at (4.05,1.75)[]{{\footnotesize $V_2$}};

   \draw[arrow](3.5,2)--(3,2.5);
   \draw[line](3,2.5)--(2.5,3);
   \draw[arrow](2.5,3)--(2.75,3.25);
   \draw[line](2.75,3.25)--(3,3.5);
  \draw[arrow](3,3.5)--(3.75,2.75);
   \draw[arrow](3.75,2.75)--(5.25,1.25);
   \draw[line](5.25,1.25)--(6,0.5);
    \draw[arrow](6,0.5)--(6.25,0.75);
    \draw[line](6.25,0.75)--(6.5,1);
    \draw[arrow](6.5,1)--(6.125,1.375);
    \draw[line](6.125,1.375)--(5.75,1.75);
    \draw[arrow](5.75,1.75)--(6.25,1.75);
    \draw[line](6.25,1.75)--(6.5,1.75);
    \draw[arrow](6.5,1.75)--(6.5,2);
    \draw[line](6.5,2)--(6.5,2.25);
    \draw[arrow](6.5,2.25)--(6.25,2.25);
    \draw[line](6.25,2.25)--(5.75,2.25);
    \draw[arrow](5.75,2.25)--(6.125,2.625);
    \draw[line](6.125,2.625)--(6.5,3);
    \draw[arrow](6.5,3)--(6.25,3.25);
    \draw[line](6.25,3.25)--(6,3.5);
    \draw[arrow](6,3.5)--(5,2.5);
    \draw[line](5,2.5)--(3.625,1.125);
   \draw[arrow](3.625,1.125)--(3.375,1.375);
    \draw[line](3.375,1.375)--(3.125,1.625);
   \draw[arrow](3.125,1.625)--(3.375,1.875);
   \draw[line](3.375,1.875)--(3.5,2);

	\node at (1.5,-0.75) {$L$};
	\node at (4.55,3) {$R$};




   \draw[dashed,very thick](2.5,2)--(3.5,1);

  \end{tikzpicture}\label{pathFeynFact}}\qquad\qquad
\subfloat[][]{\begin{tikzpicture}
	\begin{scope}[xshift=0.75cm]
	\coordinate (g1) at (22.5:2.5);
	\coordinate (g2) at (67.5:2.5);
	\coordinate (g3) at (112.5:2.5);
	\coordinate (g8) at (-22.5:2.5);
	\coordinate (g7) at (-67.5:2.5);
	
	\draw [very thick,blue] (g2) -- (g7) (g7) -- (g3);
	\draw [very thick] (g7) -- (g8) -- (g1) -- (g2)-- (g3);
	
	\path [name path=p12] ($(g1)!0.15cm!90:(g2)$) -- ++(135:2);
	\path [name path=p23] ($(g2)!0.15cm!90:(g3)$) -- ++(-2,0);
	\path [name path=p81] ($(g8)!0.15cm!90:(g1)$) -- ++(0,2);
	\path [name path=p78] ($(g7)!0.15cm!90:(g8)$) -- ++(45:2);
	\path [name path=p72a] ($(g2)!0.15cm!90:(g7)$) -- ++(0,-5);
	\path [name path=p72b] ($(g2)!-0.15cm!90:(g7)$) -- ++(0,-5);
	\path [name path=p73a] ($(g3)!0.15cm!90:(g7)$) -- ++(-67.5:5);
	\path [name path=c72] ($(g2)!0.5!(g7)$) circle (0.3cm);
	
	\path [name intersections={of=p12 and p81,name=a}] (a-1);
	\path [name intersections={of=p78 and p81,name=b}] (b-1);
	\path [name intersections={of=p78 and p72a,name=c}] (c-1);
	\path [name intersections={of=p12 and p72a,name=d}] (d-1);
	\path [name intersections={of=p23 and p72b,name=e}] (e-1);
	\path [name intersections={of=p23 and p73a,name=f}] (f-1);
	\path [name intersections={of=p72b and p73a,name=g}] (g-1);
	
	\path [name intersections={of=c72 and p72a,name=q}] (q-1);
	\path [name intersections={of=c72 and p72a,name=q}] (q-2);
	
	\path [name intersections={of=c72 and p72b,name=r}] (r-1);
	\path [name intersections={of=c72 and p72b,name=r}] (r-2);
	
	\draw [red,dashed,thick,rounded corners=3pt] (a-1) -- (b-1) -- (c-1) -- (q-2) -- (r-1) -- (e-1) -- (f-1) -- (g-1) -- (r-2) -- (q-1) -- (d-1) -- cycle;
	
	\node at ($(g1)!0.5!(g2)$) [above right=-1.5pt]{$4$};
	\node at ($(g2)!0.5!(g3)$) [above=0pt]{$3$};
	\node at ($(g7)!0.5!(g8)$) [below right=-1.5pt]{$6$};
	\node at ($(g8)!0.5!(g1)$) [right=0pt]{$5$};
	
	\node at (barycentric cs:g2=1,g3=1,g7=1) {$V_2$};
	
	\node at ($(a-1)!0.5!(b-1)$) [rotate=0,red] {\tikz\draw[-Latex](0,0);};
	\node at ($(b-1)!0.5!(c-1)$) [rotate=-45,red] {\tikz\draw[-Latex](0,0);};
	\node at ($(c-1)!0.5!(q-2)$) [rotate=180,red] {\tikz\draw[-Latex](0,0);};
	\node at ($(q-1)!0.5!(d-1)$) [rotate=180,red] {\tikz\draw[-Latex](0,0);};
	\node at ($(d-1)!0.5!(a-1)$) [rotate=45,red] {\tikz\draw[-Latex](0,0);};
	\node at ($(f-1)!0.5!(e-1)$) [rotate=-90,red] {\tikz\draw[-Latex](0,0);};
	\node at ($(e-1)!0.5!(r-2)$) [rotate=180,red] {\tikz\draw[-Latex](0,0);};
	\node at ($(r-2)!0.5!(g-1)$) [rotate=180,red] {\tikz\draw[-Latex](0,0);};
	\node at ($(g-1)!0.5!(f-1)$) [rotate=22.5,red] {\tikz\draw[-Latex](0,0);};
	\end{scope}
	
	\begin{scope}[xshift=-0.75cm]
	\coordinate (g3) at (112.5:2.5);
	\coordinate (g4) at (157.5:2.5);
	\coordinate (g5) at (-157.5:2.5);
	\coordinate (g6) at (-112.5:2.5);
	\coordinate (g7) at (-67.5:2.5);
	
	\draw [very thick,blue]  (g7) -- (g3)  (g3) -- (g5) (g5) -- (g7);
	\draw [very thick](g3) -- (g4) -- (g5) -- (g6) -- (g7);
	
	\path [name path=p73b] ($(g3)!-0.15cm!90:(g7)$) -- ++(-67.5:5);
	\path [name path=p34] ($(g3)!0.15cm!90:(g4)$) -- ++(-135:2);
	\path [name path=p45] ($(g4)!0.15cm!90:(g5)$) -- ++(0,-2);
	\path [name path=p35a] ($(g3)!0.15cm!90:(g5)$) -- ++(-112.5:5);
	\path [name path=p35b] ($(g3)!-0.15cm!90:(g5)$) -- ++(-112.5:5);
	\path [name path=p57a] ($(g5)!0.15cm!90:(g7)$) -- ++(-22.5:5);
	\path [name path=p57b] ($(g5)!-0.15cm!90:(g7)$) -- ++(-22.5:5);
	\path [name path=p56] ($(g5)!0.15cm!90:(g6)$) -- ++(-45:2);
	\path [name path=p67] ($(g6)!0.15cm!90:(g7)$) -- ++(2,0);
	
	\path [name path=c35] ($(g3)!0.5!(g5)$) circle (0.3cm);
	\path [name path=c57] ($(g7)!0.5!(g5)$) circle (0.3cm);
	
	\path [name intersections={of=p73b and p35a,name=h}] (h-1);
	\path [name intersections={of=p34 and p35b,name=i}] (i-1);
	\path [name intersections={of=p34 and p45,name=j}] (j-1);
	\path [name intersections={of=p35b and p45,name=k}] (k-1);
	\path [name intersections={of=p57a and p35a,name=l}] (l-1);
	\path [name intersections={of=p57a and p73b,name=m}] (m-1);
	\path [name intersections={of=p57b and p56,name=n}] (n-1);
	\path [name intersections={of=p67 and p56,name=o}] (o-1);
	\path [name intersections={of=p57b and p67,name=p}] (p-1);
	
	\path [name intersections={of=c35 and p35a,name=u}] (u-1);
	\path [name intersections={of=c35 and p35a,name=u}] (u-2);
	
	\path [name intersections={of=c35 and p35b,name=v}] (v-1);
	\path [name intersections={of=c35 and p35b,name=v}] (v-2);
	
	\path [name intersections={of=c57 and p57a,name=w}] (w-1);
	\path [name intersections={of=c57 and p57a,name=w}] (w-2);
	
	\path [name intersections={of=c57 and p57b,name=x}] (x-1);
	\path [name intersections={of=c57 and p57b,name=x}] (x-2);

	\draw [red,dashed,thick,rounded corners=3pt] (h-1) -- (u-1) -- (v-2) -- (k-1) -- (j-1) -- (i-1) -- (v-1) -- (u-2) -- (l-1) -- (w-2) -- (x-2) -- (p-1) -- (o-1) -- (n-1) -- (x-1) -- (w-1) -- (m-1) -- cycle;
	
	\node at ($(g3)!0.5!(g4)$) [above left=-1.5pt]{$2$};
	\node at ($(g4)!0.5!(g5)$) [left=0pt]{$1$};
	\node at ($(g5)!0.5!(g6)$) [below left=-1.5pt]{$8$};
	\node at ($(g6)!0.5!(g7)$) [below=0pt]{$7$};
	\node at (barycentric cs:g3=1,g5=1,g7=1) {$V_1$};
	
	\node at ($(m-1)!0.5!(h-1)$) [rotate=22.5,red] {\tikz\draw[-Latex](0,0);};
	\node at ($(h-1)!0.5!(u-1)$) [rotate=157.5,red] {\tikz\draw[-Latex](0,0);};
	\node at ($(u-2)!0.5!(l-1)$) [rotate=157.5,red] {\tikz\draw[-Latex](0,0);};
	\node at ($(l-1)!0.5!(w-2)$) [rotate=-112.5,red] {\tikz\draw[-Latex](0,0);};
	\node at ($(w-1)!0.5!(m-1)$) [rotate=-112.5,red] {\tikz\draw[-Latex](0,0);};
	\node at ($(k-1)!0.5!(j-1)$) [rotate=0,red] {\tikz\draw[-Latex](0,0);};
	\node at ($(j-1)!0.5!(i-1)$) [rotate=-45,red] {\tikz\draw[-Latex](0,0);};
	\node at ($(p-1)!0.5!(o-1)$) [rotate=90,red] {\tikz\draw[-Latex](0,0);};
	\node at ($(o-1)!0.5!(n-1)$) [rotate=45,red] {\tikz\draw[-Latex](0,0);};
	\end{scope}
	\draw [very thick,dashed] (-67.5:2.5) -- (112.5:2.5);
	\end{tikzpicture}}
  \caption{The zig-zag paths of subdiagrams in Feynman diagram and $n$-gon diagram.  }\label{Figzigzagbreak}
 \end{figure}
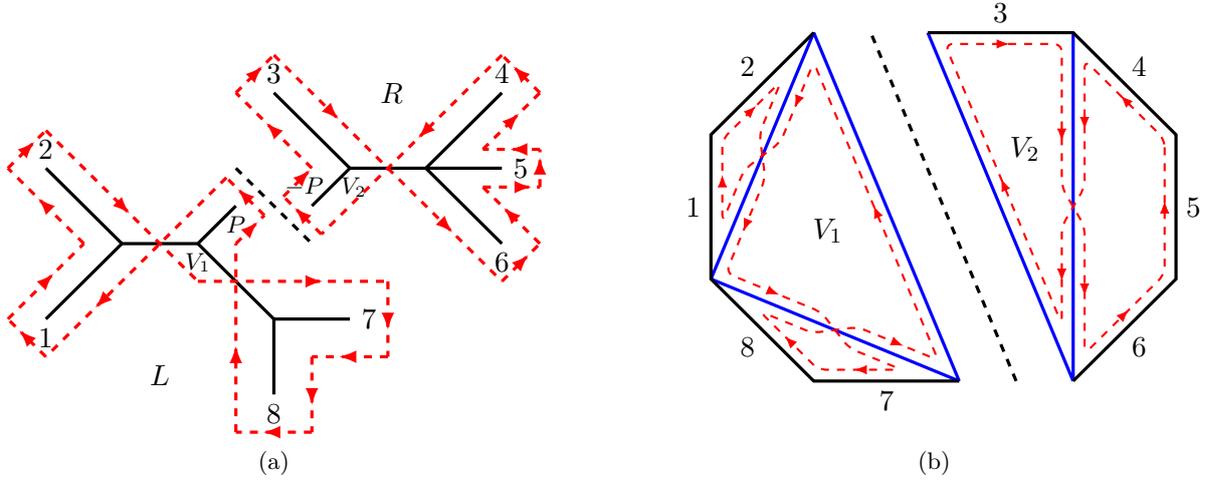

Now we consider splitting the Feynman diagram into two subdiagrams (labeled as $L$ and $R$) at the propagator $P_{7812}$ that connects the vertex $V_1$ and $V_2$, as shown in Fig.~\ref{Figzigzagbreak}.
In each subdiagram, the zig-zag path form a closed loop, from which we can read out the PT-factors. In order to define the canonical ordering of PT-factors, we read out the $\pmb\alpha$-orderings from both subdiagrams in the clockwise direction as
%
%
\bea \PT(\pmb{\alpha}_L)=\Spaa{7812P}~~~,~~~ \PT(\pmb{\alpha}_R)=\Spaa{P3456}~.~~~\eea
Next, we read out $\PT(\pmb\b_{L,R})$ from the zig-zag paths of both subdiagrams, starting from the leg $P$.
We emphasize that the zig-zag path for subdiagram $L$ is anti-clockwise with respect to the vertex $V_1$, while the zig-zag path for subdiagram $R$ is clockwise with respect to $V_2$. The orientations of two zig-zag paths are always opposite with respect to the two vertices connect by the split propagator. This is a generic and important feature since, as we mentioned before,
the orientations of zig-zag paths of two adjacent polygons in the $n$-gon diagram are always opposite. This feature will play a consequential role in determining the cycle representations for subdiagrams. Now we write down the PT-factors for the subdiagrams according to the arrows in the zig-zag paths of Fig.~\ref{pathFeynFact} as
\bea \PT(\pmb{\beta}_L)=\Spaa{P1278}~~~,~~~\PT(\pmb{\beta}_R)=\Spaa{P3654}~.~~~\label{zigzagCycle}\eea
The complete $\PT(\pmb\b)$ is given by the union of~\eqref{zigzagCycle} in a specific way: $\Spaa{1278P}\oplus \Spaa{P3654}\to \Spaa{12783654}$. The pattern will be clear if we use their cycle representation,
\begin{align}
\PT(\pmb\b_L)=\Spaa{P1278}\sim (P)(17)(28)~~~,~~~\PT(\pmb\b_R)=\Spaa{P3654}\sim (P)(3)(46)(5)~.~~~\label{eq:betaLRcycle}
\end{align}
Namely, we can obtain a V-type cycle representation of the complete Feynman diagram by the union of the above two after eliminating the single cycle $(P)$:  $(17)(28)(P)\oplus(P)(3)(46)(5)=(17)(28)(3)(46)(5)$.\footnote{Now the operation $\oplus$ can be properly defined as follows: if $\pmb\b_1$ and $\pmb\b_2$ are two permutations that have only one overlap element $P$, and in both of them $P$ sits in a single cycle, namely, $\pmb\b_{1,2}=(P)\pmb\b'_{1,2}$, we have
\begin{equation*}
\pmb\b_1\oplus\pmb\b_2=\pmb\b'_1\pmb\b'_2\,.
\end{equation*}}

To obtain the above result, we have to select two specific cycle representations for the subdiagrams out of the $10$ equivalent ones of $\pmb\b_{L,R}$,
\begin{subequations}
\label{eq:betaLR}
\begin{align}
&\pmb\b_{L}: & &\left\{ \setlength{\arraycolsep}{10pt}\begin{array}{lllll}
               (P)(17)(28) & (P8127) & (P78)(1)(2) & (P21)(7)(8) & (P1872)\\
               (P)(78)(12) & (P17)(8)(2) & (P2718) & (P7281) & (P82)(7)(1)
             \end{array}
\right\} \\
&\pmb\b_{R}: & &\left\{ \setlength{\arraycolsep}{10pt}\begin{array}{lllll}
               (P)(3)(46)(5) & (P43)(56) & (P534)(6) & (P635)(4) & (P36)(45)\\
               (P)(3456) & (P3)(4)(5)(6) & (P654)(3) & (P5)(364) & (P46)(35)
             \end{array}
\right\}~.~~~
\end{align}
\end{subequations}
First of all, we choose those that leave $P$ in a single cycle as $(P)$ in the equivalent class. This is reasonable since when gluing subdiagrams, $P$ should get eliminated without affecting other cycles, which is possible only if $P$ is in a single cycle. This limits us to the two in the first column of~\eqref{eq:betaLR}. Next, we need to answer which one to choose among the two. Looking back to~\eqref{eq:betaLRcycle}, we find that $\pmb\b_R$ is a V-type cycle representation that manifests the vertex $V_2$, while $\pmb\b_L$ is not a good cycle representation. If we color the legs according to how they are separated by $V_1$ and $V_2$, then for $\PT(\pmb\b_R)$, we have $({P})({\color{red}3})({\color{cyan}46})({\color{cyan}5})$, namely,
each cycle contains elements with the same color, i.e., elements from the same part of the $V_2$ splitting. We say that this cycle representation satisfies \emph{planar splitting} for short. In contrary, for $\PT(\pmb\b_L)$, some cycles contain elements with different colors, i.e., the elements inside one cycle are from different parts of the $V_1$ splitting. We call it cycle representation \emph{non-planar splitting} for short. One notices that among the two subdiagrams of Fig.~\ref{pathFeynFact}, one cycle representation is planar splitting while the other is non-planar splitting. For instance, if the arrows in Fig.~\ref{pathFeynFact} is reversed globally, then for $\PT(\pmb{\b}_L)$ the consequent cycle representation is $({P})({\color{red}78})({\color{cyan}12})$ while for $\PT(\pmb{\b}_R)$ it is $({P})({\color{red}3}{\color{cyan}456})$. Again, one cycle representation is planar splitting and the other is non-planar splitting. By gluing them together, we get $(78)(12)(3456)$, which is another V-type cycle representation of the complete Feynman diagram. This feature results from the fact that
the orientations of the subdiagram zig-zag paths are opposite with respect to the two vertices connected by the split propagator. It is generally true no matter how we cut the complete Feynman diagram.

The above discussion indicates clearly that the permutation representation of PT-factor can be recursively constructed by breaking a complete Feynman diagram into subdiagrams along internal propagators. We will provide a systematic and abstract construction in next subsection.

\subsection{The recursive construction of PT-factor via cycle representation}
\label{secPermutation2}

We already have the experience that, (1) in the gluing of two subdiagrams to one complete diagram, cycle representation of one subdiagram should be planar splitting, and that of the other should be non-planar splitting, (2) the planar splitting cycle representation corresponds to a zig-zag path in clockwise direction, while the non-planar splitting cycle representation corresponds to a zig-zag path in anti-clockwise direction. The orientation of zig-zag path is related to the planar or non-planar splitting of cycle representations because we use the convention that $\PT(\pmb\a)$ is obtained by traversing the external legs in clockwise direction.

\begin{figure}
\centering
\begin{tikzpicture}
    (* The main city *)
  \fill (0,0) circle (0.15);
  \fill (-6,0) circle (0.15);
  \draw [very thick](0,0)--(0,2.5);
  \filldraw [color=black,fill=white,very thick] (0,3) circle (0.5);
  \draw[very thick](-0.354,3.354)--(-1,4);
  \draw[very thick](0.354,3.354)--(1,4);
  \draw [very thick](0,0)--(0,-2.5);
  \filldraw [color=black,fill=white,very thick] (0,-3) circle (0.5);
  \draw [very thick](0,0)--(2.5,0);
  \draw[very thick](-0.354,-3.354)--(-1,-4);
  \draw[very thick](0.354,-3.354)--(1,-4);
  \filldraw [color=black,fill=white,very thick] (3,0) circle (0.5);
  \draw[very thick](3.354,-0.354)--(4,-1);
  \draw[very thick](3.354,0.354)--(4,1);
  \draw [very thick](0,0)--(-4,0);
  \draw [very thick](0,0)--(1.768,1.768);
  \filldraw [color=black,fill=white,very thick] (2.121,2.121) circle (0.5);
  \draw [very thick](0,0)--(1.768,-1.768);
  \draw[very thick](2.121,2.621)--(2.121,3.5);
  \draw[very thick](2.621,2.121)--(3.5,2.121);
  \filldraw [color=black,fill=white,very thick] (2.121,-2.121) circle (0.5);
  \draw [very thick](0,0)--(-1.768,1.768);
  \draw[very thick](2.121,-2.621)--(2.121,-3.5);
  \draw[very thick](2.621,-2.121)--(3.5,-2.121);
  \filldraw [color=black,fill=white,very thick] (-2.121,2.121) circle (0.5);
  \draw [very thick](0,0)--(-1.768,-1.768);
  \draw[very thick](-2.621,2.121)--(-3.5,2.121);
  \draw[very thick](-2.121,2.621)--(-2.121,3.5);
  \filldraw [color=black,fill=white,very thick] (-2.121,-2.121) circle (0.5);
  \draw[very thick](-2.621,-2.121)--(-3.5,-2.121);
  \draw[very thick](-2.121,-2.621)--(-2.121,-3.5);

  \node at (-4.25,0)[]{$P_i$};

  (* the suburban *)

  \draw[very thick](-4.5,0)--(-6,0);

  \draw[very thick](-6,0)--(-6,1.5);
  \filldraw [color=black,fill=white,very thick] (-6,2) circle (0.5);
  \draw[very thick](-6.354,2.354)--(-7,3);
  \draw[very thick](-5.646,2.354)--(-5,3);

  \draw[very thick](-6,0)--(-6,-1.5);
  \filldraw [color=black,fill=white,very thick] (-6,-2) circle (0.5);
  \draw[very thick](-6.354,-2.354)--(-7,-3);
  \draw[very thick](-5.646,-2.354)--(-5,-3);

  \draw[very thick](-6,0)--(-8.5,0);
  \filldraw [color=black,fill=white,very thick] (-9,0) circle (0.5);
  \draw[very thick](-9.354,0.354)--(-10,1);
  \draw[very thick](-9.354,-0.354)--(-10,-1);

  (* partial triangulation *)
  \draw[very thick, blue](-0.5,1.5)--(0.5,1.5)--(1.5,0.5)--(1.5,-0.5)--(0.5,-1.5)--(-0.5,-1.5)
  --(-1.5,-0.5)--(-1.5,0.5)--(-0.5,1.5);
  \draw[very thick,blue](-7.5,0.5)--(-1.5,0.5);
  \draw[very thick,blue](-7.5,-0.5)--(-1.5,-0.5);
  \draw[very thick,blue](-7.5,0.5)--(-7.5,-0.5);

  \draw[very thick, dotted, blue](-1.5,0.5)--(-3.1,1.5);
  \draw[very thick, dotted, blue](-1.5,-0.5)--(-3.1,-1.5);
  \draw[very thick,dotted, blue](-0.5,1.5)--(-1.35,3.25);
  \draw[very thick,dotted, blue](-0.5,-1.5)--(-1.35,-3.25);
  \draw[very thick,dotted, blue](0.5,1.5)--(1.35,3.25);
  \draw[very thick,dotted, blue](0.5,-1.5)--(1.35,-3.25);
  \draw[very thick,dotted, blue](1.5,0.5)--(3.25,1.35);
  \draw[very thick,dotted, blue](1.5,-0.5)--(3.25,-1.35);
  \draw[very thick,dotted, blue](-7.5,0.5)--(-8.25,1.75);
  \draw[very thick,dotted, blue](-7.5,-0.5)--(-8.25,-1.75);

  (* zig-zag path *)
  \tikzstyle{line}=[red,very thick, dashed]
  \tikzstyle{arrowout}=[red,->,>=latex,shorten >=-4pt, very thick,dashed]
  \tikzstyle{arrowin}=[red,-<,>=latex,shorten >=1pt, very thick,dashed]
  \tikzstyle{arrow2}=[red,>-,>=latex,shorten >=0pt, very thick,dashed]

  \draw[line](-1.25,0.25)--(-1.75,-0.25);
  \draw[line](-1.25,-0.25)--(-1.75,0.25);
  \draw[arrowout](-1.25,0.25)--(-0.875,0.25);
  \draw[line](-0.875,0.25)--(-0.5,0.25)--(-0.75,0.5);
  \draw[arrow2](-0.75,0.5)--(-1,0.75);

  \draw[line](-1,0.75)--(-1,1.25);
  \draw[line](-1.25,1)--(-0.75,1);
  \draw[arrowout](-0.75,1)--(-0.5,0.75);
  \draw[line](-0.5,0.75)--(-0.25,0.5)--(-0.25,0.875);
  \draw[arrow2](-0.25,0.875)--(-0.25,1.25);
  \draw[arrowin](-1.25,1)--(-2.5,1.5);
  \draw[arrowout](-1,1.25)--(-1.5,2.5);

  \draw[line](-0.25,1.75)--(0.25,1.25);
  \draw[line](-0.25,1.25)--(0.25,1.75);
  \draw[arrowout](0.25,1.25)--(0.25,0.875);
  \draw[line](0.25,0.875)--(0.25,0.5)--(0.5,0.75);
  \draw[arrow2](0.5,0.75)--(0.75,1);
  \draw[arrowin](-0.25,1.75)--(-0.75,3);
  \draw[arrowout](0.25,1.75)--(0.75,3);

  \draw[line](1,0.75)--(1,1.25);
  \draw[line](1.25,1)--(0.75,1);
  \draw[arrowout](1,0.75)--(0.75,0.5);
  \draw[line](0.75,0.5)--(0.5,0.25)--(0.875,0.25);
  \draw[arrow2](0.875,0.25)--(1.25,0.25);
  \draw[arrowout](1.25,1)--(2.5,1.5);
  \draw[arrowin](1,1.25)--(1.5,2.5);

  \draw[line](1.25,0.25)--(1.75,-0.25);
  \draw[line](1.25,-0.25)--(1.75,0.25);
  \draw[arrowout](1.25,-0.25)--(0.875,-0.25);
  \draw[line](0.875,-0.25)--(0.5,-0.25)--(0.75,-0.5);
  \draw[arrow2](0.75,-0.5)--(1,-0.75);
  \draw[arrowin](1.75,0.25)--(3,0.75);
  \draw[arrowout](1.75,-0.25)--(3,-0.75);

  \draw[line](1,-0.75)--(1,-1.25);
  \draw[line](1.25,-1)--(0.75,-1);
  \draw[arrowout](0.75,-1)--(0.5,-0.75);
  \draw[line](0.5,-0.75)--(0.25,-0.5)--(0.25,-0.875);
  \draw[arrow2](0.25,-0.875)--(0.25,-1.25);
  \draw[arrowin](1.25,-1)--(2.5,-1.5);
  \draw[arrowout](1,-1.25)--(1.5,-2.5);

  \draw[line](-0.25,-1.75)--(0.25,-1.25);
  \draw[line](-0.25,-1.25)--(0.25,-1.75);
  \draw[arrowout](-0.25,-1.25)--(-0.25,-0.875);
  \draw[line](-0.25,-0.875)--(-0.25,-0.5)--(-0.5,-0.75);
  \draw[arrow2](-0.5,-0.75)--(-0.75,-1);
  \draw[arrowout](-0.25,-1.75)--(-0.75,-3);
  \draw[arrowin](0.25,-1.75)--(0.75,-3);

  \draw[line](-1,-0.75)--(-1,-1.25);
  \draw[line](-1.25,-1)--(-0.75,-1);
  \draw[arrowout](-1,-0.75)--(-0.75,-0.5);
  \draw[line](-0.75,-0.5)--(-0.5,-0.25)--(-0.875,-0.25);
  \draw[arrow2](-0.875,-0.25)--(-1.25,-0.25);
  \draw[arrowout](-1.25,-1)--(-2.5,-1.5);
  \draw[arrowin](-1,-1.25)--(-1.5,-2.5);

  (* zig-zag path in suburban *)
  \draw[line](-6.25,-0.25)--(-5.75,-0.75);
  \draw[line](-6.25,-0.75)--(-5.75,-0.25);
  \draw[arrowout](-5.75,-0.75)--(-5.25,-2);
  \draw[arrowin](-6.25,-0.75)--(-6.75,-2);
  \draw[arrowout](-1.75,0.25)--(-3.75,0.25);
  \draw[line](-3.75,0.25)--(-5.75,0.25);
  \draw[arrowout](-5.75,-0.25)--(-3.75,-0.25);
  \draw[line](-3.75,-0.25)--(-1.75,-0.25);

  \draw[line](-7.75,-0.25)--(-7.25,0.25);
  \draw[line](-7.25,-0.25)--(-7.75,0.25);
  \draw[arrowout](-7.25,-0.25)--(-6.75,-0.25);
  \draw[line](-6.75,-0.25)--(-6.25,-0.25);
  \draw[arrowout](-7.75,-0.25)--(-9,-0.75);
  \draw[arrowin](-7.75,0.25)--(-9,0.75);

  \draw[line](-6.25,0.25)--(-5.75,0.75);
  \draw[line](-6.25,0.75)--(-5.75,0.25);
  \draw[arrowout](-6.25,0.25)--(-6.75,0.25);
  \draw[line](-6.75,0.25)--(-7.25,0.25);
  \draw[arrowin](-5.75,0.75)--(-5.25,2);
  \draw[arrowout](-6.25,0.75)--(-6.75,2);

  (* zig-zag path in dotted form *)
  \draw[dotted,red,very thick](-2.5,1.5) to [out=135,in=225](-2.625,2.625) to [out=45,in=135] (-1.5,2.5);
  \draw[dotted,red,very thick](-2.5,-1.5) to [out=225,in=135](-2.625,-2.625) to [out=315,in=225] (-1.5,-2.5);
  \draw[dotted,red,very thick](-0.75,3) to [out=135,in=180](0,3.75) to [out=0,in=45] (0.75,3);
  \draw[dotted,red,very thick](-0.75,-3) to [out=225,in=180](0,-3.75) to [out=0,in=315] (0.75,-3);
  \draw[dotted,red,very thick](1.5,2.5) to [out=45,in=135](2.625,2.625) to [out=315,in=45] (2.5,1.5);
  \draw[dotted,red,very thick](1.5,-2.5) to [out=315,in=225](2.625,-2.625) to [out=45,in=315] (2.5,-1.5);
  \draw[dotted,red,very thick](3,0.75) to [out=45,in=90](3.75,0) to [out=270,in=315] (3,-0.75);

  \draw[dotted,red,very thick](-9,0.75) to [out=135,in=90](-9.75,0) to [out=270,in=225] (-9,-0.75);
  \draw[dotted,red,very thick](-6.75,2) to [out=135,in=180](-6,2.75) to [out=0,in=45] (-5.25,2);
  \draw[dotted,red,very thick](-6.75,-2) to [out=225,in=180](-6,-2.75) to [out=0,in=315] (-5.25,-2);

  (* n-gon *)
  \draw[gray,very thick](-3,1.7)--(-3,2.5);
  \draw[gray, very thick](-2.5,3)--(-1.7,3);
  \draw[dotted,gray, very thick](-3,2.5)--(-2.5,3);
  \draw[dotted,gray, very thick](-1.7,3)--(-1,3.5);

  \draw[gray,very thick](-1,3.5)--(-0.5,4);
  \draw[gray, very thick](0.5,4)--(1,3.5);
  \draw[dotted,gray, very thick](-0.5,4)--(0.5,4);
  \draw[dotted,gray, very thick](1,3.5)--(1.7,3);

  \draw[gray,very thick](3,1.7)--(3,2.5);
  \draw[gray, very thick](2.5,3)--(1.7,3);
  \draw[dotted,gray, very thick](2.5,3)--(3,2.5);
  \draw[dotted,gray, very thick](3,1.7)--(3.5,1);

  \draw[gray,very thick](3.5,1)--(4,0.5);
  \draw[gray, very thick](4,-0.5)--(3.5,-1);
  \draw[dotted,gray, very thick](4,0.5)--(4,-0.5);
  \draw[dotted,gray, very thick](3.5,-1)--(3,-1.7);

  \draw[gray,very thick](3,-1.7)--(3,-2.5);
  \draw[gray, very thick](2.5,-3)--(1.7,-3);
  \draw[dotted,gray, very thick](3,-2.5)--(2.5,-3);
  \draw[dotted,gray, very thick](1.7,-3)--(1,-3.5);

  \draw[gray,very thick](-1,-3.5)--(-0.5,-4);
  \draw[gray, very thick](0.5,-4)--(1,-3.5);
  \draw[dotted,gray, very thick](0.5,-4)--(-0.5,-4);
  \draw[dotted,gray, very thick](-1,-3.5)--(-1.7,-3);

  \draw[gray,very thick](-3,-1.7)--(-3,-2.5);
  \draw[gray, very thick](-2.5,-3)--(-1.7,-3);
  \draw[dotted,gray, very thick](-2.5,-3)--(-3,-2.5);

  (* n-gon in suburban *)
  \draw[gray,very thick](-5,-2.5)--(-5.5,-3);
  \draw[gray, very thick](-6.5,-3)--(-7,-2.5);
  \draw[dotted,gray, very thick](-5.5,-3)--(-6.5,-3);
  \draw[dotted,gray, very thick](-7,-2.5)--(-9.5,-1);
  \draw[dotted,gray,very thick](-3,-1.7) to [out=90,in=45] (-5,-2.5);

  \draw[gray,very thick](-9.5,-1)--(-10,-0.5);
  \draw[gray, very thick](-10,0.5)--(-9.5,1);
  \draw[dotted,gray, very thick](-10,-0.5)--(-10,0.5);
  \draw[dotted,gray, very thick](-9.5,1)--(-7,2.5);

  \draw[gray,very thick](-5,2.5)--(-5.5,3);
  \draw[gray, very thick](-6.5,3)--(-7,2.5);
  \draw[dotted,gray, very thick](-6.5,3)--(-5.5,3);
  \draw[dotted,gray,very thick](-3,1.7) to [out=270,in=315] (-5,2.5);

  (* Labels *)
  \fill (0,3) circle (0.07);
  \fill (0.25,3) circle (0.07);
  \fill (-0.25,3) circle (0.07);
  \fill (0,-3) circle (0.07);
  \fill (0.25,-3) circle (0.07);
  \fill (-0.25,-3) circle (0.07);
  \fill (-9,0.25) circle (0.07);
  \fill (-9,0) circle (0.07);
  \fill (-9,-0.25) circle (0.07);

  \node at (-2.121,2.121) []{$\mathsf{A}_{i+1}$};
  \node at (-2.121,-2.121) []{$\mathsf{A}_{i-1}$};
  \node at (2.121,2.121) []{$\mathsf{A}_{m-1}$};
  \node at (2.121,-2.121) []{$\mathsf{A}_{1}$};
  \node at (3,0) []{$\mathsf{A}_{m}$};

  \node at (-6,-2)[]{$\mathsf{a}_{i_1}$};
  \node at (-6,2)[]{$\mathsf{a}_{i_s}$};

\end{tikzpicture}
\caption{The recursive construction of $n$-point PT-factor. Black lines represent a general $n$-point Feynman
diagram. Gray lines denotes the dual $n$-gon. Blue line represent the partial triangulations of $n$-gon dual to the Feynman diagram, while red dashed lines denote the zig-zag path. The direction of zig-zag path is labeled by arrows in preference of our convention. All dotted lines are abbreviation of their detailed structures that are
not explicitly shown in the diagram.}\label{Figzigzagrecur}
\end{figure}
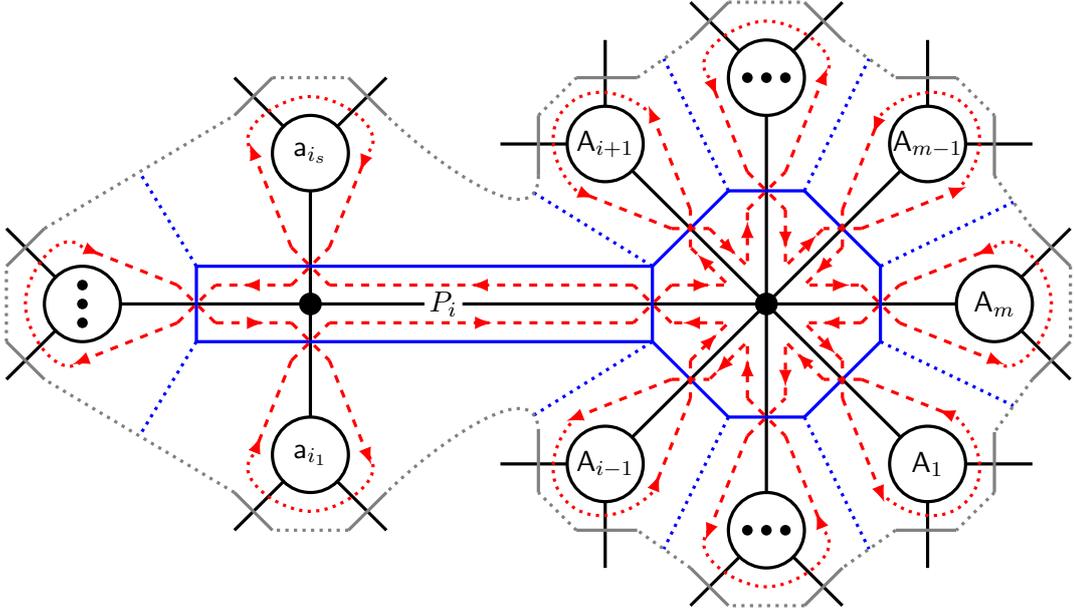

Let us head to a more systematic and abstract discussion on the recursive construction of cycle representation for a Feynman diagram. Consider a generic $n$-point effective Feynman diagram, where besides cubic vertices, the effective $(m>3)$-point vertices can also appear. An $m$-point vertex represents a collection of all possible $\frac{(2m-4)!}{(m-1)!(m-2)!}$ such $m$-point trivalent subdiagrams. If we use $v_m$ to denote the number of $m$-point vertices appearing in the effective Feynman diagram, then they should satisfy the constraint
\bea \sum_{m=3}^{n}(m-2)v_m=n-2~,~~~\label{eq:vertexrel}\eea
where the total number of vertices $\sum v_m$ falls between 1 and $n-2$. An illustration of the $n$-point Feynman diagram as well as the dual $n$-gon diagram with also the zig-zag path is shown in Fig.~\ref{Figzigzagrecur}.

Let us focus on a $m$-point vertex, marked as a black dot in the middle of the blue octagon\footnote{It should connect $m$ propagators, but we only sketch eight lines as illustration. The circles with dots inside represent many subdiagrams.} in Fig.~\ref{Figzigzagrecur}. This vertex connects $m$ subdiagrams via $m$ propagators $P_k$ with $k=1,2\ldots m$. Our goal is to write the cycle representation of $n$-point PT-factor into a form that manifests the factorization into those cycle representations of the $m$ subdiagrams connected to the $m$-point vertex.
Since according to our convention, the $\PT(\pmb{\a}_m)$ is read in clockwise direction, we have
\bea \PT(\pmb{\a}_m)=\Spaa{P_1P_2\cdots P_m}\eea
for the subdiagram inside the blue octagon. We intentionally choose the direction of zig-zag path as clockwise with respect to the considered $m$-point vertex, so that for this subdiagram, we have
\bea \PT(\pmb{\b}_m)=\Spaa{P_1P_2\cdots P_m}\sim (P_1)(P_2)\cdots (P_m)~.~~~\eea
Note that among the $2m$ equivalent cycle representations of $\PT(\pmb{\b}_m)$, this is the {\sl only} one that allows every $P_k$ appear as a single element in a cycle. Now consider the $m$ subdiagrams in the other side of $P_k$, denoted as $\mathsf{A}_1$, $\mathsf{A}_2$, $\ldots$, $\mathsf{A}_{m}$. Since $P_k$ is an external leg of $\mathsf{A}_k$, we know {\sl a priori} that there must be a cycle representation $(P_k)\pmb{\beta}^\text{cyc-rep}_{\mathsf{A}_k}$ for this subdiagram, where $\pmb{\beta}^\text{cyc-rep}_{\mathsf{A}_k}$ is to be determined. So when gluing all the $m$ subdiagrams to the one inside the octagon by propagator $P_k$'s, we obtain the follow factorization form
\bea \pmb{\beta}^\text{cyc-rep}&=&(P_1)(P_2)\cdots (P_m)\oplus(P_1)\pmb{\beta}^\text{cyc-rep}_{\mathsf{A}_1}\oplus \cdots \oplus (P_m)\pmb{\beta}^\text{cyc-rep}_{\mathsf{A}_m}\nn
&=&\pmb{\beta}^\text{cyc-rep}_{\mathsf{A}_1}\pmb{\beta}^\text{cyc-rep}_{\mathsf{A}_2}\cdots \pmb{\beta}^\text{cyc-rep}_{\mathsf{A}_{m}}~,~~~\label{cyclefactor}\eea
namely, it allows a planar separation into $m$ parts. Once the cycle representations of PT-factors for subdiagrams are known, the complete cycle representation is simply a combination of them. Note that the factorization~\eqref{cyclefactor} is based upon a given vertex. Now we can go into each subdiagram $\mathsf{A}_i$ and perform the same construction, until we reach a subdiagram with only one vertex.

A remaining problem is that suppose all the cycle representations of a subdiagram $\mathsf{A}_i$ is known, how do we choose the $\pmb\b^{\text{cyc-rep}}_{\mathsf{A}_i}$ that is used as the building block of~\eqref{cyclefactor}.
According to Fig.~\ref{Figzigzagrecur}, the subset $\mathsf{A}_i$ and propagator $P_i$ form a $(|\mathsf{A}_i|+1)$-point subdiagram, connected to the remaining parts via the propagator $P_i$. In order to connect it with the subdiagram inside octagon, we already constrain the cycle representation of this subdiagram as $(P_i)\pmb{\beta}^\text{cyc-rep}_{\mathsf{A}_{i}}$, where $(P_i)$ itself is a cycle. From the definition of equivalent class~\eqref{eq:equivperm}, we know that there are two cycle representations satisfying this condition. One of them can be constructed according to~\eqref{cyclefactor}: suppose the propagator $P_i$ is connected to an $(s+1)$-point vertex in subdiagram $\mathsf{A}_i$, marked as black dot in the interior of blue rectangular in Fig.~\ref{Figzigzagrecur}. This $(s+1)$-point vertex splits the subdiagram $\mathsf{A}_i$ into $s$ disjoint sub-subsets $\mathsf{a}_{i_1}$ to $\mathsf{a}_{i_s}$ via $s$ propagator $P_{i_\ell}$ with $\ell=1,\ldots,s$. Then following~\eqref{cyclefactor}, we should choose the zig-zag path inside the rectangle in clockwise direction, from which we can obtain a cycle representation satisfying the planar splitting,
\begin{equation}
\label{eq:cycAi}
\pmb\b_{\mathsf{A}_i+1}^{\text{cyc-rep}}=(P_i)\pmb{\beta}^\text{cyc-rep}_{\mathsf{A}_i}=(P_i)~\pmb{\beta}^\text{cyc-rep}_{\mathsf{a}_{i_1}}\pmb{\beta}^\text{cyc-rep}_{\mathsf{a}_{i_2}}\cdots\pmb{\beta}^\text{cyc-rep}_{\mathsf{a}_{i_s}}~.~~~
\end{equation}
However, since we have already set the zig-zag path around the octagon to be clockwise, the zig-zag path around the rectangle must be anti-clockwise, and what we should use is the cycle representation other than~\eqref{eq:cycAi} with $P_i$ in a single cycle. Thus we can obtain $(P_i)\pmb{\beta}^\text{cyc-rep}_{\mathsf{A}_i}$ by acting reversing and cyclic rotation onto~\eqref{eq:cycAi}. Moreover, the $\pmb\b_{\mathsf{A}_i}^{\text{cyc-rep}}$ obtained this way must be non-planar splitting, namely,
at least one cycle of  $\pmb{\beta}^\text{cyc-rep}_{\mathsf{A}_i}$ contains elements from different subsets $\mathsf{a}_{i_\ell}$. In other words, there must exist at least one cycle that can not be a part of any $\pmb{\beta}^\text{cyc-rep}_{\mathsf{a}_{i_\ell}}$.
To prove this point, it is suffices to study the following problem. Given an identity element in the permutation group,
\bea \Spaa{P_i,{\color{red}a_1,a_2,\ldots, a_i},{\color{cyan} b_1, b_2, \ldots , b_j}}~,~~~\label{permumapped}\eea
which splits into three planar parts $\{P_i\}$, $\{a_1,\ldots,a_i\}$ and $\{b_1,\ldots,b_j\}$. Let us consider the planar splitting cycle representation $(P_i)\pmb{a}^\text{cyc-rep}\pmb{b}^\text{cyc-rep}$ that maps the identity element~\eqref{permumapped} into another permutation as
\begin{equation}
({P_i})~{\color{red}\pmb{a}}^\text{cyc-rep}{\color{cyan}\pmb{b}}^\text{cyc-rep}=\begin{pmatrix}
{P_i} & {\color{red}a_1} & {\color{red}a_2} & {\color{red}\cdots} & {\color{red}a_i} & {\color{cyan}b_1} & {\color{cyan}b_2 }& {\color{cyan}\cdots} & {\color{cyan}b_j} \\
\downarrow & \downarrow & \downarrow & \cdots & \downarrow & \downarrow & \downarrow & \cdots & \downarrow \\
{P_i} & {\color{red}a'_1} & {\color{red}a'_2} & {\color{red}\cdots }& {\color{red}a'_i} & {\color{cyan}b'_1} & {\color{cyan}b'_2} & {\color{cyan}\cdots} & {\color{cyan}b'_j}
\end{pmatrix}~,
\end{equation}
where $\{a'_1,\ldots,a'_i\}$ is a permutation of $\{\a_1,\ldots,\a_i\}$ and $\{b'_1,\ldots,b'_j\}$ is a permutation of $\{b_1,\ldots,b_j\}$. The other cycle representation with $(P_i)$ as a single cycle in equivalent class is thus given by reversing the ordering and dragging $P_i$ to the first position as
\begin{equation}
\begin{pmatrix}
{P_i} & {\color{red}a_1} & {\color{red}a_2} & {\color{red}\cdots} & {\color{red}a_i} & {\color{cyan}b_1} & {\color{cyan}\cdots} & {\color{cyan}b_{j-1}} &{\color{cyan} b_j }\\
\downarrow & \downarrow & \downarrow & \cdots & \downarrow & \downarrow & \cdots & \downarrow & \downarrow \\
{P_i} & {\color{cyan}b'_j} & {\color{cyan}b'_{j-1}} & {\color{cyan}\cdot}{\color{red}\cdot}{\color{cyan}\cdot}{\color{red}\cdot} & {\color{cyan}\cdot}{\color{red}\cdot}{\color{cyan}\cdot}{\color{red}\cdot} & {\color{cyan}\cdot}{\color{red}\cdot}{\color{cyan}\cdot}{\color{red}\cdot} & {\color{cyan}\cdot}{\color{red}\cdot}{\color{cyan}\cdot}{\color{red}\cdot} & {\color{red} a'_2} & {\color{red}a'_1}
\end{pmatrix}=({P_i})({\color{red}a_1} {\color{cyan}b'_j}{\color{cyan}\cdot}{\color{red}\cdot}{\color{cyan}\cdot}{\color{red}\cdot})({\color{cyan}\cdot}{\color{red}\cdot}{\color{cyan}\cdot}{\color{red}\cdot})\cdots({\color{cyan}\cdot}{\color{red}\cdot}{\color{cyan}\cdot}{\color{red}\cdot})~.~~~\label{permumapto}
\end{equation}
Since $b'_j\in \{b_1,\ldots,b_j\}$ and $a_1\in \{a_1,\ldots,a_i\}$, it is clear in (\ref{permumapto}) that the legs in two subsets $\{a_1,\ldots,a_i\}$, $\{b_1,\ldots,b_j\}$ must appear together in at least one cycle in the cycle representation.

To recap, the cycle representation of a Feynman diagram can be written as a simple combination of cycle representations of subdiagrams as presented in~\eqref{cyclefactor}. This leads to a recursive construction of cycle representation from those of lower point Feynman subdiagrams. Crucially, \emph{the cycle representation of subdiagram $\mathsf{A}_i$ that used in the complete cycle representation~\eqref{cyclefactor} should be the one with $P_i$ as a single cycle $(P_i)$ and be non-planar splitting with respect to its vertex connected to $P_i$.} Then we can work out the permutation from cycle representation and eventually the PT-factor $\PT(\pmb{\beta})$. Note that, it is possible to start the recursive construction from any vertex of a Feynman diagram, and different choice leads to different cycle representation but they are all in the same equivalent class. We will show in Appendix~\ref{secAppendix} that different planar splittings characterize the shapes of the associahedron boundaries.

\subsection{Examples}

Let us now present some nontrivial examples to illustrate the recursive construction of cycle representation. First we consider the three point diagram.
The cubic vertex splits diagram into three subdiagrams, each one  is trivially a single external leg. This is also true for the diagrams with only a single vertex. So following~\eqref{cyclefactor}, we get
\begin{align}
\adjustbox{raise=-0.75cm}{\begin{tikzpicture}
	\draw [] (0,0) -- (1,0) node[above=0pt]{$3$} (0,0) -- (135:1) node[left=0pt]{$2$} (0,0) -- (-135:1) node[left=0pt]{$1$};
	\fill[color=red] (0,0) circle (0.1);
	\end{tikzpicture}}\;\Longleftrightarrow\;\pmb{\beta}^\text{cyc-rep}=(1)(2)(3)~~~,~~~ \adjustbox{raise=-0.75cm}{\begin{tikzpicture}
	\draw [] (0,0) -- (1,0) node[above=0pt]{$n$} (0,0) -- (135:1) node[left=0pt]{$3$} (-1,0) node[left=0pt]{$2$} -- (0,0) -- (-135:1) node[left=0pt]{$1$};
	\fill[color=red] (0,0) circle (0.1);
	\foreach \x in {15,30,45,60,75,90,105,120} {
	\fill (\x:0.7) circle (0.75pt);	
	}
	\end{tikzpicture}}\;\Longleftrightarrow\;\pmb{\beta}^\text{cyc-rep}=(1)(2)(3)\ldots (n)~.~~~
\end{align}
Note that $(12)(3)$ is in the same equivalent class of $(1)(2)(3)$ for this three point diagram. When it appears as a subdiagram, one leg becomes an internal line $P$, namely
\begin{equation}
\label{eq:3psub}
\adjustbox{raise=-0.75cm}{\begin{tikzpicture}
	\draw [] (0,0) -- (1,0) node[pos=0.5,above=0pt]{$P$} (0,0) -- (135:1) node[left=0pt]{$a_2$} (0,0) -- (-135:1) node[left=0pt]{$a_1$};
	\fill (0,0) circle (2pt);
	\foreach \x in {-135,-90,-45,0,45,90,135} {
		\draw [] (1,0) -- ++(\x:0.3);}
	\fill (1,0) circle (2pt);
	\end{tikzpicture}}~.~~~
\end{equation}
In this case, we should use the non-planar splitting cycle representation $(a_1a_2)(P)$ instead of $(a_1)(a_2)(P)$. To see this, let us proceed to a four point Feynman diagram as shown below. If we start from the vertex marked by red dot, the diagram splits to three subdiagrams, two of which are single external legs and one is three point subdiagram. The non-planar splitting for three point subdiagram appears as $(12)(P_{12})$, so according to~\eqref{cyclefactor}, the recursive procedure is described as follows,
\begin{equation}
\label{eq:4pv1}
\adjustbox{raise=-0.85cm}{\begin{tikzpicture}[every node/.style={font=\fontsize{8pt}{8pt}\selectfont,inner sep=1pt}]
	\draw [thick] (135:1) node[left=0pt]{$2$} -- (0,0) -- (-135:1) node[left=0pt]{$1$};
	\draw [thick] (0,0) -- (1,0) -- ++(45:1) node[right=0pt]{$3$} (1,0) -- ++(-45:1) node[right=0pt]{$4$};
	\filldraw [draw=red,fill=red] (1,0) circle (0.1);
	\node at (2.5,0) {$\longrightarrow$};
	\begin{scope}[xshift=4cm]
	\draw [thick] (135:1) node[left=0pt]{$2$} -- (0,0) -- (-135:1) node[left=0pt]{$1$};
	\node (A) at (0.75,0.5) [above=1pt]{$(12)(P_{12})$};
	\node (B) at (0.75,1.5) {$(1)(2)(P_{12})$};
	\node (C) at (1.25,0) {$P_{12}$};
	\draw [-stealth] (B.south) -- (A.north);
	\draw [stealth-] (45:0.15) -- (A.south);
	\draw [thick] (0,0) -- (C.west) (C.east) -- (2.5,0) -- ++(45:1) node[right=0pt]{$3$} (2.5,0) -- ++(-45:1) node[right=0pt]{$4$};
	\node (E) at (2.5,0) [circle,fill=red,inner sep=2pt] {};
	\node (F) at (1.75,-0.5) [below=1pt] {$(P_{34})(3)(4)$};
	\draw [stealth-] (2.5,0) ++(-135:0.15) -- (F.north) ;
	\end{scope}
	\end{tikzpicture}}\,\Longrightarrow\quad\pmb{\beta}^\text{cyc-rep}=(12)(3)(4)~.~~~
\end{equation}
From above results, we can recursively compute the cycle representation of $\PT(\pmb{\beta})$ for five point CHY-integrand. Here we present an example as follows,
\begin{align}
{\begin{tikzpicture}[baseline={(current bounding box.center)}]
	\draw [] (135:1) node[left=0pt]{$2$} -- (0,0) -- (1,0) -- ++(0,1) node[above=0pt]{$3$} (1,0) -- (2,0) -- ++(45:1) node [above=0pt]{$4$} (2,0) -- ++(-45:1) node[left=0pt]{$5$} (-135:1) node[left=0pt]{$1$} -- (0,0);
	\fill (0,0) circle (0.05);
	\fill[color=red] (1,0) circle (0.1) node [below=1pt,black]{$V_2$};
	\fill[color=red] (2,0) circle (0.1) node [below=1pt,black]{$V_1$};
	\node at (3,0) {$,$};
	\end{tikzpicture}
}
\end{align}
and construct the cycle representation starting from two different
vertices respectively. If we start from vertex $V_1$, then the
diagram is split to three parts: the external leg $4$, $5$ and a four point subdiagram. As mentioned above, $(12)(3)(P_{123})$ is a planar splitting cycle representation with respect to $V_2$, and we should take the non-planar splitting one in the equivalent class, namely, $(132)(P_{123})$, to form the complete
cycle representation, which is obtained from $(12)(3)(P)$ by acting the reversing permutation $(13)(2)$, namely, $[(12)(3)]\cdot[(13)(2)]=(132)$. Hence, the final result is $(132)(4)(5)$.

Alternatively, we can start from vertex $V_2$. Then the diagram is
split to another three parts, two of which are three point subdiagrams with non-planar splitting cycle representation $(12)(P_{12})$ and $(45)(P_{45})$, while the other is the single external leg $3$. Connecting them via the vertex $(P_{12})(P_3)(P_{45})$, we obtain $(12)(3)(45)$. We see that different splitting of diagram leads to different cycle representations. However, all of them are in the same equivalent class. In fact, both $(132)(4)(5)$ and  $(12)(3)(45)$ lead to the PT-factor $\PT(\pmb{\beta})=\Spaa{12453}$.

Next, we give a seven point example. The Feynman diagram is shown below, together with the resultant cycle representations when we carry out the recursive construction at different vertices,
\begin{align}
&{\begin{tikzpicture}[baseline={(current bounding box.center)},every node/.style={black}]
	\draw [] (135:1) node[left=0pt]{$2$} -- (0,0) -- (-135:1) node[left=0pt]{$1$} (0,0) -- (1,0) -- ++(0,1) node[above=0pt]{$3$} (1,0) -- (2,0) -- ++(0,1) node[above=0pt]{$4$} (2,0) -- (3,0) -- ++(0,-1) node[below=0pt]{$7$} (3,0) -- (4,0) -- ++(45:1) node[right=0pt]{$5$} (4,0) -- ++(-45:1) node[below=0pt]{$6$};
	\fill [red](0,0) node [below=0.1cm] {$V_1$} circle (0.1) (1,0) node[below=0.1cm]{$V_2$} circle (0.1) ++(1,0)  node[below=0.1cm]{$V_3$} circle (0.1) ++(1,0) node [above=0.1cm] {$V_4$} circle (0.1) ++(1,0)  node[below=0.1cm] {$V_5$} circle (0.1);
	\end{tikzpicture}} & &\begin{array}{l}
(V_1)\qquad\pmb\beta=(1)(2)(347)(5)(6) \\
(V_2)\qquad\pmb\beta=(12)(3)(467)(5) \\
(V_3)\qquad\pmb\beta=(132)(4)(576) \\
(V_4)\qquad\pmb\beta=(143)(2)(56)(7) \\
(V_5)\qquad\pmb\beta=(1)(2)(347)(5)(6)
\end{array}
\end{align}
Some brief explanation is in order. For the vertex $V_4$, the planar splitting cycle representation is $(P)(56)(7)$, while its non-planar splitting one is $[(7)(56)]\cdot[(75)(6)]=(576)$. The former can be used in the recursive construction starting from vertex $V_4$, while the latter can be used in the recursive construction starting from vertex $V_3$. Similarly, for the vertex $V_3$, the planar splitting cycle representation is $(P)(4)(576)$, while the non-planar splitting one is $[(4)(576)]\cdot[(47)(56)]=(467)(5)$. The latter can be used in the recursive construction starting from vertex $V_2$.

This recursive construction can be easily taken to higher points, and we have run
extensive checking up to eight point diagrams.

\section{Relations between different PT-factors}
\label{secRelation}

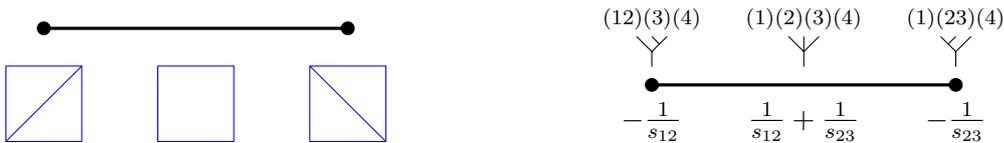
\begin{figure}[t]
	\centering
	\begin{tikzpicture}
	\filldraw [very thick] (0,0) circle (2pt) -- (4,0) circle (2pt);
	\draw[blue] (-0.5,-0.5)--(0.5,-0.5)--(0.5,-1.5)--(-0.5,-1.5)--(-0.5,-0.5);
	\draw[blue] (0.5,-0.5)--(-0.5,-1.5);
	\draw[blue] (1.5,-0.5)--(2.5,-0.5)--(2.5,-1.5)--(1.5,-1.5)--(1.5,-0.5);
	\draw[blue] (3.5,-0.5)--(4.5,-0.5)--(4.5,-1.5)--(3.5,-1.5)--(3.5,-0.5);
	\draw[blue] (3.5,-0.5)--(4.5,-1.5);
	
	\begin{scope}[xshift=8cm,yshift=-0.75cm]
	\node at (0,0) [above=0.1cm] {
		\begin{tikzpicture}[scale=0.2]
		\draw (0,0) -- (0,1) -- (-1,2) (0,1) -- (1,2) (0,2) -- (0.5,1.5);
		\end{tikzpicture}
	};
	\node at (4,0) [above=0.1cm]{
		\begin{tikzpicture}[scale=0.2]
		\draw (0,0) -- (0,1) -- (-1,2) (0,1) -- (1,2) (0,2) -- (-0.5,1.5);
		\end{tikzpicture}
	};
	\node at (2,0) [above=0.1cm]{
		\begin{tikzpicture}[scale=0.2]
		\draw (0,0) -- (0,1) -- (-1,2) (0,1) -- (1,2) (0,2) -- (0,1);
		\end{tikzpicture}
	};
	\filldraw [very thick] (0,0) circle (2pt) -- (4,0) circle (2pt);
	\node at (0,0) [below=0.1cm] {$-\frac{1}{s_{12}}$};
	\node at (4,0) [below=0.1cm] {$-\frac{1}{s_{23}}$};
	\node at (2,0) [below=0.1cm] {$\frac{1}{s_{12}}+\frac{1}{s_{23}}$};
	\node at (0,0) [above=0.6cm] {\scriptsize$(12)(3)(4)$};
	\node at (4,0) [above=0.6cm] {\scriptsize$(1)(23)(4)$};
	\node at (2,0) [above=0.6cm] {\scriptsize$(1)(2)(3)(4)$};
	\node at (5,0) {$.$};
\end{scope}
\end{tikzpicture}
\caption{The associahedron for four point amplitudes and PT-factors.}\label{4passo}
\end{figure}

After clarifying the relations between permutations of PT-factors and the Feynman diagrams, we move on to the relations between different PT-factors in the language of permutation and cycle representation. This topic has been discussed from the \emph{associahedron} point of view \cite{Arkani-Hamed:2017mur} and before proceeding let us briefly review their result. The major conclusion is that, the canonical form of an $(n-3)$-dimensional associahedron is the $n$-particle tree-level amplitude of bi-adjoint scalar theory with identical ordering. A consequence is that, the codimension $d$ faces of an associahedron are in one-to-one correspondence with the partial triangulations with $d$ diagonals, while the partial triangulations are dual to cuts on planar cubic diagrams with each diagonal corresponding to a cut. Hence the faces of the associahedron are dual to the singularities of cubic scalar amplitude. In this sense, PT-factors can also be related to corresponding faces of associahedron. For instance, a three point amplitude is dual to a triangle, allowing only one trivial triangulation. Thus the corresponding associahedron is just a zero dimensional point, on which sits the only independent PT-factor $\PT(\pmb\b)=\Spaa{123}$.
A four point amplitude is dual to a box, while the associahedron formed by its (partial) triangulations is a one dimensional line as shown in Fig.~\ref{4passo}.
The vertices correspond to all complete triangulations of the box, while the edge corresponds to partial triangulations.
The edge is related to $\PT(\pmb{\beta})=\Spaa{1234}$ with amplitude $\frac{1}{s_{12}}+\frac{1}{s_{23}}$, while the two ending vertices are related to $\PT(\pmb{\beta})=\Spaa{1243}$ and $\Spaa{1324}$ respectively with amplitude $\frac{1}{s_{12}}$ and $\frac{1}{s_{23}}$. The relations between different PT-factors are manifest in this geometric picture.

\begin{figure}[t]
	\centering
	\begin{tikzpicture}
	
	\filldraw [very thick] (0,0) circle (2pt) -- (1,-2) circle (2pt)-- (3,-2) circle (2pt)-- (4,0) circle (2pt)-- (2,1.5) circle (2pt)--(0,0);
	
	\draw [blue] (-1.25,0.25)--(-1.,-0.25)--(-0.5,-0.25)--(-0.25,0.25)--(-0.75,0.625)--(-1.25,0.25);
	\draw[blue](-1,-0.25)--(-0.75,0.625)--(-0.5,-0.25);
	
	\draw[blue](4.25,0.25)--(4.5,-0.25)--(5.,-0.25)--(5.25,0.25)--(4.75,0.625)--(4.25,0.25);
	\draw[blue] (4.25,0.25)--(5.25,0.25)--(4.5,-0.25);
	
	\draw[blue](0,-2.5)--(0.25,-3.)--(0.75,-3.)--(1,-2.5)--(0.5,-2.125)--(0,-2.5);
	\draw[blue] (0,-2.5) -- (0.75,-3) -- (0.5,-2.125);
	
	\draw[blue](1.5,-2.5)--(1.75,-3.)--(2.25,-3.)--(2.5,-2.5)--(2.,-2.125)--(1.5,-2.5);
	\draw[blue] (2.25,-3)--(1.5,-2.5);
	
	\draw[blue](3,-2.5)--(3.25,-3.)--(3.75,-3.)--(4,-2.5)--(3.5,-2.125)--(3,-2.5);
	\draw[blue](4,-2.5)--(3,-2.5)--(3.75,-3);
	
	\draw[blue](3.75,-1)--(4.,-1.5)--(4.5,-1.5)--(4.75,-1)--(4.25,-0.625)--(3.75,-1);
	\draw[blue](3.75,-1)--(4.75,-1);
	
	\draw[blue](3,1.25)--(3.25,0.75)--(3.75,0.75)--(4,1.25)--(3.5,1.625)--(3,1.25);
	\draw[blue](4,1.25)--(3.25,0.75);
	
	\draw[blue](1.5,2.25)--(1.75,1.75)--(2.25,1.75)--(2.5,2.25)--(2.,2.625)--(1.5,2.25);
	\draw[blue](2,2.625)--(1.75,1.75)--(2.5,2.25);
	
	\draw[blue](0,1.25)--(0.25,0.75)--(0.75,0.75)--(1,1.25)--(0.5,1.625)--(0,1.25);
	\draw[blue](0.25,0.75)--(0.5,1.625);
	
	\draw[blue](1.5,-0.25)--(1.75,-0.75)--(2.25,-0.75)--(2.5,-0.25)--(2.,0.125)--(1.5,-0.25);
	
	\draw[blue](-0.75,-1)--(-0.5,-1.5)--(0.,-1.5)--(0.25,-1)--(-0.25,-0.625)--(-0.75,-1);
	\draw[blue](0,-1.5)--(-0.25,-0.625);

	\node at (2,-0.885) []{\scriptsize $1$};
	\node at (2.5,-0.55) []{\scriptsize $5$};
	\node at (2.375,0.05) []{\scriptsize $4$};
	\node at (1.625,0.05) []{\scriptsize $3$};
	\node at (1.5,-0.55) []{\scriptsize $2$};
	
	\begin{scope}[xshift=10cm]
	\node at (0,0) {
		\begin{tikzpicture}[scale=0.2]
		\draw (0,0) -- (0,1) -- (1.5,2) (0,1) -- (0.5,2) (0,1) -- (-0.5,2) (0,1) -- (-1.5,2);
		\end{tikzpicture}
	};
	\node at (0,0) [above=0.15cm] {\scriptsize$(1)(2)(3)(4)(5)$};
	\node (A) at (0,3) [inner sep=0] {};
	\node (B) at (18:3) [inner sep=0] {};
	\node (C) at (-54:3) [inner sep=0] {};
	\node (D) at (-126:3) [inner sep=0] {};
	\node (E) at (162:3) [inner sep=0] {};
	\node (A1) at ($(A)!0.5!(B)$) [inner sep=0] {};
	\node (B1) at ($(B)!0.5!(C)$) [inner sep=0] {};
	\node (C1) at ($(C)!0.5!(D)$) [inner sep=0] {};
	\node (D1) at ($(D)!0.5!(E)$) [inner sep=0] {};
	\node (E1) at ($(E)!0.5!(A)$) [inner sep=0] {};
	\draw [very thick] (A) -- (B) -- (C) -- (D) -- (E) -- (A);
	\fill (A) circle (2pt) (B) circle (2pt) (C) circle (2pt) (D) circle (2pt) (E) circle (2pt);
	\node at (A) [above=0cm] {
		\begin{tikzpicture}[scale=0.2]
		\draw (0,0) -- (0,1) -- (1.5,2) (0.5,2) -- (-0.5,1.333) (-0.5,2) -- (-1,1.666) (0,1) -- (-1.5,2);
		\end{tikzpicture}
	};
	\node at (A) [above=0.5cm] {\scriptsize$(15)(23)(4)$};
	\node at (A) [below=0.25cm] {$\frac{1}{s_{23}s_{15}}$};
	\node at (B) [right=-0.25cm] {
		\begin{tikzpicture}[scale=0.2]
		\node at (0,2) [above=0pt] {\scriptsize$(15)(2)(34)$};
		\draw (0,0) -- (0,1) -- (1.5,2) (0.5,2) -- (-0.5,1.333) (-0.5,2) -- (0,1.666) (0,1) -- (-1.5,2);
		\end{tikzpicture}
	};
	\node at (B) [left=1pt] {$\frac{1}{s_{15}s_{34}}$};
	\node at (C) [right=0cm]{
		\begin{tikzpicture}[scale=0.2]
		\node at (0,2) [above=0pt] {\scriptsize$(12)(34)(5)$};
		\draw (0,0) -- (0,1) -- (1.5,2) (-0.5,2) -- (0.5,1.333) (0.5,2) -- (0,1.666) (0,1) -- (-1.5,2);
		\end{tikzpicture}
	};
	\node at (C) [above left=0pt] {$\frac{-1}{s_{12}s_{34}}$};
	\node at (D) [left=0cm] {
		\begin{tikzpicture}[scale=0.2]
		\node at (0,2) [above=0pt] {\scriptsize$(12)(3)(45)$};
		\draw (0,0) -- (0,1) -- (1.5,2) (-0.5,2) -- (0.5,1.333) (0.5,2) -- (1,1.666) (0,1) -- (-1.5,2);
		\end{tikzpicture}
	};
	\node at (D) [above right=0pt] {$\frac{-1}{s_{12}s_{45}}$};
	\node at (E) [left=0pt] {
		\begin{tikzpicture}[scale=0.2]
		\node at (0,2) [above=0pt] {\scriptsize$(1)(23)(45)$};
		\draw (0,0) -- (0,1) -- (1.5,2) (-0.5,2) -- (-1,1.666) (0.5,2) -- (1,1.666) (0,1) -- (-1.5,2);
		\end{tikzpicture}
	};
	\node at (E) [right=0pt] {$\frac{1}{s_{23}s_{45}}$};
	\node at (A1) [above right=-8pt] {
		\begin{tikzpicture}[scale=0.2]
		\node at (0,2) [above=0pt] {\scriptsize$(15)(2)(3)(4)$};
		\draw (0,0) -- (0,1) -- (1.5,2) (-0.5,2) -- (-0.5,1.333) (0.5,2) -- (-0.5,1.333) (0,1) -- (-1.5,2);
		\node at (-5,-1) [right=1pt] {\tiny$\frac{1}{s_{15}}(\frac{1}{s_{23}}+\frac{1}{s_{34}})$};
		\end{tikzpicture}
	};
	\node at (B1) [right=-3pt] {
		\begin{tikzpicture}[scale=0.2]
		\node at (0,2) [above=0pt] {\scriptsize$(1)(2)(34)(5)$};
		\draw (0,0) -- (0,1) -- (1.5,2) (-0.5,2) -- (0,1.666) (0.5,2) -- (0,1.666) -- (0,1) (0,1) -- (-1.5,2);
		\node at (-5,-1) [right=1pt] {\tiny$\frac{-1}{s_{34}}(\frac{1}{s_{12}}+\frac{1}{s_{15}})$};
		\end{tikzpicture}
	};
	\node at (C1) [below=0pt]{
		\begin{tikzpicture}[scale=0.2]
		\node at (0,2) [above=0pt] {\scriptsize$(12)(3)(4)(5)$};
		\draw (0,0) -- (0,1) -- (1.5,2) (-0.5,2) -- (0.5,1.333) (0.5,2) -- (0.5,1.333) (0,1) -- (-1.5,2);
		\node at (0,0) [below=0pt] {\tiny$\frac{1}{s_{12}}(\frac{1}{s_{34}}+\frac{1}{s_{45}})$};
		\end{tikzpicture}
	};
	\node at (D1) [left=-3pt] {
		\begin{tikzpicture}[scale=0.2]
		\node at (0,2) [above=0pt] {\scriptsize$(1)(2)(3)(45)$};
		\draw (0,0) -- (0,1) -- (1.5,2) (-0.5,2) -- (0,1) (0.5,2) -- (1,1.666) (0,1) -- (-1.5,2);
		\node at (5,-1) [left=1pt] {\tiny$\frac{1}{s_{45}}(\frac{1}{s_{12}}+\frac{1}{s_{23}})$};
		\end{tikzpicture}
	};
	\node at (E1) [above left=-8pt] {
		\begin{tikzpicture}[scale=0.2]
		\node at (0,2) [above=0pt] {\scriptsize$(1)(23)(4)(5)$};
		\draw (0,0) -- (0,1) -- (1.5,2) (-0.5,2) -- (-1,1.666) (0.5,2) -- (0,1) (0,1) -- (-1.5,2);
		\node at (5,-1) [left=1pt] {\tiny$\frac{1}{s_{23}}(\frac{1}{s_{45}}+\frac{1}{s_{51}})$};
		\end{tikzpicture}
	};
\end{scope}
\end{tikzpicture}
\caption{The associahedron for five point amplitudes and the PT-factors.}\label{5passo}
\end{figure}
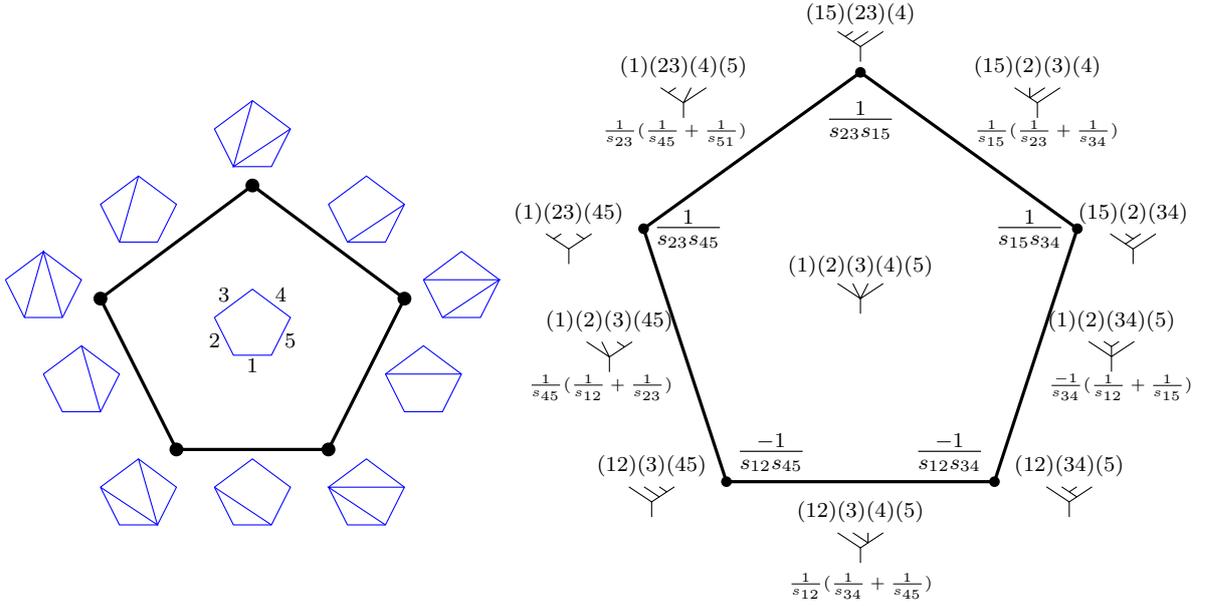

A five point amplitude is dual to a pentagon, and the associahedron constructed from all its (partial) triangulations is also a pentagon, as shown in Fig.~\ref{5passo},
where the thick black lines form the associahedron and the blue line is the triangulations of pentagon. The face corresponds to $\Spaa{12345}$, while the edges correspond to $\Spaa{12543}$, $\Spaa{12354}$, $\Spaa{13245}$, $\Spaa{14325}$, $\Spaa{12435}$, and the vertices correspond to $\Spaa{12453}$, $\Spaa{13254}$, $\Spaa{14235}$, $\Spaa{13425}$, $\Spaa{12534}$. The PT-factor of each vertex evaluates to a single Feynman diagram. An edge connects two vertices, which means that the PT-factor of the edge evaluates to two Feynman diagrams. Two edges share a common point, which means that there is common Feynman diagram shared by them. The face contains five vertices, such that its PT-factor evaluates to five Feynman diagrams. Therefore the relation among different PT-factors is obvious in the diagram. For higher point amplitude, the associahedron would be much more complicated geometric objects, however the correspondence is similar.

\subsection{Relation analysis via cycle representation}

From the $n$-gon picture, we
see that by adding one more triangulation line (i.e.,
fix one more propagator), we get the immediate child amplitude.
In contrary, by removing one triangulation line (i.e., unfix a propagator), we recover the mother amplitude. In terms of zig-zag path, we have the following picture,
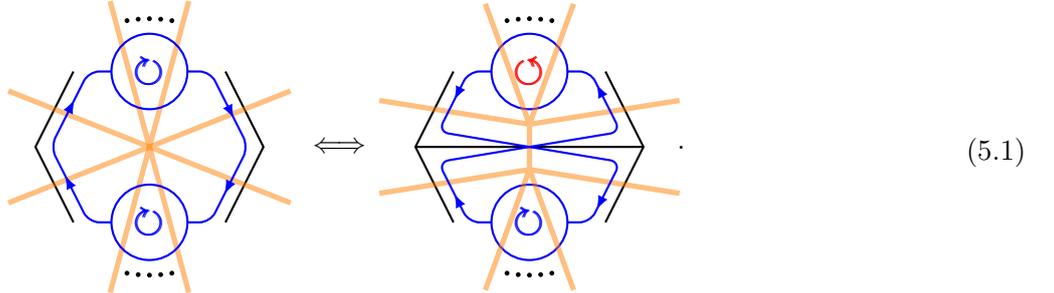
\begin{equation}
\label{eq:split}
\adjustbox{raise=-1.8cm}{\begin{tikzpicture}
	\draw [thick] (0.5,-1) -- (0,0) -- (0.5,1);
	\draw [thick] (2.5,-1) -- (3,0) -- (2.5,1);
	\draw [line width=2pt,orange,draw opacity=0.5] (1.5,0) -- ++(158.2:2) (1.5,0) -- ++(-21.8:2) (1.5,0) -- ++(-158.2:2) (1.5,0) -- ++(21.8:2);
	\draw [line width=2pt,orange,draw opacity=0.5] (1.5,0) -- ++(75:2) (1.5,0) -- ++(105:2);
	\draw [line width=2pt,orange,draw opacity=0.5] (1.5,0) -- ++(-75:2) (1.5,0) -- ++(-105:2);
	\foreach \x in {80,85,90,95,100} {\filldraw (1.5,0) ++(\x:1.7) circle (0.7pt);}
	\foreach \x in {-80,-85,-90,-95,-100} {\filldraw (1.5,0) ++(\x:1.7) circle (0.7pt);}
	\draw [thick,blue] (1.5,1) circle (0.5cm) (1.5,-1) circle (0.5cm);
	\draw [thick,blue,rounded corners=4pt,postaction={decorate,decoration={markings,mark=at position 0.25 with {\arrowreversed{latex}},
		mark=at position 0.65 with {\arrowreversed{latex}}}}]
		(1,1) -- (0.7,1) -- (0.2,0) -- (0.7,-1) -- (1,-1);
	\draw [thick,blue,rounded corners=4pt,postaction={decorate,decoration={markings,mark=at position 0.35 with {\arrow{latex}},
			mark=at position 0.75 with {\arrow{latex}}}}]
		(2,1) -- (2.3,1) -- (2.8,0) -- (2.3,-1) -- (2,-1);
	\node at (1.5,1) [blue]{\Large$\pmb\circlearrowright$};
	\node at (1.5,-1) [blue]{\Large$\pmb\circlearrowright$};
	\node at (4,0) {$\Longleftrightarrow$};
	\begin{scope}[xshift=5cm]
		\draw [thick] (0.5,-1) -- (0,0) -- (0.5,1);
		\draw [thick] (2.5,-1) -- (3,0) -- (2.5,1);
		\draw [thick] (0,0) -- (3,0);
		\draw [thick,blue] (1.5,1) circle (0.5cm) (1.5,-1) circle (0.5cm);
		\draw [line width=2pt,orange,draw opacity=0.5] (1.5,-0.3) -- (1.5,0.3) (1.5,0.3) -- ++(170.91:2) (1.5,0.3) -- ++(9.09:2) (1.5,0.3) -- ++(70:1.7) (1.5,0.3) -- ++(110:1.7) (1.5,-0.3) -- ++(-9.09:2) (1.5,-0.3) -- ++(-170.91:2) (1.5,-0.3) -- ++(-70:1.7) (1.5,-0.3) -- ++(-110:1.7);
		\foreach \x in {80,85,90,95,100} {\filldraw (1.5,0) ++(\x:1.7) circle (0.7pt);}
		\foreach \x in {-80,-85,-90,-95,-100} {\filldraw (1.5,0) ++(\x:1.7) circle (0.7pt);}
		\draw [thick,blue,rounded corners=4pt,postaction={decorate,decoration={markings,mark=at position 0.15 with {\arrow{latex}},
				mark=at position 0.9 with {\arrow{latex}}}}]
			(1,-1) -- (0.7,-1) -- (0.3,-0.2) -- (2.7,0.2) -- (2.3,1) -- (2,1);
		\draw [thick,blue,rounded corners=4pt,postaction={decorate,decoration={markings,mark=at position 0.15 with {\arrow{latex}},
				mark=at position 0.9 with {\arrow{latex}}}}] (1,1) -- (0.7,1) -- (0.3,0.2) -- (2.7,-0.2) -- (2.3,-1) -- (2,-1);
		\node at (1.5,1) [red]{\Large$\pmb\circlearrowleft$};
		\node at (1.5,-1) [blue]{\Large$\pmb\circlearrowright$};
		\node at (3.5,0){.};
	\end{scope}
	\end{tikzpicture}}
\end{equation}
The extra triangulation separates the $n$-gon into two subpolygons. Notably, in one of them we need to reverse the ordering. In this section, we study how the above picture is realized by good cycle representations, namely, how to merge or split certain parts in good cycle representations to fix or unfix a propagator.

We start with the general discussion. The left hand side of Eq.~\eqref{eq:split} indicates that there exists a good cycle representation with the form
\begin{equation}
\label{eq:split_prop}
\pmb{\beta}=\pmb{\beta}_{\text{lower}}\pmb{\beta}_{\text{upper}}=\underbrace{(\ldots)\pmb{|}(\ldots)(\ldots)\pmb{|}\ldots(\ldots)}_{\text{lower}}\,\pmb{\Bigg|}\,\underbrace{(\ldots)\pmb{|}(\ldots)(\ldots)\pmb{|}\ldots\ldots}_{\text{upper}}~,~~~
\end{equation}
where subscripts $_{\text{lower}}$ and $_{\text{upper}}$ denote the external legs below and above the extra triangulation line in~\eqref{eq:split}. This cycle representation can either be a V-type or P-type one. Suppose the upper set consists of
\begin{equation}
\text{upper}=\{i,i+1\ldots j\}~,~~~
\end{equation}
then the reversing process of~\eqref{eq:split} can be realized by
\begin{equation}
\label{eq:partreverse}
\pmb{\beta}_{\text{upper}}^{\text{reversed}}=\pmb{\beta}_{\text{upper}}\pmb{\beta}_{r}~,~~~
\end{equation}
where $\pmb{\beta}_{r}$ simply flips the ordering $\pmb{\beta}_r|i,i+1\ldots j\rangle=|j\ldots i+1,i\rangle$. Similar to Eq.~\eqref{eq:grgc}, $\pmb{\beta}_r$ has the cycle representation
\begin{equation}
\label{eq:beta_r}
\pmb{\beta}_{r}=\left\{\begin{array}{lcl}
(ij)(i+1,j-1)\ldots(\frac{i+j-1}{2},\frac{i+j+1}{2}) & \qquad\qquad &j-i=\text{odd} \\
(ij)(i+1,j-1)\ldots(\frac{i+j-2}{2},\frac{i+j+2}{2})(\frac{i+j}{2}) & \qquad\qquad & j-i=\text{even}
\end{array}\right.~.~~~
\end{equation}
Therefore, the process of~\eqref{eq:split} can be realized in an algebraic way as\footnote{Equivalently, we can reverse the lower part. The result will only differ by an overall reversing.}
\begin{equation}
\pmb{\beta}_{\text{lower}}\pmb{\beta}_{\text{upper}}\;\Longleftrightarrow\;\pmb{\beta}_{\text{lower}}\pmb{\beta}_{\text{upper}}\pmb{\beta}_{r}~.~~~
\end{equation}
If the propagator manifested in~\eqref{eq:split_prop} is an \emph{overall} one, then the above process gives the immediate mother amplitude that has the original amplitude as a part. \emph{Otherwise}, the above process gives the immediate child amplitude that contains this propagator as an overall factor, and is a part of the original amplitude.

Next, we use the example given in~\eref{8p-bad-1-1} with cycle representations~\eref{8p-bad-1-2} to demonstrate our idea,
\begin{equation}
\label{eq:example1}
\PT(\pmb\b)=\langle 12846573\rangle\;\Longrightarrow\;\frac{1}{s_{12}s_{56}s_{8123}}\left(\frac{1}{s_{812}}+\frac{1}{s_{123}}\right)\left(\frac{1}{s_{456}}+\frac{1}{s_{567}}\right)~.~~~
\end{equation}
We first consider its immediate child amplitudes by fixing the poles $s_{812}$, $s_{123}$, $s_{456}$ and $s_{567}$ one by one.

\paragraph{Fix the pole $s_{812}$:} To achieve this goal, we need to use the good cycle representations that manifest the separation $\{8,1,2\}$ and $\{3,4,5,6,7\}$. In~\eref{8p-bad-1-2}, only $(12)(3)(47)(5)(6)(8)$ and $(128)(3467)(5)$ satisfy the condition. We can achieve the child amplitude by using either one. Starting with $(12)(3)(47)(5)(6)(8)$, we split all cycles into two parts according to the pole, namely, $(8)(12)\pmb{|}(3)(47)(5)(6)$. Now we can keep one part invariant and perform the prescription~\eqref{eq:partreverse} to the other. For example, we keep the part $(8)(12)$, such that for another part we should do the following manipulation as
\bea [(3)(47)(5)(6)]\cdot[(37)(46)(5)]=(3467)(5)~,~~~\eea
where $(37)(46)(5)$ is the reversing permutation $\pmb{\beta}_r$ obtained from~\eqref{eq:beta_r}. Putting the two parts together, we get the cycle representation $(8)(12)(3467)(5)$, which corresponds to the PT-factor $\PT(\pmb{\beta})=\Spaa{12837564}$. It indeed gives the desired child amplitude,
\bea \PT(\pmb{\beta})=\Spaa{12837564}\;\Longrightarrow\;\frac{1}{s_{12} s_{56}s_{8123}} \left(\frac{1}{s_{812}}\right)\left(\frac{1}{s_{456}}+\frac{1}{s_{567}}\right)~.~~~\label{8p-bad-1-daughter-812}\eea
Alternatively, we keep the part $(3)(47)(5)(6)$ intact and act the prescription~\eqref{eq:partreverse} onto the part $(8)(12)$,
\bea [(8)(12)]\cdot[(82)(1)]=(812)~.~~~\eea
Putting them together, we get another cycle representation $(812)(3)(47)(5)(6)$, corresponding to the same PT-factor as~\eqref{8p-bad-1-daughter-812}. If we use the other cycle representation $(128)\pmb{|}(3467)(5)$, we get the same result,
\bea [(128)]\cdot[(82)(1)](3476)(5)\ & = &
	(8)(12)(3467)(5)~,\nonumber\\
	(128)[(3467)(5)]\cdot[(37)(46)(5)] & = &
	(128)(3)(47)(5)(6)~.~~~\eea

\paragraph{Fix the pole $s_{123}$:} For this case, the cycle representations $(12)(3)(47)(5)(6)(8)$ and $(132)(4875)(6)$ from~\eref{8p-bad-1-2} can be used. By  similar manipulations, we get
\begin{subequations}
\bea
	(12)(3)(47)(5)(6)(8) & \Longrightarrow & \left\{\begin{array}{rcl}
	[(12)(3)]\cdot[(13)(2)](47)(5)(6)(8) & = & (132)(47)(5)(6)(8) \\
	(12)(3)[(47)(5)(6)(8)]\cdot[(48)(57)(6)] & = & (12)(3)(4875)(6)
	\end{array}\right. \,,\\
    (132)(4875)(6) & \Longrightarrow & \left\{\begin{array}{rcl}
		(132)]\cdot[(13)(2)](4875)(6) & = & (12)(3)(4875)(6) \\
		(132)[(4875)(6)]\cdot[(48)(57)(6)] & = & (132)(47)(5)(6)(8)
	\end{array}\right.~.~~~
\eea
\end{subequations}
Both results correspond to the PT-factor $\PT(\pmb{\beta})=\Spaa{12756483}$, which is evaluated to
\bea \frac{1}{s_{12} s_{56}s_{8123}} \left(\frac{1}{s_{123}}\right)\left(\frac{1}{s_{456}}+\frac{1}{s_{567}}\right)~.~~~\label{8p-bad-1-daughter-123}\eea
\paragraph{Fix the pole $s_{456}$:} For this case, the cycle representations $(1)(2)(38)(4)(56)(7)$ and	$(1)(2378)(456)$ from~\eref{8p-bad-1-2} can be used. By similar manipulations we get
\begin{subequations}
\bea
	(1)(2)(38)(4)(56)(7) & \Longrightarrow & \left\{\begin{array}{rcl}
	(1)(2)(38)(7)[(4)(56)]\cdot[(46)(5)] & =  & (1)(2)(38)(7)(456) \\
	\left[(1)(2)(38)(7)\right]\cdot[(73)(82)(1)](4)(56) & = & (1)(2378)(4)(56)
	\end{array}\right.\,,\\
	(1)(2378)(456) & \Longrightarrow & \left\{\begin{array}{rcl}
	(1)(2378)[(456)]\cdot[(46)(5)] & =  & (1)(2378)(4)(56) \\
	\left[(1)(2378)\right]\cdot[(73)(82)(1)](456) & = &(1)(2)(38)(7)(456)
	\end{array}\right.~.~~~
\eea
\end{subequations}
Both results correspond to the PT-factor $\PT(\pmb{\beta})=\Spaa{12856473}$, which is evaluated to
\bea \frac{1}{s_{12} s_{56}s_{8123}} \left(\frac{1}{s_{123}}+\frac{1}{s_{812}}\right)\left(\frac{1}{s_{456}}\right)~.~~~\label{8p-bad-1-daughter-456}\eea

\paragraph{Fix the pole $s_{567}$:} For this case, the cycle representations $(1)(2)(38)(4)(56)(7)$ and $(1843)(2)(576)$ from~\eref{8p-bad-1-2} can be used. By similar manipulations we get
\begin{subequations}
\bea
	(1)(2)(38)(4)(56)(7) & \Longrightarrow &\left\{\begin{array}{rcl}
	(1)(2)(38)(4)[(56)(7)]\cdot[(57)(6)] & =  & (1)(2)(38)(4) (576) \\
	\left[(1)(2)(38)(4)\right]\cdot[(84)(13)(2)](56)(7) & = & (8431)(2)(56)(7)
	\end{array}\right.\,, \\
	(1843)(2)(576) & \Longrightarrow & \left\{\begin{array}{rcl}
		(1843)(2)[(576)]\cdot[(57)(6)] & =  & (1843)(2)(56)(7)\\
		\left[(1843)(2)\right]\cdot[(84)(13)(2)](576) & = &	(1)(2)(38)(4)(576)
	\end{array}\right.~.~~~
\eea
\end{subequations}
Both results correspond to the PT-factor $\PT(\pmb{\beta})=\Spaa{12847563}$, which is evaluated to
\bea \frac{1}{s_{12} s_{56}s_{1238}} \left(\frac{1}{s_{123}}+\frac{1}{s_{812}}\right)\left(\frac{1}{s_{567}}\right)~.~~~\label{8p-bad-1-daughter-567}\eea

After showing how to get the child amplitudes, we now discuss the case of mother amplitudes. This happens when we perform the prescription~\eqref{eq:partreverse} to a separation that manifests an \emph{overall} propagator. For the case~\eqref{eq:example1}, there are three common poles $s_{12}$, $s_{56}$ and $s_{8123}$. Relaxing any one of them, we can get a mother amplitude. The procedure is similar to the above discussion, and we again do it one by one.
\paragraph{Unfix the pole $s_{12}$:} For this case, the following two cycle representations $(1)(2)(38)(4)(56)(7)$ and $(12)(3)(47)(5)(6)(8)$ manifest the pole $s_{12}$.
If we take $(1)(2)(38)(4)(56)(7)$, we have the following calculation according to~\eqref{eq:partreverse},
\begin{subequations}
\begin{align} (1)(2)(38)(4)(56)(7) & \Longrightarrow & \left\{\begin{array}{rcl}
	[(1)(2)]\cdot[(12)](38)(4)(56)(7) & = & (12)(38)(4)(56)(7) \\
	(1)(2)[(38)(4)(56)(7)]\cdot[(38)(47)(56)] & = & (1)(2)(3)(47)(5)(6)(8)
	\end{array}\right.~,~~~ \\
	(12)(3)(47)(5)(6)(8) & \Longrightarrow & \left\{\begin{array}{rcl}
	[(12)]\cdot[(12)](3)(47)(5)(6)(8) & = &	(1)(2)(3)(47)(5)(6)(8) \\
	(12)[(3)(47)(5)(6)(8)]\cdot[(38)(47)(56)] & = & (12)(38)(4)(56)(7)
	\end{array}\right.~,~~~
\end{align}
\end{subequations}
which correspond to $\PT(\pmb{\beta})=\Spaa{12375648}$. It gives ten terms, among which four are just~\eqref{eq:example1}.
	
\paragraph{Unfix the pole $s_{56}$:} For this case, the cycle representations $(1)(2)(38)(4)(56)(7)$ and $(12)(3)(47)(5)(6)(8)$ manifest the pole $s_{56}$. We have the following calculation according to~\eqref{eq:partreverse},
\begin{subequations}
\bea
	(1)(2)(38)(4)(56)(7) & \Longrightarrow & \left\{\begin{array}{rcl}
		[(56)]\cdot[(56)](1)(2) (38)(4)(7) & = & (1)(2) (38)(4)(5)(6)(7) \\
		(56)[(1)(2)(38)(4)(7)]\cdot[(47)(38)(12)] & = & (56)(12)(3)(8)(47)
	\end{array}\right.\,, \\
	(12)(3)(47)(5)(6)(8) & \Longrightarrow &\left\{\begin{array}{rcl}
		[(5)(6)]\cdot[(56)](12)(3)(47)(8) & = & (12)(3)(47)(56)(8) \\
		(5)(6)[(12)(3)(47)(8)]\cdot[(47)(38)(12)] & = & (5)(6)(1)(2)(4)(7)(38)
	\end{array}\right.~.~~~
\eea
\end{subequations}
One can check that they correspond to $\PT(\pmb{\beta})=\Spaa{12845673}$. It also gives ten terms, among which four are just~\eqref{eq:example1}.

\paragraph{Unfix the pole $s_{8123}$:} For this case, we have two good cycle representations $(1)(2)(38)(4)(56)(7)$ and $(12)(3)(47)(5)(6)(8)$ that manifest the pole $s_{8123}$. We have the following calculation according to~\eqref{eq:partreverse},
\begin{subequations}
\bea
	(1)(2)(38)(4)(56)(7) & \Longrightarrow & \left\{\begin{array}{rcl}
		[(1)(2)(38)]\cdot[(12)(38)](4)(56)(7) & = & (12)(3)(8)(4)(56)(7) \\
		(1)(2)(38)[(4)(56)(7)]\cdot[(47)(56)] & = & (1)(2)(38)(47)(5)(6)
		\end{array}\right.\,,\\
	(12)(3)(47)(5)(6)(8) & \Longrightarrow & \left\{\begin{array}{rcl}
		[(12)(3)(8)]\cdot[(12)(38)](47)(5)(6) & = &(1)(2)(38)(47)(5)(6)\\
		(12)(3)(8)[(47)(5)(6)]\cdot[(47)(56)] & \to &(12)(3)(8)(4)(56)(7)
	\end{array}\right.~,~~~
\eea
\end{subequations}
which correspond to $\PT(\pmb{\beta})=\Spaa{12875643}$. It gives 14 terms, among which four are just~\eqref{eq:example1}.

\subsection{Relation analysis via cross-ratio factor}

We can also study the relations of PT-factor from another approach.
As we have already known,  the relations between different
PT-factors can be seen as selecting terms corresponding to specific
pole structures in the evaluated results. For instance,
$\Spaa{1234}\Spaa{1234}$ evaluates to $\frac{1}{s_{12}}+\frac{1}{s_{23}}$ while $\Spaa{1234}\Spaa{1243}$ evaluates to $\frac{1}{s_{12}}$. It means that, by selecting terms with pole
$\frac{1}{s_{12}}$ in $\Spaa{1234}\Spaa{1234}$, we can reproduce the
result of $\Spaa{1234}\Spaa{1243}$.
To achieve this goal at the CHY-integrand level, we can use the
cross-ratio factor given  in paper \cite{Feng:2016nrf}, which we
will call it the \emph{selecting factor}
\bea f^{\text{select}}[a,b,c,d]:=\frac{[ab][cd]}{[ac][bd]}~~~,~~~[ab]:=\sigma_{ab}~.~~~\eea
%
To pick up the Feynman diagrams with a pole $\frac{1}{s_A}$ from a given CHY-integral result, where $A$ follows a certain color ordering, we propose to multiply the CHY-integrand with a selecting factor $f^{\text{select}}[\overline{A}_{-1},A_1,A_{-1},\overline{A}_1]$, where $A_1,A_{-1}$ are the first and last elements of the subset $A$ respectively, and $\overline{A}$ is the complement subset of $A$, i.e., in order to pick up terms with pole $\frac{1}{s_A}$, the
arguments $b,c$ in the selecting factor should be the two ending
legs in the set $A$, while the arguments $a,d$ are the nearby two legs
of $b,c$ outside the set $A$ respectively. As an illustration, let
us consider the above mentioned PT-factors $\Spaa{1234}$ and
$\Spaa{1243}$. We want to select terms with $\frac{1}{s_{12}}$ pole
in the evaluated result of $\Spaa{1234}\Spaa{1234}$, which means
that we need to take the selecting factor $f^{\text{select}}[4,1,2,3]$,
\bea \Spaa{1234}f^{\text{select}}[4,1,2,3]=\frac{1}{\sigma_{12}\sigma_{23}\sigma_{34}\sigma_{41}}\frac{\sigma_{41}\sigma_{23}}{\sigma_{42}\sigma_{13}}=-\Spaa{1243}~.~~~\eea
It indeed produces the PT-factor $\Spaa{1243}$, despite of the overall sign.

There is a subtlety in the choice of the selecting factor. The bi-adjoint scalar theory has two color orderings, given by $\PT(\pmb \a)$ and $\PT(\pmb \b)$. We should choose $A$ to follow one of the orderings. If $\pmb\a=\pmb\b$, there is no ambiguity in defining the selecting factor, which is the
situation discussed in~\cite{Feng:2016nrf}. However if $\pmb \a\neq \pmb \b$, the multiplication
\bea \PT(\pmb \a)\times \PT(\pmb\b)\times f^{\text{select}}[\overline{A}_{-1},A_1,A_{-1},\overline{A}_1]\eea
has two choices for a given set $A$. The arguments of $f^{\text{select}}$ depend on the color ordering of legs, and we have two color orderings to rely on. It can be shown that although we would get two different selecting factors, the resulting CHY-integrands are equivalent in the sense that the difference of two CHY-integrands evaluates to zero. We note that this happens only when $\frac{1}{s_A}$ is indeed a physical pole of the integrated result of $\PT(\pmb\a)\times\PT(\pmb\b)$, but not an \emph{overall} one.

Now let us present a brief explanation about why the selecting
factor  $f^{\text{select}}$ is able to pick up terms with
specific poles. It is known that from
\cite{Baadsgaard:2015voa,Baadsgaard:2015ifa,Baadsgaard:2015hia}, the
order of pole $\frac{1}{s_A}$ in the evaluated result is
characterized by the \emph{pole index}
\bea \chi[A]=\mathbb{L}[A]-2(|A|-1)~,~~~\eea
where $\mathbb{L}[A]$ is the linking number of subset $A$, and $|A|$ is the length of subset $A$.\footnote{The linking number $\mathbb{L}[A]$ can be read out from the so-called {\sl 4-regular} diagrams, which is discussed in details in~\cite{Baadsgaard:2015voa,Baadsgaard:2015ifa,Baadsgaard:2015hia}.} If $\chi [A]<0$, there is no ${s_A}$ pole, while if $\chi[A]\geqslant 0$, the pole would appear in the result as $\frac{1}{s_A^{\chi+1}}$. For CHY-integrands with two PT-factors, we only have simple poles, namely, $\chi[A]\leqslant 0$ for any subset $A$. With this in mind, let us take a further look on the selecting factor $f^{\text{select}}$, assuming that $\frac{1}{s_A}$ is not an overall pole. The combinations in both the numerator and denominator of $f^{\text{select}}$ represent lines connecting elements in subset $A$ and its complement $\overline{A}$, so that $f^{\text{select}}$ will not change the linking number of $A$ itself: after multiplying $f^{\text{select}}$, we still have $\chi[A]=0$ and the pole $\frac{1}{s_A}$ remains unchanged. Now suppose there is another pole $\frac{1}{s_B}$, where $B$ has nonempty overlap with both $A$ and $\overline{A}$, then it will be removed by $f^{\text{select}}$. The reason is that
in the denominator of PT-factor there are $[A_{-1}\overline{A}_1]$ and $[\overline{A}_{-1}A_{1}]$, while in the numerator of the selecting factor there are $[A_{-1}\overline{A}_1]$ and $[\overline{A}_{-1}A_1]$, which at least reduces the linking number of subset $B$ by one, such that $\chi[B]$ is reduced from $0$ to $-1$. Thus all the poles that are not compatible with $\frac{1}{s_A}$ are removed. Finally, we need to show that the terms with $\frac{1}{s_A}$ are not altered by the selecting factor, namely, we should confirm that in a term with pole $\frac{1}{s_A}$, $f^{\text{select}}$ do not change the pole indices of all the other poles.
By the compatibility condition, these poles should either be a subset of $A$ or a subset of $\overline{A}$. For the four factors $[\overline{A}_{-1}A_1]$, $[A_{-1}\overline{A}_1]$, $[\overline{A}_{-1}A_{-1}]$ and $[A_1\overline{A}_1]$ in $f^{\text{select}}$, each one contains an element from subset $A$ and another from $\overline{A}$, so that none of them contributes to the linking number of either $A$ or $\overline{A}$.

In general, when using $f^{\text{select}}$ to pick up a pole $\frac{1}{s_A}$, we will encounter three situations.
In the first situation, the original theory does not contain such a pole. Then multiplying the selecting factor does not make any sense. For instance, the CHY-integrand $\Spaa{123456}\Spaa{124563}$ evaluates to
\bea { \Spaa{123456}}\times { \Spaa{124563}}\to \frac{1}{s_{12}s_{123}}\left(\frac{1}{s_{45}}+\frac{1}{s_{56}}\right)~,~~~\label{selectExam}\eea
which does not contain the pole $\frac{1}{s_{34}}$.
If we insist to take the selecting factor
$f^{\text{select}}[2,3,4,5]$ following the color ordering of the
first PT-factor, we get
\bea \Big({ \Spaa{123456}}
f^{\text{select}}[2,3,4,5]\Big)\times { \Spaa{124563}}\to
-\frac{1}{s_{56}s_{124}}\left(\frac{1}{s_{12}}+\frac{1}{s_{56}}\right)~,~~~\eea
which is a completely irrelevant answer.

In the second situation,  the pole $\frac{1}{s_A}$ we pick is overall to all the terms. By multiplying the selecting factor, we
produce the mother amplitude, obtained by pinching the propagator $\frac{1}{s_A}$ in the Feynman diagram. For example, there are two overall poles $\frac{1}{s_{12}}$ and $\frac{1}{s_{123}}$ in~\eqref{selectExam}. If
we follow the color ordering of the first PT-factor, and multiply $f^{\text{select}}[6,1,2,3]$ that corresponds to $\frac{1}{s_{12}}$, we get
\bea \Big({
	\Spaa{123456}}f^{\text{select}}[6,1,2,3]\Big)\times {
	\Spaa{124563}} \to \frac{1}{s_{12}s_{123}}\left(\frac{1}{s_{45}}+\frac{1}{s_{56}}\right)+ \frac{1}{s_{13}s_{123}}\left(\frac{1}{s_{45}}+\frac{1}{s_{56}}\right)~,~~~\eea
which is a mother amplitude with additional terms produced. Similarly, if we
follow the color ordering of the second PT-factor, and multiply $f^{\text{select}}[3,1,2,4]$ that corresponds to $\frac{1}{s_{12}}$, we get
\bea { \Spaa{123456}}\times \Big({
	\Spaa{124563}}f^{\text{select}}[3,1,2,4]\Big)\to \frac{1}{s_{12}s_{123}}\left(\frac{1}{s_{45}}+\frac{1}{s_{56}}\right)+ \frac{1}{s_{23}s_{123}}\left(\frac{1}{s_{45}}+\frac{1}{s_{56}}\right)~,~~~\eea
which is another mother amplitude with different additional terms produced.
Similar calculation can be done for the overall pole $s_{123}$ and the
two different mother amplitudes are given by
\begin{align} \Big({
	\Spaa{123456}}f^{\text{select}}[6,1,3,4]\Big)\times {
	\Spaa{124563}} & \to  \frac{1}{s_{12} s_{123}}\left(\frac{1}{s_{45}}+\frac{1}{s_{56}}\right)+\frac{1}{s_{12} s_{124}}\left(
\frac{1}{s_{36}}+\frac{1}{s_{56}}\right)+\frac{1}{s_{12} s_{36}s_{45}}~,~~~\nonumber\\
 { \Spaa{123456}}\times \Big( {
	\Spaa{124563}}f^{\text{select}}[6,3,2,4]\Big) & \to
\frac{1}{s_{12} s_{123}}\left( \frac{1}{s_{45}}+\frac{1}{s_{56}}\right)+\frac{1}{s_{12} s_{126}}\left( \frac{1}{s_{34}}+\frac{1}{s_{45}}\right)+\frac{1}{s_{12} s_{34} s_{56}}~,~~~\nonumber
\end{align}
indicating that~\eqref{selectExam} can be part of the mother amplitudes with different color orderings.

In the third situation, the pole we pick is physical but an
overall one, thus multiplying the selecting factor we get the
immediate child amplitude. This is the case we have considered in detail in the beginning of this subsection. Consider an eight point CHY-integrand with
$\PT(\pmb \a)=\Spaa{12345678}$ and $\PT(\pmb\b)=\Spaa{12348765}$. The amplitude is
\begin{align}
\PT(\pmb\a)\PT(\pmb \b)&\to\frac{1}{s_{1234}}\left( \frac{1}{s_{12} s_{34}}
+\frac{1}{s_{12} s_{123}}+\frac{1}{s_{23} s_{123}}+\frac{1}{s_{23} s_{234}}+\frac{1}{s_{34} s_{234}}\right)\nonumber\\
&~~~~~~~~~~~~\times\left( \frac{1}{s_{56} s_{78}}
+\frac{1}{s_{56} s_{567}}+\frac{1}{s_{67} s_{567}}+\frac{1}{s_{67} s_{678}}+\frac{1}{s_{78} s_{678}}\right)\nonumber\\
&=\frac{1}{s_{1234}}\Bigg(\begin{array}{c}
    \begin{tikzpicture}[scale=0.5]
                     \draw[](0,0)--(1,0) (0.5,1)--(0,0)--(0.5,-1) (-0.7,0.7)--(0,0)--(-0.7,-0.7);
                     \node at (1.2,0.35) []{{\footnotesize $P_{1234}$}};
                     \node at (0.5,-1.35) []{{\footnotesize $1$}};
                     \node at (-0.75,-1.1) []{{\footnotesize $2$}};
                     \node at (-0.75,1.1) []{{\footnotesize $3$}};
                     \node at (0.5,1.35) []{{\footnotesize $4$}};
                   \end{tikzpicture} \end{array}\Bigg)\Bigg(
                   \begin{array}{c}
                     \begin{tikzpicture}[scale=0.5]
                     \draw[](0,0)--(-1,0) (-0.5,1)--(0,0)--(-0.5,-1) (0.7,0.7)--(0,0)--(0.7,-0.7);
                     \node at (-1.2,0.35) []{{\footnotesize $P_{5678}$}};
                     \node at (-0.5,-1.35) []{{\footnotesize $8$}};
                     \node at (0.75,-1.1) []{{\footnotesize $7$}};
                     \node at (0.75,1.1) []{{\footnotesize $6$}};
                     \node at (-0.5,1.35) []{{\footnotesize $5$}};
                   \end{tikzpicture}
 \end{array}\Bigg)~,~~~\label{selectExam1}
\end{align}
where we used the five point vertices to represent the terms in two parentheses. For instance, if we want to pick up terms with pole $\frac{1}{s_{78}}$, either $f^{\text{select}}[6,7,8,1]$ following the color ordering of $\PT(\pmb \a)$ or $f^{\text{select}}[4,8,7,6]$ following the color ordering of $\PT(\pmb \b)$ can do the job. Indeed, by direct computation, we confirm that
\begin{equation}
\left.\renewcommand*{\arraystretch}{1.5}\begin{array}{l}
	\Big(\PT(\pmb \a)f^{\text{select}}[6,7,8,1]\Big)\times \PT(\pmb \b) \\
	\PT(\pmb \a) \times \Big(\PT(\pmb \b)f^{\text{select}}[4,8,7,6]\Big)
\end{array}\right\}\to 	\frac{1}{s_{1234}}\times\frac{1}{s_{78}}\left( \frac{1}{s_{56}}+\frac{1}{s_{678}}\right)\times \Bigg(\begin{array}{c}
    \begin{tikzpicture}[scale=0.5]
                     \draw[](0,0)--(1,0) (0.5,1)--(0,0)--(0.5,-1) (-0.7,0.7)--(0,0)--(-0.7,-0.7);
                     \node at (1.2,0.35) []{{\footnotesize $P_{1234}$}};
                     \node at (0.5,-1.35) []{{\footnotesize $1$}};
                     \node at (-0.75,-1.1) []{{\footnotesize $2$}};
                     \node at (-0.75,1.1) []{{\footnotesize $3$}};
                     \node at (0.5,1.35) []{{\footnotesize $4$}};
                   \end{tikzpicture} \end{array}\Bigg)~.~~~
\label{selectExam2}
\end{equation}
In fact, we can multiply more than one selecting factors to pick up terms with several specific poles. For example, if we want to pick up terms with pole $\frac{1}{s_{78}}\frac{1}{s_{678}}$, we can start from the result~\eqref{selectExam2} and take the selecting factor $f^{\text{select}}[5,6,8,1]$ following the color ordering of $\PT(\pmb \a)$ in the second row of~\eqref{selectExam2}, or $f^{\text{select}}[4,8,6,5]$ following the color ordering of $\PT(\pmb \b)$ in the first row of~\eqref{selectExam2}. It leads to four possible multiplications, and by direct computation, they produce the same result as
\small
\bea \left.{\renewcommand*{\arraystretch}{1.5} \begin{array}{l}
		\big(\PT(\pmb \a)f^{\text{select}}[6,7,8,1]f^{\text{select}}[5,6,8,1]\big)\times \PT(\pmb \b)\\
		\big(\PT(\pmb \a)f^{\text{select}}[6,7,8,1]\big)\times \big(\PT(\pmb \b)f^{\text{select}}[4,8,6,5]\big)\\
		\big(\PT(\pmb \a)f^{\text{select}}[5,6,8,1]\big)\times \big(\PT(\pmb \b)f^{\text{select}}[4,8,7,6]\big)\\
		\PT(\pmb \a)\times \big(\PT(\pmb \b)f^{\text{select}}[4,8,7,6]f^{\text{select}}[4,8,6,5]\big)
\end{array}}\right\} \to \frac{1}{s_{1234}}\times\frac{1}{s_{78}s_{678}}\times \Bigg(\begin{array}{c}
    \begin{tikzpicture}[scale=0.5]
                     \draw[](0,0)--(1,0) (0.5,1)--(0,0)--(0.5,-1) (-0.7,0.7)--(0,0)--(-0.7,-0.7);
                     \node at (1.2,0.35) []{{\footnotesize $P_{1234}$}};
                     \node at (0.5,-1.35) []{{\footnotesize $1$}};
                     \node at (-0.75,-1.1) []{{\footnotesize $2$}};
                     \node at (-0.75,1.1) []{{\footnotesize $3$}};
                     \node at (0.5,1.35) []{{\footnotesize $4$}};
                   \end{tikzpicture} \end{array}\Bigg)~.~~~
\label{selecttwopoles}
\eea\normalsize
We note that the first and last of them do not end up directly with pure PT-factors. Nontrivial identities are required to further reduce the results.
Thus, multiplying the selecting factor is a little bit broader than the
situation discussed in the previous subsection.

We can further pick up terms with, say, $\frac{1}{s_{12}}$ pole, from
the previous result. It can be checked that the following eight
multiplications of selecting factors
\small
\begin{align*}
		&\left(\PT(\pmb{\a})f[6,7,8,1]f[5,6,8,1]f[8,1,2,3]\right)\times \PT(\pmb{\b})& &\left(\PT(\pmb{\a})f[6,7,8,1]f[5,6,8,1]\right)\times \left(\PT(\pmb{\b}) f[5,1,2,3]\right)\\
		&\left(\PT(\pmb{\a})f[6,7,8,1]f[8,1,2,3]\right)\times \left(\PT(\pmb{\b})f[4,8,6,5]\right)& &\left(\PT(\pmb{\a})f[6,7,8,1]\right)\times \left(\PT(\pmb{\b})f[4,8,6,5]f[5,1,2,3]\right)\\
		&\left(\PT(\pmb{\a})f[5,6,8,1]f[8,1,2,3]\right)\times \left(\PT(\pmb{\b})f[4,8,7,6]\right)& &\left(\PT(\pmb{\a})f[5,6,8,1]\right)\times \left(\PT(\pmb{\b})f[4,8,7,6]f[5,1,2,3]\right)\\
		&\left( \PT(\pmb{\a})f[8,1,2,3]\right)\times \left(\PT(\pmb{\b})f[4,8,7,6]f[4,8,6,5]\right)& &\left( \PT(\pmb{\a})\right)\times \left(\PT(\pmb{\b})f[4,8,7,6]f[4,8,6,5]f[5,1,2,3]\right)
\end{align*}\normalsize
indeed evaluate to $\frac{1}{s_{1234}}\left( \frac{1}{s_{12} s_{34}}
+\frac{1}{s_{12} s_{123}}\right)\times \frac{1}{s_{78} s_{678}}$. So we have extracted all terms with poles $\frac{1}{s_{12}}\frac{1}{s_{78}}\frac{1}{s_{678}}$ from the result~\eqref{selectExam1}.

As another illustration, let us apply the above discussion to six point CHY-integrands with $\PT(\pmb \a)=\Spaa{123456}$ fixed, such that we can examine relations between the independent $\PT(\pmb \beta)$'s. Starting from the identity PT-factor $\PT(\pmb\beta)=\Spaa{123456}$, we can choose the selecting factors to corresponds to
the independent Mandelstam variables $\{s_{12}, s_{23},s_{34},s_{45},s_{56},s_{61}\}$ and $\{s_{123},s_{234},s_{345}\}$. Explicitly, we have

\bea \Spaa{123456}\times \left\{ \begin{array}{c}
	f^{\text{select}}[6,1,2,3]_{s_{12}}\to \Spaa{126543}\\
	f^{\text{select}}[1,2,3,4]_{s_{23}}\to \Spaa{132456}\\
	f^{\text{select}}[2,3,4,5]_{s_{34}}\to \Spaa{124356}\\
	f^{\text{select}}[3,4,5,6]_{s_{45}}\to \Spaa{123546}\\
	f^{\text{select}}[4,5,6,1]_{s_{56}}\to \Spaa{123465}\\
	f^{\text{select}}[5,6,1,2]_{s_{61}}\to \Spaa{154326}\\
\end{array}\right.~~~,~~~\Spaa{123456}\times \left\{ \begin{array}{c}
	f^{\text{select}}[6,1,3,4]_{s_{123}}\to \Spaa{123654}\\
	f^{\text{select}}[1,2,4,5]_{s_{234}}\to \Spaa{143256}\\
	f^{\text{select}}[2,3,5,6]_{s_{345}}\to \Spaa{125436}\\
\end{array}\right.~,~~~\label{cross6pt1}\eea
where the subscripts are to remind which pole we are picking up. The left column shows the results of picking up the terms with a common two-particle pole $\frac{1}{s_{i,i+1}}$, while the right column shows the results of picking up terms with a common three-particle pole $\frac{1}{s_{i,i+1,i+2}}$. It can be simply checked that the six resulting PT-factors in the left column are those evaluating to five diagrams, while the three in the right column are those evaluating to four diagrams, as presented in \S\ref{secFeynmanSub3}.

Based on above results, we can further consider multiplying one more selecting factor. For example, let us take the resulting PT-factor  $\Spaa{123465}$ from the left column and $\Spaa{123654}$ from the right column. For $\Spaa{123465}$, it has overall pole $\frac{1}{s_{56}}$ in the evaluated result. From compatibility condition, the selecting factors we can take are those corresponding to poles $\frac{1}{s_{12}}$, $\frac{1}{s_{23}}$, $\frac{1}{s_{34}}$, $\frac{1}{s_{123}}$ and $\frac{1}{s_{234}}$. For $\Spaa{123654}$, it has overall pole $\frac{1}{s_{123}}$ in the evaluated result, and from compatibility condition the selecting factors we can take are those corresponding to poles $\frac{1}{s_{12}}$, $\frac{1}{s_{23}}$, $\frac{1}{s_{45}}$ and $\frac{1}{s_{56}}$. Explicitly, we have
\bea \Spaa{123465}\times \left\{ \begin{array}{c}
	f^{\text{select}}[5,1,2,3]_{s_{12}}\to \Spaa{125643}\\
	f^{\text{select}}[1,2,3,4]_{s_{23}}\to \Spaa{132465}\\
	f^{\text{select}}[2,3,4,6]_{s_{34}}\to \Spaa{124365}\\
	f^{\text{select}}[5,1,3,4]_{s_{123}}\to \Spaa{123564}\\
	f^{\text{select}}[1,2,4,6]_{s_{234}}\to \Spaa{143265}\\
\end{array}\right.~~~,~~~\Spaa{123654}\times \left\{\begin{array}{c}
	f^{\text{select}}[4,1,2,3]_{s_{12}}\to \Spaa{124563}\\
	f^{\text{select}}[1,2,3,6]_{s_{23}}\to \Spaa{132654}\\
	f^{\text{select}}[6,5,4,1]_{s_{45}}\to \Spaa{123645}\\
	f^{\text{select}}[3,6,5,4]_{s_{56}}\to \Spaa{123564}\\
\end{array}\right.~.~~~\label{cross6pt2}\eea
It can be checked that all the resulting PT-factors evaluate to two Feynman diagrams. In fact, all the six resulting PT-factors in the left column of~\eqref{cross6pt1} can be treated in the same manner as $\Spaa{123465}$, which lead to $6\times 5=30$ PT-factors. All the three resulting PT-factors in the right column of~\eqref{cross6pt1} can be treated in the same manner as $\Spaa{123654}$, which in all produce $3\times 4=12$ PT-factors. However, we have over-counted each PT-factor by one since the result is independent of the order of picking up poles.
Therefore, the independent PT-factors that can be produced from all resulting PT-factors in~\eqref{cross6pt1} by multiplying another selecting factor should be $\frac{6\times 5+3\times 4}{2!}=21$, which is exactly the number of independent PT-factors that evaluated to two Feynman diagrams.\footnote{Alternatively, we can pick up the two poles from $\Spaa{123456}$ at the same time, similar to our example~\eqref{selecttwopoles}. The resultant integrands will be different at the first glance, but still evaluate to the same amplitudes.} Indeed, it can be checked that the $21$ resulting PT-factors as partly shown in~\eqref{cross6pt2}, together with the other not written down, are just those evaluated to two diagrams as presented in~\S\ref{secFeynmanSub3}.

Again, based on above result, we can multiply one more selecting factor that is compatible with the previous two.
The resulting PT-factors from~\eqref{cross6pt2} can be shown as
\small
\begin{align}
 \begin{array}{l}
	\Spaa{125643}\times \left\{ \begin{array}{c}
		f^{\text{select}}[6,4,3,1]_{s_{34}}\to \Spaa{125634}\\
		f^{\text{select}}[4,3,2,5]_{s_{123}}\to \Spaa{124653}
	\end{array}\right.\\
	\Spaa{132465}\times \left\{\begin{array}{c}
		f^{\text{select}}[5,1,2,4]_{s_{123}}\to \Spaa{132564}\\
		f^{\text{select}}[1,3,4,6]_{s_{234}}\to \Spaa{142365}
	\end{array}\right.\\
	\Spaa{124365}\times \left\{\begin{array}{c}
		f^{\text{select}}[5,1,2,4]_{s_{12}}\to \Spaa{125634}\\
		f^{\text{select}}[1,2,3,6]_{s_{234}}\to \Spaa{134265}
	\end{array}\right.\\
	\Spaa{123564}\times \left\{\begin{array}{c}
		f^{\text{select}}[4,1,2,3]_{s_{12}}\to \Spaa{124653}\\
		f^{\text{select}}[1,2,3,5]_{s_{23}}\to \Spaa{132564}
	\end{array}\right.\\
	\Spaa{143265}\times \left\{\begin{array}{c}
		f^{\text{select}}[4,3,2,6]_{s_{23}}\to \Spaa{142365}\\
		f^{\text{select}}[1,4,3,2]_{s_{34}}\to \Spaa{134265}
	\end{array}\right.\\
\end{array}~~~,~~~\begin{array}{l}
	\Spaa{124563}\times \left\{\begin{array}{c}
		f^{\text{select}}[2,4,5,6]_{s_{45}}\to \Spaa{125463}\\
		f^{\text{select}}[4,5,6,3]_{s_{56}}\to \Spaa{124653}
	\end{array}\right.\\
	\Spaa{132654}\times \left\{\begin{array}{c}
		f^{\text{select}}[6,5,4,1]_{s_{45}}\to \Spaa{132645}\\
		f^{\text{select}}[2,6,5,4]_{s_{56}}\to \Spaa{132564}
	\end{array}\right.\\
	\Spaa{123645}\times \left\{\begin{array}{c}
		f^{\text{select}}[5,1,2,3]_{s_{12}}\to \Spaa{125463}\\
		f^{\text{select}}[1,2,3,6]_{s_{23}}\to \Spaa{132645}
	\end{array}\right.\\
	\Spaa{123564}\times \left\{\begin{array}{c}
		f^{\text{select}}[4,1,2,3]_{s_{12}}\to \Spaa{124653}\\
		f^{\text{select}}[1,2,3,5]_{s_{23}}\to \Spaa{132564}
	\end{array}\right.\\
\end{array}~.~~~\label{cross6pt3}
\end{align}\normalsize
We notice that each PT-factor in~\eqref{cross6pt2} has two compatible selecting factors, leading to two new PT-factors, all of which evaluate to one Feynman diagram. Since from~\eqref{cross6pt1} to~\eqref{cross6pt2}, we get $42$ PT-factors (of which $21$ are independent), in this step we will get $42\times 2=84$ PT-factors. After excluding the double counting, we get $\frac{42\times 2}{3!}=14$ independent PT-factors,
which are exactly the PT-factors that evaluated to one Feynman diagram. Our results~\eqref{cross6pt3}, together with those not written down, agree with the $14$ independent PT-factors that evaluated to one Feynman diagram as presented in \S\ref{secFeynmanSub3}, after deleting the double counting.
Since now every pole is overall, if we multiply another selecting factor, we will either get an irrelevant result or return to the mother amplitude.

Before closing, we give a criterion on whether a pole $s_A$ is overall with fixed $\PT(\pmb\a)$. First, $s_A$ is a physical pole {\sl iff} $A$ is consecutive with both $\PT(\pmb\a)$ and $\PT(\pmb\b)$. Next, we define $\{\overline{A}_{-1},A_1,A_{-1},\overline{A}_{1}\}$ according to $\PT(\pmb\b)$, and put $\PT(\pmb\a)$ into the unique form $\PT(\pmb\a)=\Spaa{A_1\ldots A_{-1}\ldots}$. Then $s_A$ is an overall pole {\sl iff} in $\PT(\pmb\a)$ we have $A_{-1}\prec \overline{A}_{-1}\prec \overline{A}_1$, namely, $\overline{A}_{-1}$ precedes $\overline{A}_1$. Otherwise, $s_A$ is not an overall pole. This can be easily understood by the zig-zag path in $n$-gon. Since an overall pole corresponds to a partial triangulation line in the $n$-gon, a reverse of ordering must happen when the zig-zag path cross this triangulation line.

\section{Conclusion}
\label{secConclusion}

The CHY-integrand of bi-adjoint cubic scalar theory consists of two PT-factors as $\PT(\pmb{\a})\times \PT(\pmb{\b})$. Once we fix the color ordering of first PT-factor $\PT(\pmb{\a})$ as the natural ordering $\pmb{e}=\Spaa{12\cdots n}$, the second PT-factor $\PT(\pmb{\b})$ then can be interpreted as a permutation acting on the identity element. It is shown in this paper that, the pole structure and vertex information of Feynman diagrams evaluated by a CHY-integrand is completely encoded in the permutations of corresponding PT-factors. The cycle representation of permutation, which neatly organizes the external legs into disjoint cycles, manifests the pole and vertex information. More concretely, since a PT-factor is invariant under cyclic rotations and gains at most a sign $(-)^{n}$ under reversing of color ordering, we are actually considering $2n$ equivalent permutations of a PT-factor. We then write all the equivalent permutations of $\PT(\pmb\b)$ into the cycle representation, and pick out the good ones.
Those that can be separated to at least three consecutive parts with respect of $\PT(\pmb\a)$ are called V-type cycle representations. Those that can only be separated into two parts, while each part contains more than two elements, are called P-type cycle representations. We show that the CHY-integrand $\PT(\pmb{e})\times \PT(\pmb{\b})$ gives nonzero contributions if and only if the ways of planar separations allowed by all the V-type representations satisfy the constraint~\eqref{V-cond}.
The Feynman diagram of a CHY-integrand can be completely determined by one V-type cycle representation by going into its substructures, or collectively determined by all P-type cycle representations of a PT-factor. The vertex structure can be obviously seen from the planar separation of V-type or P-type cycle representations. We presented the algorithm to read out the physical poles and vertices from them.

On the other hand, given an effective Feynman diagram, with possible effective higher point vertices, we have proposed a recursive algorithm to obtain directly the correct cycle representation of corresponding PT-factor $\PT(\pmb{\b})$. We show that cycle representations of any Feynman diagram allow a factorization as Eq.~\eqref{cyclefactor} with respect to an arbitrary $m$-point vertex, called a planar splitting. We have figured out that, the cycle representations of subdiagrams that used in the factorization~\eqref{cyclefactor} are the non-planar splitting ones in the equivalent class of PT-factors of subdiagrams. The same algorithm applies to the subdiagrams as well, so we can reconstruct the cycle representation of any $n$-point PT-factor basically from three point PT-factor. We show that all the discussions are parallel in the Feynman diagram and $n$-gon diagram, while the latter also takes its role in the associahedron discussion.

It is shown in~\cite{Arkani-Hamed:2017mur} that different PT-factors
are  neatly connected in the associahedron picture. In this paper,
we also investigate the relations among different PT-factors via the
reversing permutation on cycles, which corresponds to adding or
removing triangulation line in the $n$-gon diagrams. The merging and
splitting of cycles in a cycle representation mainly select terms
with the same poles in a result. From the same thought, we further
study the relations among PT-factors via the multiplication of
certain cross-ratio factor which we call selecting factor. They all
give similar topology about how the PT-factors are connected.

Finally, since the planar diagram possess a natural interpretation as the vertices and boundaries of an associahedron, the structure of good cycle representation introduced in this paper can be used to characterize certain boundaries. We have shown how this can be achieved by merging of cycles, in the equivalent class of a $\PT(\pmb\beta)$, the number of different factorizations into disjoint permutations describes the shape of boundaries of the associahedron.

\acknowledgments

We thank Song He for helpful discussions. We also thank the hospitality of the Institute of Theoretical Physics at Chinese Academy of Science, where this work was initiated. B.F. is supported by Qiu-Shi Funding and the National Natural
Science Foundation of China (NSFC) with Grant No.11575156 and
No.11135006. R.H. is supported by the NSFC with Grant No.11575156
and the starting grant from Nanjing Normal University. F.T. is
supported in part by the Knut and Alice Wallenberg Foundation under
grant KAW 2013.0235 and the Ragnar S\"{o}derberg Foundation under
grant S1/16.

\appendix

\section{Associahedron and cycle representation of permutation}
\label{secAppendix}

As we have briefly mentioned, all the cubic scalar diagrams are related to the associahedron. More concretely, all the $n$-point PT-factors can be elegantly encoded into an associahedron denoted as $\mathcal{K}_{n-1}$, which is an $(n-3)$-dimensional convex polytope whose vertices are labeled by $n$-point cubic diagrams. The two vertices connected by an edge represent the two  diagrams that share the same $(n-4)$ internal edges, such that we can label the edges with those diagrams with $(n-4)$ cubic vertices and one quartic vertex. In general, an $k$-dimensional boundary consists of vertices that share the same $(n-3-k)$ internal edges. The number of $k$-dimensional boundaries in $\mathcal{K}_{n-1}$ is given by the equation\footnote{See \url{https://oeis.org/A033282}}
\begin{equation}
T(n,k)=\begin{pmatrix}
k \\ n-3 \\
\end{pmatrix}\begin{pmatrix}
n-k-3 \\ 2n-k-4 \\
\end{pmatrix}\frac{1}{n-k-2}~.
\end{equation}
All these boundaries correspond to diagrams with higher point vertices. They are thus characterized by the vertex number vector
\begin{equation}
\pmb{v}\equiv(v_3,v_4,v_5,\ldots, v_n)~,
\end{equation}
where $v_i$ satisfies the constraint of \eqref{eq:vertexrel}. For instance, at $n=6$, the three-dimensional associahedron $\mathcal{K}_5$ has two kinds of faces (two-dimensional boundaries): pentagon ($\mathcal{K}_4$) and rectangle ($\mathcal{K}_3\times \mathcal{K}_3$). They correspond to the diagrams with $\pmb{v}=(1,0,1,0)$ and $(0,2,0,0)$ respectively. All the boundaries of $\mathcal{K}_{n-1}$ can be obtained by direct products of lower dimensional associahedrons. We use $N_{n}(v_3,v_4,\ldots, v_n)$ to denote the number of boundaries $\mathcal{K}_{n-1}$ that are of the form
\begin{equation}
\underbrace{\mathcal{K}_2\times\cdots\times\mathcal{K}_2}_{v_3}\times\underbrace{\mathcal{K}_3\times\cdots\times\mathcal{K}_3}_{v_4}\times\underbrace{\mathcal{K}_4\times\cdots\times\mathcal{K}_4}_{v_5}\times\cdots~.
\end{equation}
In particular, $N_n(n-2,0,\ldots, 0)$ gives the number of vertices of $\mathcal{K}_{n-1}$, which is the Catalan number $C_n=\frac{(2n-4)!}{(n-1)!(n-2)!}$, and $N_{n}(0,0,\ldots, 0,1)=1$ just stands for the $\mathcal{K}_{n-1}$ itself.

Knowing the relation between cycle representations and Feynman diagrams, we can give the number $N_n(v_3,v_4,\ldots, v_n)$ another interpretation. From \S\ref{secPermutation2}, we know that a PT-factor $\PT(\pmb\b)$ corresponds to a Feynman diagram with $\pmb{v}=(v_3,v_4\ldots v_n)$ only if
among the $(2n)$ elements of its equivalent class, there are exactly $v_i$ different cycle partitions $\{\pmb\beta_{\mathsf{A}_1}^{\text{cyc-rep}},\ldots,\pmb\beta_{\mathsf{A}_i}^{\text{cyc-rep}}\}$ such that
$$\pmb\beta_{\mathsf{A}_1}^{\text{cyc-rep}}\pmb\beta_{\mathsf{A}_2}^{\text{cyc-rep}}\cdots\pmb\beta_{\mathsf{A}_i}^{\text{cyc-rep}}\in \mathfrak{b}[\pmb\beta]~~,~~ 3\leqslant i\leqslant n\,,$$
where each $\pmb{\beta}_{\mathsf{A}_{i}}^{\text{cyc-rep}}$ can not be further split planarly. Then the number $N_n(\pmb{v})$ simply counts the number of such permutations. To clarify this statement, a few typical examples are in order. First, at $n=6$, the PT-factor $\langle 123465\rangle$ corresponds to a diagram with two quartic vertices, namely, $\pmb{v}=(0,2,0,0)$,
\begin{equation}
\adjustbox{raise=-1cm}{\begin{tikzpicture}[every node/.style={font=\fontsize{8pt}{5pt}\selectfont},scale=0.8]
	\draw (0,0) node[left=0pt]{$2$} -- (3,0) node[right=0pt]{$5$};
	\draw (1,-1) node[below=0pt]{$1$} -- (1,1) node[above=0pt]{$3$};
	\draw (2,-1) node[below=0pt]{$6$} -- (2,1) node[above=0pt]{$4$};
	\end{tikzpicture}}\;\Longleftrightarrow\;\PT(\pmb\beta)=\langle 123465\rangle\,.
\end{equation}
In the equivalent class of $\langle 123465\rangle$, there are two cycle representations that can be split planarly, and there are in all two ways to split them into four planar parts,
\bea \begin{array}{l}
       (1)\pmb{|}(2)\pmb{|}(3)\pmb{|}(46)(5)\Longrightarrow\pmb{\beta}_{\mathsf{A}_1}^{\text{cyc-rep}}=(1)~~,~~\pmb{\beta}_{\mathsf{A}_2}^{\text{cyc-rep}}=(2)~~,~~\pmb{\beta}_{\mathsf{A}_3}^{\text{cyc-rep}}=(3)~~,~~\pmb{\beta}_{\mathsf{A}_4}^{\text{cyc-rep}}=(46)(5) \\
       (13)(2)\pmb{|}(4)\pmb{|}(5)\pmb{|}(6)\Longrightarrow\pmb{\beta}_{\mathsf{A}_1}^{\text{cyc-rep}}=(13)(2)~~,~~\pmb{\beta}_{\mathsf{A}_2}^{\text{cyc-rep}}=(4)~~,~~\pmb{\beta}_{\mathsf{A}_3}^{\text{cyc-rep}}=(5)~~,~~\pmb{\beta}_{\mathsf{A}_4}^{\text{cyc-rep}}=(6)
     \end{array}~.~~~\label{eq:6p4g}
\eea
Now comparing with Eq.~\eqref{eq:6p4g}, we see that the two four-part planar splitting is due to the fact that the corresponding Feynman diagram consist of two quartic vertices.
Since each subdiagram can also be factorized to sub-subdiagrams, this allows a recursive computation for $N_n(\pmb{v})$, down to the pieces where only cubic vertices exist and only length-1 or length-2 cycles appear in the cycle representation. Thus $\langle 123465\rangle$ contributes to the counting $N_6(0,2,0,0)$. Next, we consider an eight point example with $\pmb{v}=(2,2,0,0,0,0)$,
\begin{equation}
\adjustbox{raise=-1cm}{\begin{tikzpicture}[every node/.style={font=\fontsize{8pt}{5pt}\selectfont},scale=0.8]
	\draw (0.5,0) -- (2.5,0);
	\draw (0.5,0) -- ++(135:0.5) node[left=0pt]{$2$} (0.5,0) -- ++(-135:0.5) node[left=0pt]{$1$};
	\draw (2.5,0) -- ++(45:0.5) node[right=0pt]{$5$} (2.5,0) -- ++(-45:0.5) node[right=0pt]{$6$};
	\draw (1,-1) node[below=0pt]{$8$} -- (1,1) node[above=0pt]{$3$};
	\draw (2,-1) node[below=0pt]{$7$} -- (2,1) node[above=0pt]{$4$};
	\end{tikzpicture}}\;\Longleftrightarrow\;\PT(\pmb\beta)=\langle 12846573\rangle\,.
\end{equation}
In the equivalent class of $\langle 12846573\rangle$, there are again two cycle representations that have planar splittings, but this time there are two ways to split them into three parts (blue partitions), and two ways to split them into four parts (red partitions),
\begin{align}
& (1)\textcolor{blue}{\pmb{|}}(2)\textcolor{blue}{\pmb{|}}(38)\textcolor{red}{\pmb{|}}(4)\textcolor{red}{\pmb{|}}(56)\textcolor{red}{\pmb{|}}(7)~~~,~~~(8)\textcolor{red}{\pmb{|}}(12)\textcolor{red}{\pmb{|}}(3)\textcolor{red}{\pmb{|}}(47)\textcolor{blue}{\pmb{|}}(5)\textcolor{blue}{\pmb{|}}(6)~.~~~
\end{align}
Again, the cycle splitting pattern agrees exactly with the vertex number vector $(2,2,0,0,0,0)$. Thus the PT-factor $\langle 12846573\rangle$ contributes to the counting $N_8(2,2,0,0,0,0)$. The above machinery can also be used to pick out those PT-factors that do not give any Feynman diagram. We again illustrate this point by a few examples first. At $n=6$, we consider the PT-factor $\langle 124635\rangle$. In its equivalent class, there is only one three-part cycle partition,
\begin{equation}
(1)\pmb{|}(2)\pmb{|}(3465)~.~~~
\end{equation}
Had this PT-factor given any Feynman diagram, the vertex number vector would be $\pmb{v}=(1,0,0,0)$. Since this $\pmb{v}$ does not satisfy the constraint~\eqref{eq:vertexrel}, $\langle 124635\rangle$ does not correspond to any Feynman diagram compatible with the planar order $\Spaa{123456}$. Similarly, the PT-factor $\langle 135264\rangle$ does not have any valid cycle partition (namely, more than three parts) in its equivalent class, it must also give zero Feynman diagram. In general, by inspecting how the cycle representations of a PT-factor split, we can get a vertex number vector $\pmb{v}$. If this $\pmb{v}$ fails to satisfy the constraint~\eqref{eq:vertexrel}, the PT-factor under consideration must give zero Feynman diagram.

In order to write down a recursive formula for $N_n$, let us consider the following factorization of Feynman diagram. Starting from an arbitrary external leg, it would connect to, say, a $(s+1)$-point vertex. This vertex splits the diagram into $(s+1)$ subdiagrams. One of the subdiagram is just the single external leg we started from, which does not contain any other external legs of the Feynman diagram. The remaining $(n-1)$ external legs are separated into the other $s$ subdiagrams. Let us assume that the number of legs is $x_i$ for each subdiagram $i$, with $1\leqslant i\leqslant s$. Since each subdiagram consists of external legs and an internal propagator, we would have the following constraint
\begin{equation}
\sum_{i=1}^{s}x_i=(n-1)+s~.\label{constraint1}
\end{equation}
If the vertex number vector for each subdiagram is $\pmb{u}_i$, then we have the following constraint
\bea \sum_{i=1}^{s}\pmb{u}_i=(v_3,v_4,\ldots,v_s,v_{s+1}-1,v_{s+2},\ldots,v_n)~,~~~\label{constraint2}\eea
where $\pmb{v}=(v_3,v_4,\ldots,v_n)$ is the vertex number vector of the complete Feynman diagram, and $v_m$ is the number of the $m$-point vertices. The meaning of Eq.~\eqref{constraint2} is clear: summing over the numbers of $m$-point vertices in all the subdiagrams should give the number of $m$-point vertices in the complete Feynman diagram, except that the number of $(s+1)$-point vertex should be reduced by one, since the $(s+1)$-point vertex that connects the starting external leg has been split and does not belong to any subdiagram\footnote{The $\pmb{u}_i$ in (\ref{constraint2}) is the vertex number vector of $x_i$-point subdiagram, so it has $x_i-2$ components $\pmb{u}_i=(u_{i3},u_{i4},\ldots,u_{ix_i})$. When we equal the left and right side of (\ref{constraint2}), all the $u_{ik}$ with $x_i+1\leqslant k\leqslant n$ are automatically understood to be zero.}. Then $N_n$ can be recursively computed from those $N_{x_i}$'s of the subdiagrams as,
\begin{align}
\label{eq:N_nRecursion}
N_{n}(v_3,v_4,\ldots, v_n)=\sum_{s=2}^{n-1}\,\sum_{\{x_i\}}\,\sum_{\{\pmb{u}_i\}}N_{x_1}(\pmb{u}_1)N_{x_2}(\pmb{u}_2)\cdots N_{x_s}(\pmb{u}_s)~,~~~
\end{align}
where for each $2\leqslant s\leqslant n-2$, the summation over $\{x_i\}$ and $\{\pmb{u}_i\}$ are constrained respectively by Eq.~\eqref{constraint1} and \eqref{constraint2}, and $s$ takes the value $k$ whenever $v_{k+1}$ is not zero. We note that by definition the vector $\pmb{u}_i$ satisfies the constraint,
\bea \sum_{m=3}^{x_i}(m-2)u_{im}=x_i-2~.\eea
\begin{table}[t]
	\centering
	\begin{tabular}{|c|c|c|c|c|ccccc}
		\hline
		\multicolumn{5}{|c|}{$n=6$}                                                          & \multicolumn{5}{c|}{$n=8$}                                                                                                                                                                        \\ \hline
		$\pmb{v}$     & $F$  & $N_n(\pmb{v})$ & $T(n,k)$              & $k$                  & \multicolumn{1}{c|}{$\pmb{v}$}       & \multicolumn{1}{c|}{$F$}   & \multicolumn{1}{c|}{$N_n(\pmb{v})$} & \multicolumn{1}{c|}{$T(n,k)$}               & \multicolumn{1}{c|}{$k$}                  \\ \hline
		$(4,0,0,0)$   & $1$  & $14$           & $14$                  & $0$                  & \multicolumn{1}{c|}{$(6,0,0,0,0,0)$} & \multicolumn{1}{c|}{$1$}   & \multicolumn{1}{c|}{$132$}          & \multicolumn{1}{c|}{$132$}                  & \multicolumn{1}{c|}{$0$}                  \\ \hline
		$(2,1,0,0)$   & $2$  & $21$           & $21$                  & $1$                  & \multicolumn{1}{c|}{$(4,1,0,0,0,0)$} & \multicolumn{1}{c|}{$2$}   & \multicolumn{1}{c|}{$330$}          & \multicolumn{1}{c|}{$330$}                  & \multicolumn{1}{c|}{$1$}                  \\ \hline
		$(1,0,1,0)$   & $5$  & $6$            & \multirow{2}{*}{$9$}  & \multirow{2}{*}{$2$} & \multicolumn{1}{c|}{$(3,0,1,0,0,0)$} & \multicolumn{1}{c|}{$5$}   & \multicolumn{1}{c|}{$120$}          & \multicolumn{1}{c|}{\multirow{2}{*}{$300$}} & \multicolumn{1}{c|}{\multirow{2}{*}{$2$}} \\ \cline{1-3} \cline{6-8}
		$(0,2,0,0)$   & $4$  & $3$            &                       &                      & \multicolumn{1}{c|}{$(2,2,0,0,0,0)$} & \multicolumn{1}{c|}{$4$}   & \multicolumn{1}{c|}{$180$}          & \multicolumn{1}{c|}{}                       & \multicolumn{1}{c|}{}                     \\ \hline
		$(0,0,0,1)$   & $14$ & $1$            & $1$                   & $3$                  & \multicolumn{1}{c|}{$(2,0,0,1,0,0)$} & \multicolumn{1}{c|}{$14$}  & \multicolumn{1}{c|}{$36$}           & \multicolumn{1}{c|}{\multirow{3}{*}{$120$}} & \multicolumn{1}{c|}{\multirow{3}{*}{$3$}} \\ \cline{1-8}
		\multicolumn{5}{|c|}{$n=7$}                                                          & \multicolumn{1}{c|}{$(1,1,1,0,0,0)$} & \multicolumn{1}{c|}{$10$}  & \multicolumn{1}{c|}{$72$}           & \multicolumn{1}{c|}{}                       & \multicolumn{1}{c|}{}                     \\ \cline{1-8}
		$\pmb{v}$     & $F$  & $N_n(\pmb{v})$ & $T(n,k)$              & $k$                  & \multicolumn{1}{c|}{$(0,3,0,0,0,0)$} & \multicolumn{1}{c|}{$8$}   & \multicolumn{1}{c|}{$12$}           & \multicolumn{1}{c|}{}                       & \multicolumn{1}{c|}{}                     \\ \hline
		$(5,0,0,0,0)$ & $1$  & $42$           & $42$                  & $0$                  & \multicolumn{1}{c|}{$(1,0,0,0,1,0)$} & \multicolumn{1}{c|}{$42$}  & \multicolumn{1}{c|}{$8$}            & \multicolumn{1}{c|}{\multirow{3}{*}{$20$}}  & \multicolumn{1}{c|}{\multirow{3}{*}{$4$}} \\ \cline{1-8}
		$(3,1,0,0,0)$ & $2$  & $84$           & $84$                  & $1$                  & \multicolumn{1}{c|}{$(0,2,0,1,0,0)$} & \multicolumn{1}{c|}{$28$}  & \multicolumn{1}{c|}{$8$}            & \multicolumn{1}{c|}{}                       & \multicolumn{1}{c|}{}                     \\ \cline{1-8}
		$(2,0,1,0,0)$ & $5$  & $28$           & \multirow{2}{*}{$56$} & \multirow{2}{*}{$2$} & \multicolumn{1}{c|}{$(0,0,2,0,0,0)$} & \multicolumn{1}{c|}{$25$}  & \multicolumn{1}{c|}{$4$}            & \multicolumn{1}{c|}{}                       & \multicolumn{1}{c|}{}                     \\ \cline{1-3} \cline{6-10}
		$(1,2,0,0,0)$ & $4$  & $28$           &                       &                      & \multicolumn{1}{c|}{$(0,0,0,0,0,1)$} & \multicolumn{1}{c|}{$132$} & \multicolumn{1}{c|}{$1$}            & \multicolumn{1}{c|}{$1$}                    & \multicolumn{1}{c|}{$5$}                  \\ \hline
		$(1,0,0,1,0)$ & $14$ & $7$            & \multirow{2}{*}{$14$} & \multirow{2}{*}{$3$} &                                      &                            &                                     & \multicolumn{1}{l}{}                        & \multicolumn{1}{l}{}                      \\ \cline{1-3}
		$(0,1,1,0,0)$ & $10$ & $7$            &                       &                      &                                      &                            &                                     & \multicolumn{1}{l}{}                        & \multicolumn{1}{l}{}                      \\ \cline{1-5}
		$(0,0,0,0,1)$ & $42$ & $1$            & $1$                   & $4$                  &                                      &                            &                                     & \multicolumn{1}{l}{}                        & \multicolumn{1}{l}{}                      \\ \cline{1-5}
	\end{tabular}
	\caption{\label{table1} $N_n(\pmb{v})$ and the corresponding $T(n,k)$ for $n=6$, $7$ and $8$. In addition, $F$ gives the number of trivalent Feynman diagrams. }
\end{table}

In addition, we also require that each $x_i\geqslant 2$ and each component of $\pmb{u}_{i}$ is non-negative. The recursion starts at $N_2(0)=1$.
If $\pmb{v}=(n-2,0,\ldots, 0)$, i.e., all $v_i=0$ except that $v_3=n-2$, then $s$ can only take $s=2$ in~\eqref{eq:N_nRecursion}, which gives nothing but the recusion for the Catalan numbers. For low multiplicity, one can easily find that,
\begin{align}
& N_3(1)=1~, & & N_4(2,0)=2~, & & N_5(3,0,0)=5~, \nonumber\\
& & & N_4(0,1)=1~, & & N_5(1,1,0)=5~, \\
& & & & & N_5(0,0,1)=1~.~~~\nonumber
\end{align}
The cases of $n=6$, $7$ and $8$ are shown in Table~\ref{table1},
where we can easily check that the values of $N_6(\pmb{v})$, for example, agree with the counting of Feynman diagrams. By knowing the values for lower multiplicities, we can easily calculate, for example,
\begin{align}
N_7(0,1,1,0,0)&=N_2(0)N_2(0)N_5(0,0,1)+N_2(0)N_5(0,0,1)N_2(0)+N_5(0,0,1)N_2(0)N_2(0)\nonumber\\
&\quad+N_2(0)N_2(0)N_2(0)N_4(0,1)+N_2(0)N_2(0)N_4(0,1)N_2(0)\nonumber\\
&\quad+N_2(0)N_4(0,1)N_2(0)N_2(0)+N_4(0,1)N_2(0)N_2(0)N_2(0)\nonumber\\
&=7~.~~~
\end{align}
The value of $N_n$, as a complement of $T(n,k)$, further distinguishes different shapes of $k$-dimensional boundary of associahedron $\mathcal{K}_{n-1}$.


\bibliographystyle{JHEP}
\bibliography{permutation}

\providecommand{\href}[2]{#2}\begingroup\raggedright\begin{thebibliography}{10}

\bibitem{Cachazo:2013iaa}
F.~Cachazo, S.~He and E.~Y. Yuan, \emph{{Scattering in Three Dimensions from
  Rational Maps}}, \href{https://doi.org/10.1007/JHEP10(2013)141}{\emph{JHEP}
  {\bfseries 10} (2013) 141},
  [\href{https://arxiv.org/abs/1306.2962}{{\ttfamily 1306.2962}}].

\bibitem{Cachazo:2013gna}
F.~Cachazo, S.~He and E.~Y. Yuan, \emph{{Scattering equations and
  Kawai-Lewellen-Tye orthogonality}},
  \href{https://doi.org/10.1103/PhysRevD.90.065001}{\emph{Phys. Rev.}
  {\bfseries D90} (2014) 065001},
  [\href{https://arxiv.org/abs/1306.6575}{{\ttfamily 1306.6575}}].

\bibitem{Cachazo:2013hca}
F.~Cachazo, S.~He and E.~Y. Yuan, \emph{{Scattering of Massless Particles in
  Arbitrary Dimensions}},
  \href{https://doi.org/10.1103/PhysRevLett.113.171601}{\emph{Phys.Rev.Lett.}
  {\bfseries 113} (2014) 171601},
  [\href{https://arxiv.org/abs/1307.2199}{{\ttfamily 1307.2199}}].

\bibitem{Cachazo:2013iea}
F.~Cachazo, S.~He and E.~Y. Yuan, \emph{{Scattering of Massless Particles:
  Scalars, Gluons and Gravitons}},
  \href{https://doi.org/10.1007/JHEP07(2014)033}{\emph{JHEP} {\bfseries 1407}
  (2014) 033}, [\href{https://arxiv.org/abs/1309.0885}{{\ttfamily 1309.0885}}].

\bibitem{Mason:2013sva}
L.~Mason and D.~Skinner, \emph{{Ambitwistor strings and the scattering
  equations}}, \href{https://doi.org/10.1007/JHEP07(2014)048}{\emph{JHEP}
  {\bfseries 07} (2014) 048},
  [\href{https://arxiv.org/abs/1311.2564}{{\ttfamily 1311.2564}}].

\bibitem{Geyer:2014fka}
Y.~Geyer, A.~E. Lipstein and L.~J. Mason, \emph{{Ambitwistor Strings in Four
  Dimensions}},
  \href{https://doi.org/10.1103/PhysRevLett.113.081602}{\emph{Phys. Rev. Lett.}
  {\bfseries 113} (2014) 081602},
  [\href{https://arxiv.org/abs/1404.6219}{{\ttfamily 1404.6219}}].

\bibitem{Casali:2016atr}
E.~Casali and P.~Tourkine, \emph{{On the null origin of the ambitwistor
  string}}, \href{https://doi.org/10.1007/JHEP11(2016)036}{\emph{JHEP}
  {\bfseries 11} (2016) 036},
  [\href{https://arxiv.org/abs/1606.05636}{{\ttfamily 1606.05636}}].

\bibitem{Siegel:2015axg}
W.~Siegel, \emph{{Amplitudes for left-handed strings}},
  \href{https://arxiv.org/abs/1512.02569}{{\ttfamily 1512.02569}}.

\bibitem{Li:2017emw}
Y.~Li and W.~Siegel, \emph{{Chiral Superstring and CHY Amplitude}},
  \href{https://arxiv.org/abs/1702.07332}{{\ttfamily 1702.07332}}.

\bibitem{Berkovits:2013xba}
N.~Berkovits, \emph{{Infinite Tension Limit of the Pure Spinor Superstring}},
  \href{https://doi.org/10.1007/JHEP03(2014)017}{\emph{JHEP} {\bfseries 03}
  (2014) 017}, [\href{https://arxiv.org/abs/1311.4156}{{\ttfamily 1311.4156}}].

\bibitem{Gomez:2013wza}
H.~Gomez and E.~Y. Yuan, \emph{{N-point tree-level scattering amplitude in the
  new Berkovits` string}},
  \href{https://doi.org/10.1007/JHEP04(2014)046}{\emph{JHEP} {\bfseries 04}
  (2014) 046}, [\href{https://arxiv.org/abs/1312.5485}{{\ttfamily 1312.5485}}].

\bibitem{Bjerrum-Bohr:2014qwa}
N.~E.~J. Bjerrum-Bohr, P.~H. Damgaard, P.~Tourkine and P.~Vanhove,
  \emph{{Scattering Equations and String Theory Amplitudes}},
  \href{https://doi.org/10.1103/PhysRevD.90.106002}{\emph{Phys. Rev.}
  {\bfseries D90} (2014) 106002},
  [\href{https://arxiv.org/abs/1403.4553}{{\ttfamily 1403.4553}}].

\bibitem{Cachazo:2015nwa}
F.~Cachazo and H.~Gomez, \emph{{Computation of Contour Integrals on ${\cal
  M}_{0,n}$}}, \href{https://doi.org/10.1007/JHEP04(2016)108}{\emph{JHEP}
  {\bfseries 04} (2016) 108},
  [\href{https://arxiv.org/abs/1505.03571}{{\ttfamily 1505.03571}}].

\bibitem{Baadsgaard:2015voa}
C.~Baadsgaard, N.~E.~J. Bjerrum-Bohr, J.~L. Bourjaily and P.~H. Damgaard,
  \emph{{Integration Rules for Scattering Equations}},
  \href{https://doi.org/10.1007/JHEP09(2015)129}{\emph{JHEP} {\bfseries 09}
  (2015) 129}, [\href{https://arxiv.org/abs/1506.06137}{{\ttfamily
  1506.06137}}].

\bibitem{Baadsgaard:2015ifa}
C.~Baadsgaard, N.~E.~J. Bjerrum-Bohr, J.~L. Bourjaily and P.~H. Damgaard,
  \emph{{Scattering Equations and Feynman Diagrams}},
  \href{https://doi.org/10.1007/JHEP09(2015)136}{\emph{JHEP} {\bfseries 09}
  (2015) 136}, [\href{https://arxiv.org/abs/1507.00997}{{\ttfamily
  1507.00997}}].

\bibitem{Baadsgaard:2015hia}
C.~Baadsgaard, N.~E.~J. Bjerrum-Bohr, J.~L. Bourjaily, P.~H. Damgaard and
  B.~Feng, \emph{{Integration Rules for Loop Scattering Equations}},
  \href{https://doi.org/10.1007/JHEP11(2015)080}{\emph{JHEP} {\bfseries 11}
  (2015) 080}, [\href{https://arxiv.org/abs/1508.03627}{{\ttfamily
  1508.03627}}].

\bibitem{Lam:2016tlk}
C.~S. Lam and Y.-P. Yao, \emph{{Evaluation of the Cachazo-He-Yuan gauge
  amplitude}}, \href{https://doi.org/10.1103/PhysRevD.93.105008}{\emph{Phys.
  Rev.} {\bfseries D93} (2016) 105008},
  [\href{https://arxiv.org/abs/1602.06419}{{\ttfamily 1602.06419}}].

\bibitem{Huang:2016zzb}
R.~Huang, B.~Feng, M.-x. Luo and C.-J. Zhu, \emph{{Feynman Rules of
  Higher-order Poles in CHY Construction}},
  \href{https://doi.org/10.1007/JHEP06(2016)013}{\emph{JHEP} {\bfseries 06}
  (2016) 013}, [\href{https://arxiv.org/abs/1604.07314}{{\ttfamily
  1604.07314}}].

\bibitem{Cardona:2016gon}
C.~Cardona, B.~Feng, H.~Gomez and R.~Huang, \emph{{Cross-ratio Identities and
  Higher-order Poles of CHY-integrand}},
  \href{https://doi.org/10.1007/JHEP09(2016)133}{\emph{JHEP} {\bfseries 09}
  (2016) 133}, [\href{https://arxiv.org/abs/1606.00670}{{\ttfamily
  1606.00670}}].

\bibitem{Bjerrum-Bohr:2016axv}
N.~E.~J. Bjerrum-Bohr, J.~L. Bourjaily, P.~H. Damgaard and B.~Feng,
  \emph{{Manifesting Color-Kinematics Duality in the Scattering Equation
  Formalism}}, \href{https://doi.org/10.1007/JHEP09(2016)094}{\emph{JHEP}
  {\bfseries 09} (2016) 094},
  [\href{https://arxiv.org/abs/1608.00006}{{\ttfamily 1608.00006}}].

\bibitem{Huang:2017ydz}
R.~Huang, Y.-J. Du and B.~Feng, \emph{{Understanding the Cancelation of Double
  Poles in the Pfaffian of CHY-formulism}},
  \href{https://doi.org/10.1007/JHEP06(2017)133}{\emph{JHEP} {\bfseries 06}
  (2017) 133}, [\href{https://arxiv.org/abs/1702.05840}{{\ttfamily
  1702.05840}}].

\bibitem{Zhou:2017mfj}
K.~Zhou, J.~Rao and B.~Feng, \emph{{Derivation of Feynman Rules for Higher
  Order Poles Using Cross-ratio Identities in CHY Construction}},
  \href{https://doi.org/10.1007/JHEP06(2017)091}{\emph{JHEP} {\bfseries 06}
  (2017) 091}, [\href{https://arxiv.org/abs/1705.04783}{{\ttfamily
  1705.04783}}].

\bibitem{Stieberger:2016lng}
S.~Stieberger and T.~R. Taylor, \emph{{New relations for Einstein-Yang-Mills
  amplitudes}},
  \href{https://doi.org/10.1016/j.nuclphysb.2016.09.014}{\emph{Nucl. Phys.}
  {\bfseries B913} (2016) 151--162},
  [\href{https://arxiv.org/abs/1606.09616}{{\ttfamily 1606.09616}}].

\bibitem{Nandan:2016pya}
D.~Nandan, J.~Plefka, O.~Schlotterer and C.~Wen, \emph{{Einstein-Yang-Mills
  from pure Yang-Mills amplitudes}},
  \href{https://doi.org/10.1007/JHEP10(2016)070}{\emph{JHEP} {\bfseries 10}
  (2016) 070}, [\href{https://arxiv.org/abs/1607.05701}{{\ttfamily
  1607.05701}}].

\bibitem{Fu:2017uzt}
C.-H. Fu, Y.-J. Du, R.~Huang and B.~Feng, \emph{{Expansion of
  Einstein-Yang-Mills Amplitude}},
  \href{https://doi.org/10.1007/JHEP09(2017)021}{\emph{JHEP} {\bfseries 09}
  (2017) 021}, [\href{https://arxiv.org/abs/1702.08158}{{\ttfamily
  1702.08158}}].

\bibitem{Teng:2017tbo}
F.~Teng and B.~Feng, \emph{{Expanding Einstein-Yang-Mills by Yang-Mills in CHY
  frame}}, \href{https://doi.org/10.1007/JHEP05(2017)075}{\emph{JHEP}
  {\bfseries 05} (2017) 075},
  [\href{https://arxiv.org/abs/1703.01269}{{\ttfamily 1703.01269}}].

\bibitem{Du:2017kpo}
Y.-J. Du and F.~Teng, \emph{{BCJ numerators from reduced Pfaffian}},
  \href{https://doi.org/10.1007/JHEP04(2017)033}{\emph{JHEP} {\bfseries 04}
  (2017) 033}, [\href{https://arxiv.org/abs/1703.05717}{{\ttfamily
  1703.05717}}].

\bibitem{Du:2017gnh}
Y.-J. Du, B.~Feng and F.~Teng, \emph{{Expansion of All Multitrace Tree Level
  EYM Amplitudes}}, \href{https://doi.org/10.1007/JHEP12(2017)038}{\emph{JHEP}
  {\bfseries 12} (2017) 038},
  [\href{https://arxiv.org/abs/1708.04514}{{\ttfamily 1708.04514}}].

\bibitem{Bern:2008qj}
Z.~Bern, J.~J.~M. Carrasco and H.~Johansson, \emph{New relations for
  gauge-theory amplitudes},
  \href{https://doi.org/10.1103/PhysRevD.78.085011}{\emph{Phys. Rev. D}
  {\bfseries 78} (2008) 085011},
  [\href{https://arxiv.org/abs/0805.3993}{{\ttfamily 0805.3993}}].

\bibitem{Bern:2010ue}
Z.~Bern, J.~J.~M. Carrasco and H.~Johansson, \emph{{Perturbative Quantum
  Gravity as a Double Copy of Gauge Theory}},
  \href{https://doi.org/10.1103/PhysRevLett.105.061602}{\emph{Phys. Rev. Lett.}
  {\bfseries 105} (2010) 061602},
  [\href{https://arxiv.org/abs/1004.0476}{{\ttfamily 1004.0476}}].

\bibitem{Arkani-Hamed:2017mur}
N.~Arkani-Hamed, Y.~Bai, S.~He and G.~Yan, \emph{{Scattering Forms and the
  Positive Geometry of Kinematics, Color and the Worldsheet}},
  \href{https://arxiv.org/abs/1711.09102}{{\ttfamily 1711.09102}}.

\bibitem{Lam:2015sqb}
C.~S. Lam and Y.-P. Yao, \emph{{Role of M\"{o}bius constants and scattering
  functions in Cachazo-He-Yuan scalar amplitudes}},
  \href{https://doi.org/10.1103/PhysRevD.93.105004}{\emph{Phys. Rev.}
  {\bfseries D93} (2016) 105004},
  [\href{https://arxiv.org/abs/1512.05387}{{\ttfamily 1512.05387}}].

\bibitem{Gao:2017dek}
X.~Gao, S.~He and Y.~Zhang, \emph{{Labelled tree graphs, Feynman diagrams and
  disk integrals}}, \href{https://doi.org/10.1007/JHEP11(2017)144}{\emph{JHEP}
  {\bfseries 11} (2017) 144},
  [\href{https://arxiv.org/abs/1708.08701}{{\ttfamily 1708.08701}}].

\bibitem{delaCruz:2017zqr}
L.~de~la Cruz, A.~Kniss and S.~Weinzierl, \emph{{Properties of scattering forms
  and their relation to associahedra}},
  \href{https://arxiv.org/abs/1711.07942}{{\ttfamily 1711.07942}}.

\bibitem{Kenyon}
R.~{Kenyon} and J.-M. {Schlenker}, \emph{{Rhombic embeddings of planar graphs
  with faces of degree 4}}, {\emph{ArXiv Mathematical Physics e-prints} (May,
  2003) }, [\href{https://arxiv.org/abs/math-ph/0305057}{{\ttfamily
  math-ph/0305057}}].

\bibitem{Hanany:2005ss}
A.~Hanany and D.~Vegh, \emph{{Quivers, tilings, branes and rhombi}},
  \href{https://doi.org/10.1088/1126-6708/2007/10/029}{\emph{JHEP} {\bfseries
  10} (2007) 029}, [\href{https://arxiv.org/abs/hep-th/0511063}{{\ttfamily
  hep-th/0511063}}].

\bibitem{Feng:2005gw}
B.~Feng, Y.-H. He, K.~D. Kennaway and C.~Vafa, \emph{{Dimer models from mirror
  symmetry and quivering amoebae}},
  \href{https://doi.org/10.4310/ATMP.2008.v12.n3.a2}{\emph{Adv. Theor. Math.
  Phys.} {\bfseries 12} (2008) 489--545},
  [\href{https://arxiv.org/abs/hep-th/0511287}{{\ttfamily hep-th/0511287}}].

\bibitem{Feng:2016nrf}
B.~Feng, \emph{{CHY-construction of Planar Loop Integrands of Cubic Scalar
  Theory}}, \href{https://doi.org/10.1007/JHEP05(2016)061}{\emph{JHEP}
  {\bfseries 05} (2016) 061},
  [\href{https://arxiv.org/abs/1601.05864}{{\ttfamily 1601.05864}}].

\end{thebibliography}\endgroup

\end{document}